\documentclass[12pt,reqno]{amsart}

\usepackage{latexsym}
\usepackage{amsmath}
\usepackage{amscd}
\usepackage{amssymb}
\usepackage{mathrsfs}
\usepackage{comment}
\usepackage{color,framed}
\usepackage{enumerate}
\usepackage{soul}
\usepackage{stmaryrd}
\usepackage[ruled]{algorithm2e} 
\usepackage{braket}
\usepackage{tensor}

\usepackage[T1]{fontenc}
\usepackage{lmodern}
\usepackage{textcomp}
\usepackage{stmaryrd}
\SetSymbolFont{stmry}{bold}{U}{stmry}{m}{n}

\usepackage{graphicx}
\usepackage{subfigure}
\usepackage[colorlinks]{hyperref}
\usepackage{url}

\def\RR{{\mathbb R}}
\def\M{{\mathcal M}}
\def\SO{{\mathrm{SO}}}

\DeclareMathOperator{\ad}{\mathrm{ad}}
\DeclareMathOperator{\Ad}{\mathrm{Ad}}

\usepackage{listings}
\usepackage[usenames,dvipsnames,svgnames,x11names]{xcolor} 
\lstdefinestyle{mystyle}{
    language=Python,
    commentstyle=\color{DarkSlateGray4}\itshape,
    keywordstyle=\color{Green},
    morecomment=[s][\color{Red3}\itshape\ttfamily]{"""}{"""},
    basicstyle=\small\ttfamily, 
    breakatwhitespace=false,         
    breaklines=true,                 
    captionpos=b,
    columns=fullflexible,                    
    showspaces=false,                
    showstringspaces=false,
    showtabs=false,                  
    tabsize=2,
    moredelim=**[is][\color{blue}]{@}{@} 
}

\newtheorem{theorem}{Theorem}[section]

\newtheorem{example}[theorem]{Example}

\numberwithin{equation}{section}

\DeclareMathOperator*{\argmin}{arg\,min}

\begin{document}
\lstset{language=Python,style=mystyle}


\title{Differential Geometry and Stochastic Dynamics with Deep Learning Numerics}
\author[L. K\"uhnel]{Line K\"uhnel}
\author[A. Arnaudon]{Alexis Arnaudon}
\author[S. Sommer]{Stefan Sommer}

\address{LK, SS: Department of Computer Science (DIKU), University of Copenhagen,
  DK-2100 Copenhagen E, Denmark}
\address{AA: Department of Mathematics, Imperial College, London SW7 2AZ, UK}

\maketitle

\begin{abstract}
  In this paper, we demonstrate how deterministic and stochastic dynamics on manifolds, as well as differential geometric constructions can be implemented concisely and efficiently using modern computational frameworks that mix symbolic expressions with efficient numerical computations. 
  In particular, we use the symbolic expression and automatic differentiation features of the python library Theano, originally developed for high-performance computations in deep learning. 
  We show how various aspects of differential geometry and Lie group theory, connections, metrics, curvature, left/right invariance, geodesics and parallel transport can be formulated with Theano using the automatic computation of derivatives of any order.
  We will also show how symbolic stochastic integrators and concepts from non-linear statistics can be formulated and optimized with only a few lines of code.
  We will then give explicit examples on low-dimensional classical manifolds for visualization and demonstrate how this approach allows both a concise implementation and efficient scaling to high dimensional problems.
\end{abstract}


\section{Introduction}

Differential geometry extends standard calculus on Euclidean spaces to nonlinear spaces described by a manifold structure, i.e. spaces locally isomorphic to the Euclidean space \cite{lee_introduction_2003}. 
This generalisation of calculus turned out to be extremely rich in the study of manifolds and dynamical systems on manifolds. 
In the first case, being able to compute distances, curvature, and even torsion provides local information on the structure of the space. 
In the second case, the question is rather on how to write a dynamical system intrinsically on a nonlinear space, without relying on external constraints from a larger Euclidean space.
Although these constructions are general and can be rather abstract, many specific examples of both cases are used for practical applications. 
We will touch upon such examples later. 

Numerical evaluation of quantities such as curvatures and obtaining solutions of nonlinear dynamical systems constitute important problems in applied mathematics.
Indeed, high dimensional manifolds or just complicated nonlinear structures make explicit closed-form computations infeasible, even if they remain crucial for applications. 
The challenge one usually faces is not even in solving the nonlinear equations but in writing them explicitly. 
Nonlinear structures often consist of several coupled nonlinear equations obtained after multiple differentiations of elementary objects such as nontrivial metrics. 
In these cases, there is no hope of finding explicit solutions. Instead, the standard solution is to implement the complicated equations in a mathematical software packages such as Matlab or Python using numerical libraries.

In this work, we propose to tackle both issues - being able to solve the equations and being able to implement the equations numerically - by using automatic differentiation software which takes symbolic formulae as input and outputs their numerical solutions. 
Such libraries in Python includes Theano~\cite{theano}, TensorFlow~\cite{45381} and PyTorch (\url{http://pytorch.org}). 
It is important to stress that these libraries are not symbolic computer algebra packages such as Mathematica or Sympy, as they do not have any symbolic output, but rather numerical evaluation of a symbolic input. 
In this work, we chose to use Theano but similar codes can be written with other packages with automatic differentiation feature.  
The main interest for us in using Theano is that it is a fully developed package which can handle derivatives of any order, it has internal compilation and computational graph optimization features that can optimize code for multiple computer architectures (CPU, GPU), and it outputs efficient numerical code. 

It is the recent explosion of interest and impact of deep learning that has lead to the development of deep learning libraries such as Theano that mix automatic differentiation with the ability to generate extremely efficient numerical code. The work presented in this paper thus takes advantage of the significant software engineering efforts to produce robust and efficient libraries for deep learning to benefit a separate domain, computational differential geometry and dynamical systems.  We aim to present the use of Theano for these applications in a similar manner as the Julia framework was recently presented in~\cite{bezanson_julia:_2017}.

We now wish to give a simple example of Theano code to illustrate this process of symbolic input and numerical output via compiled code.
We consider the symbolic implementation of the scalar product, that is the vector function $f(\boldsymbol x,\boldsymbol y) = \boldsymbol x^T\boldsymbol y$, and want to evaluate its derivative with respect to the first argument. 
In Theano, the function $f$ is first defined as a symbolic function, \lstinline!f = lambda x,y: T.dot(x,y)!, where \lstinline!T! calls functions of the library \lstinline!theano.tensor!.
Then, the gradient of $f$ with respect to $x$ is defined by calling the gradient function \lstinline!T.grad!, as \lstinline!df = lambda x,y: T.grad(f(x,y),x)!. 
Both functions \lstinline!f! and \lstinline!df! are still symbolic but can be evaluated on any numerical arrays after the compilation process, or construction of an evaluation function. 
For our function \lstinline!f!, the compilation is requested by \lstinline!ff = theano.function([x,y], f(x,y))!, where we have previously declared the variables $x$ and $y$ as \lstinline!x = T.vector()! and \lstinline!y = T.vector()!. 
The function \lstinline!ff! is now a compiled version of the function \lstinline!f! that can be evaluated on any pair of vectors. 
As we will see later in the text, such code can be written for many different functions and combination of derivatives, in particular for derivatives with respect to initial conditions in a \lstinline!for! loop. 

In this work, we want to illustrate this transparent use of Theano in various numerical computations based on objects from differential geometry. 
We will only cover a few topics here, and many other such applications will remain for future works. 
Apart from our running example, the sphere, or the rotation group, we will use higher dimensional examples, in particular the manifold of landmarks as often used in computational anatomy. 
In both cases, we will show how to compute various geometrical quantities arising from Riemannian metrics on the spaces.
In most cases, the metric is the only information on the manifold that is needed, and it allows for computing geodesics, Brownian motion, parallel transport etc. In some cases, it will be convenient to extend to computations in a fiber bundle of the manifold to have more freedom and allow for e.g. anisotropic diffusion processes. 
Also, when the manifold has a group structure, we can perform for example reduction by symmetry for dynamical systems invariant under the group action. 
All of these mechanical constructions can be used to real-world applications such as in control or robotics. 
We refer to the books \cite{bloch2003nonholonomic, chirikjian2009stochastic,chirikjian2011stochastic} for more theories and applications in these directions. 
We will not directly consider these applications here, but rather focus on applications of computational anatomy. 
We refer the interested reader to the book \cite{younes_shapes_2010} and references therein for a good overview of this topic. 
We also refer to the conference paper \cite{kuehnel2017} for a short introduction of the use of Theano in computational anatomy. 
Computational anatomy is a vast topic, and we will only focus here on a few aspects when shapes or images are represented as sets of points, or landmarks, that are used as tracers of the original shape. 
With these landmarks, we show how many algorithms related to matching of shapes, statistics of shapes or random deformations, can be implemented concisely and efficiently using Theano. 

\begin{figure}[ht]
    \centering
    \includegraphics[height=5.3cm, trim = 0 2 0 0,clip]{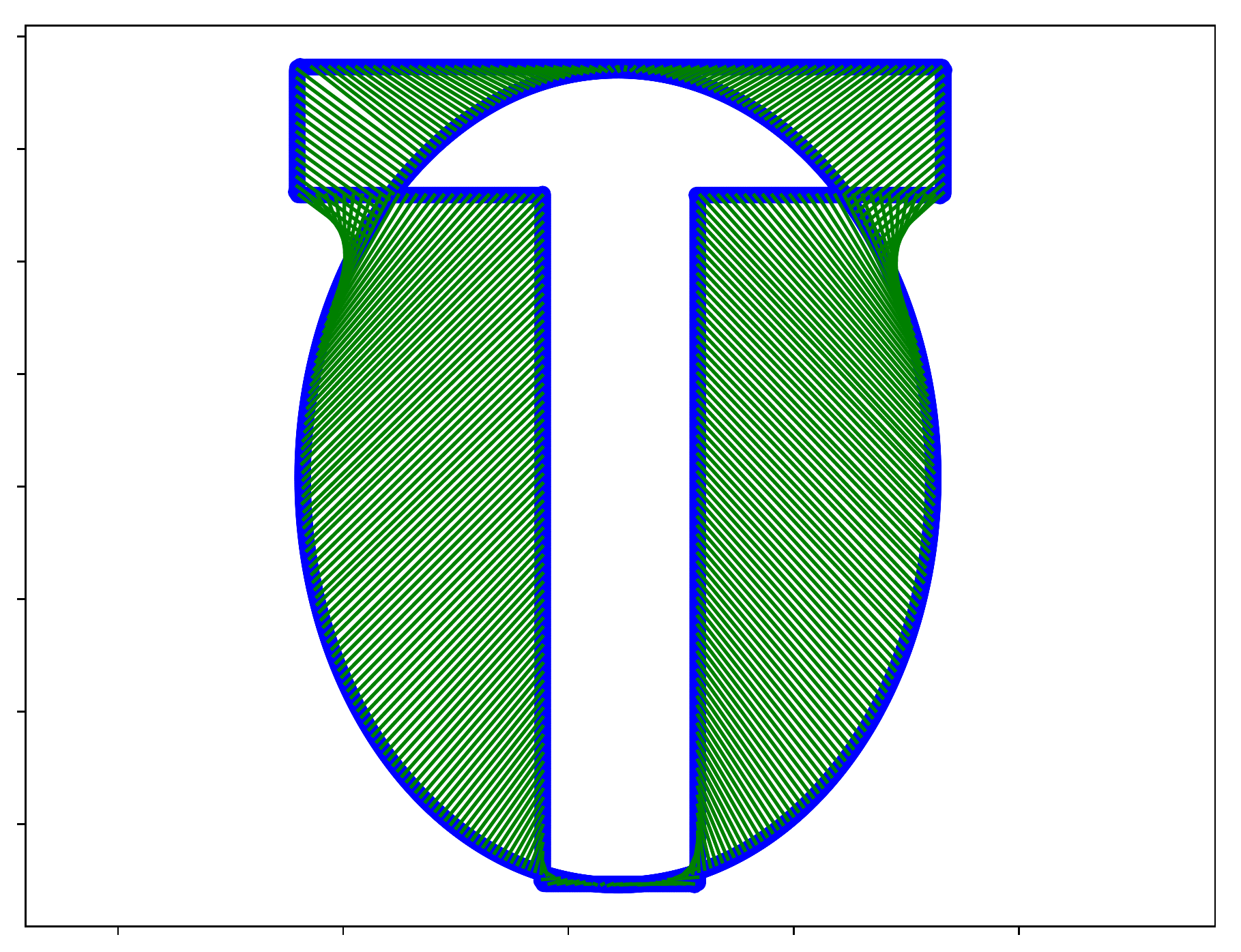}
    \includegraphics[height=5.3cm, trim = 50 31.5 0 33,clip]{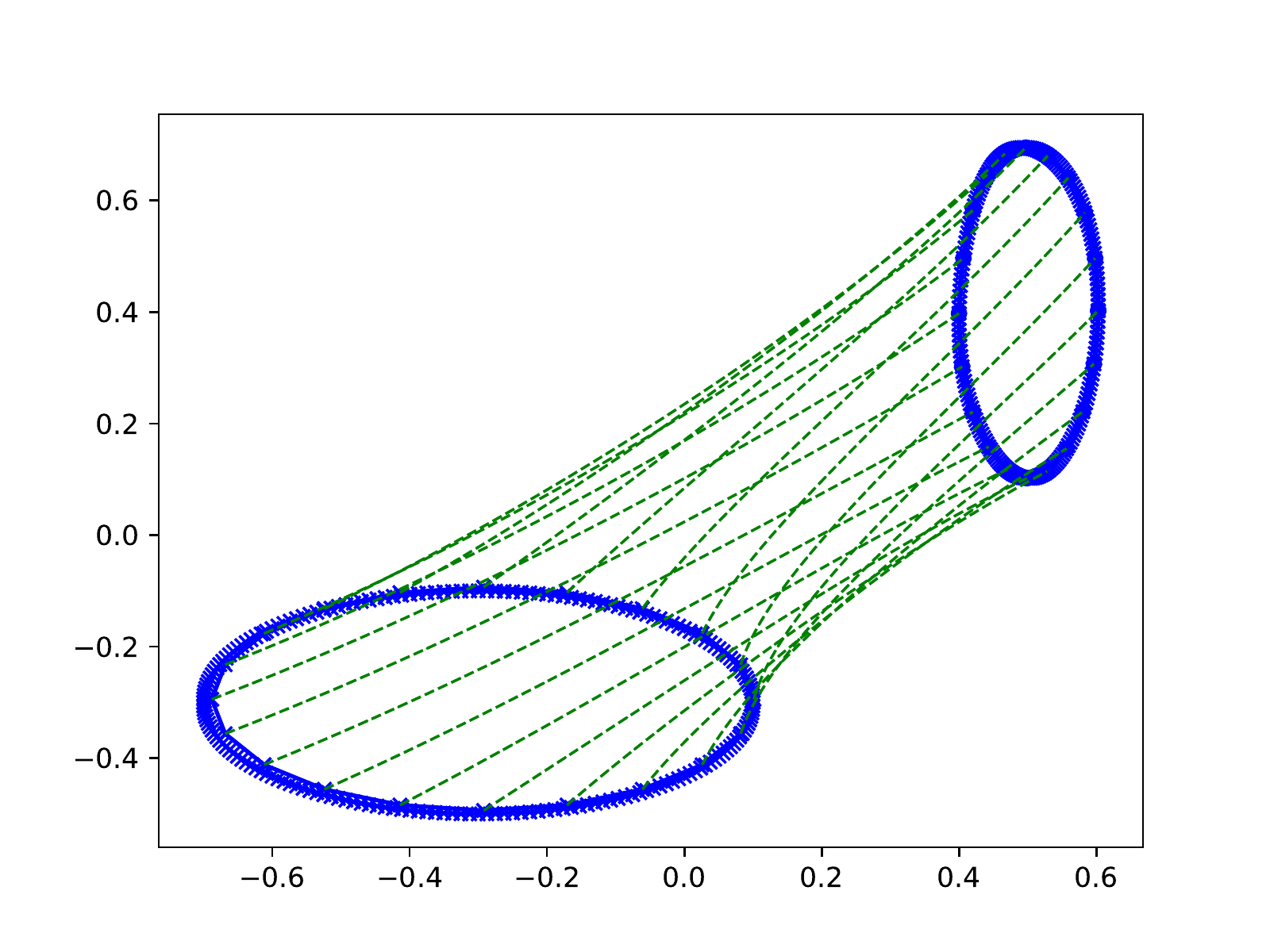}
    \caption{(left) Matching of ~2,500 landmarks on the outline of a letter 'T' to a letter 'O'. The matching is performed by computing the logarithm map $\text{Log}$ considering the ~5,000 dimensional landmark space a Riemannian manifold. (right) Similar matching of landmark configurations using $\text{Log}$ while now using the transparent GPU features of Theano to scale to configurations with 20,000 landmarks on a 40,000 dimensional manifold. Theano generates highly efficient numerical code and allows GPU acceleration transparently to the programmer. For both matches, only a subset of the geodesic landmark trajectories are display.}
    \label{fig:landmatch}
\end{figure}

As an example, we display in Figure \ref{fig:landmatch} two examples of solving the inverse problem of estimating the initial momenta for a geodesic matching landmark configurations on high-dimensional manifolds of landmarks on the plane. 
On the left panel of Figure \ref{fig:landmatch}, we solved the problem of matching a letter 'T' to a letter 'O', or more precisely an ellipse, with 2,500 landmarks. On the right panel, we solved the problem of matching two simple shapes, ellipses, however with 20,000 landmarks. 
The shapes represented by landmarks are considered elements of the LDDMM landmark manifold of dimension 5,000 and 40,000, see \cite{younes_shapes_2010}. 
The geodesics equation and inverse problem are implemented using the few lines of code presented in this paper and the computation is transparently performed on GPUs. 

Parts of the code will be shown throughout the paper with corresponding examples. The full code is available online in the Theano Geometry repository \url{http://bitbucket.org/stefansommer/theanogeometry}.
The interested reader can find a more extensive description of the mathematical notions used in this paper in the books \textit{Riemannian Manifolds: an introduction to curvature} by J. Lee~\cite{lee2006riemannian}, \textit{Stochastic Analysis on Manifolds} by E. P. Hsu~\cite{hsu_stochastic_2002} and \textit{Introduction to Mechanics and Symmetry} by Marsden, Ratiu~\cite{marsden_introduction_1999}. 

\subsection*{Content of the paper}

The paper will be structured as follows. 
Section \ref{sec:Riem} gives an account of how central concepts in Riemannian geometry can be described symbolically in Theano, including the exponential and logarithm maps, geodesics in Hamiltonian form, parallel transport and curvature. 
Concepts from Lie group theory are covered in section \ref{sec:Lie}, and section \ref{sec:FM} continues with sub-Riemannian frame bundle geometry. 
In addition to the running example of surfaces embedded in $\RR^3$, we will show in section \ref{sec:landmark} applications on landmark manifolds defined in the LDDMM framework. At the end, concepts from non-linear statistics are covered in section \ref{sec:non-lin}. 

\section{Riemannian Geometry}
\label{sec:Riem}

In this section, we will show how to implement some of the theoretical concepts from Riemannian geometry. This includes geodesics equation, parallel transport and curvature. 
The focus is to present simple and efficient implementation of these concepts using Theano~\cite{theano}.

Though the code applies to any smooth manifolds $\M$ of dimension $d$, we will only visualize the results of numerical computations on manifolds embedded in $\RR^3$. 
We represent these manifolds by a smooth injective map $F:\RR^2\rightarrow\RR^3$ and the associated metric on $\M$ inherited from $\RR^3$, that is  
\begin{equation}
  g=(dF)^TdF\, , 
  \label{eq:metric}
\end{equation}
where $dF$ denotes the Jacobian of $F$. 
One example of such representation is the sphere $S^2$ in stereographic coordinates. In this case, $F\colon\RR^2\to S^2\subset\RR^3$ is
\begin{equation}
\label{eq:steo}
    F(x,y) = 
    \begin{pmatrix}
        \frac{2x}{1+x^2+y^2}&\frac{2y}{1+x^2+y^2}&\frac{-1+x^2+y^2}{1+x^2+y^2}
    \end{pmatrix}\, . 
\end{equation}

\subsection{Geodesic Equation}
\label{sec:geoeq}

We begin by computing solutions to the Riemannian geodesic equations on a smooth $d$-dimensional manifold $\M$ equipped with an affine connection $\nabla$ and a metric $g$. 
A connection on a manifold defines the relation between tangent spaces at different points on $\M$. Let $(U,\varphi)$ denote a local chart on $\M$ with coordinate basis $\partial_i = \frac{\partial}{\partial x^i}$, $i=1,\ldots,d$. The connection $\nabla$ is related to the Christoffel symbols $\Gamma_{ij}^k$ by the relation
\begin{equation}
    \nabla_{\partial_i} \partial_j = \Gamma_{ij}^k\partial_k\, .
\end{equation}
An example of a frequently used connection on a Riemannian manifold $(\M,g)$ is the Levi-Civita connection. The Christoffel symbols for the Levi-Civita connection is uniquely defined by the metric $g$. 
Let $g_{ij}$ denote the coefficients of the metric $g$, i.e. $g = g_{ij}dx^idx^j$, and $g^{ij}$ be the inverse of $g_{ij}$. The Christoffel symbols for the Levi-Civita connection are then
\begin{align}
    \Gamma_{ij}^k = \frac{1}{2}g^{kl}(\partial_i g_{jl} + \partial_j g_{il} - \partial_l g_{ij})\, .
\label{eq:Chris}
\end{align}
The implementation of the Christoffel symbols in Theano are shown in the code snippet below.
\begin{lstlisting}
"""
Christoffel symbols for the Levi-Civita connection

Args:
    x: Point on the manifold
    g(x): metric g evaluated at position x on the manifold.

Returns:
    Gamma_g: 3-tensor with dimensions k,i,j in the respective order
"""
# Derivative of metric:
Dg = lambda x: T.jacobian(g(x).flatten(),x).reshape((d,d,d)) 
# Inverse metric (cometric):
gsharp = lambda x: T.nlinalg.matrix_inverse(g(x))

# Christoffel symbols:
Gamma_g = lambda x: 0.5*(T.tensordot(gsharp(x),Dg(x),axes = [1,0])\
       +T.tensordot(gsharp(x),Dg(x),axes = [1,0]).dimshuffle(0,2,1)\
       -T.tensordot(gsharp(x),Dg(x),axes = [1,2]))
\end{lstlisting}

 Straight lines in $\RR^n$ are lines with no acceleration and path minimizers between two points. Geodesics on a manifold are defined in a similar manner. The acceleration of a geodesic $\gamma$ is zero, i.e. $D_t\dot\gamma = 0$, in which $D_t$ denotes the covariant derivative. Moreover, geodesics determines the shortest distances between points on $\M$. Let $x_0\in\M$, $(U,\varphi)$ be a chart around $x_0$ and consider $v_0\in T_{x_0}\M$, a tangent vector at $x_0$. A geodesic $\gamma\colon I\to\M$, $I = [0,1]$, $\gamma_t = (x^i_t)_{i=1,\ldots,d}$, satisfying $\gamma_0=x_0$, $\dot\gamma_0=v_0$ can be obtained by solving the geodesic equations
\begin{equation}
\label{eq:geoeq}
    \ddot{x}^k_t + \dot{x}^i_t\dot{x}^j_t\Gamma_{ij}^k(\gamma_t) = 0\, .
\end{equation}
The goal is to solve this second order ordinary differential equation (ODE) with respect to $x^k_t$. We first rewrite the ODE in term of $w^k_t = \dot{x}^k_t$ and $x_t^k$ to instead have a system of first order ODE of the form
\begin{align*}
    \dot{w}^k_t = - w^i_tw^j_t\Gamma_{ij}^k(\gamma_t)\,,\  
    \dot{x}^k_t = w^k_t\, ,
\end{align*}
which can be solved by numerical integration. 
For this, we can use the Euler method
\begin{align}
    y_{n+1} = y_n + f(t_n,y_n)\Delta t, \quad \Delta t = t_{n+1} - t_n\, ,
\end{align}
or by higher-order integrators such as a fourth-order Runge-Kutta method.
Both integrators are available in symbolic form in the code repository. 
In Theano, we use the symbolic \lstinline!for!-loop \lstinline!theano.scan! for the loop over time-steps. 
As a consequence, symbolic derivatives of the numerical integrator can be evaluated. 
For example, we will later use derivatives with respect to the initial values when solving the geodesic matching problem in the definition of the Logarithm map. In addition, it is possible to solve stochastic differential equations in a similar way, see Appendix~\ref{sec:stoc}. 
The symbolic implementation of the integrator method is shown below.

\begin{lstlisting}
"""
Numerical Integration Method

Args:
    ode: Symbolic ode function to be solved
    integrator: Integration scheme (Euler, RK4, ...)
    x: Initial values of variables to be updated by integration method
    *y: Additional variables for ode.

Returns:
    Tensor (t,xt)
               t: Time evolution
               xt: Evolution of x
"""
def integrate(ode,integrator,x,*y):
    (cout, updates) = theano.scan(fn=integrator(ode),
            outputs_info=[T.constant(0.),x],
            sequences=[*y],
            n_steps=n_steps)
    return cout
\end{lstlisting}

Based on the symbolic implementation of the integrators, solutions to the geodesic equations are obtained by the following code. 

\begin{lstlisting}
"""
Geodesic Equation

Args:
    xq: Tensor with x and xdot components.

Returns:
    ode_geodesic: Tensor (dx,dxdot).
    geodesic: Tensor (t,xt)
                   t: Time evolution
                   xt: Geodesic path
"""
def ode_geodesic(t,xq):
    
    dxdott = - T.tensordot(T.tensordot(xq[1], Gamma_g(xq[0]), axes=[0,1]),
                             xq[1],axes=[1,0])
    dxt = xq[1]
    return T.stack((dxt,dxdott))

# Geodesic:
geodesic = lambda x,xdot: integrate(ode_geodesic, T.stack((x,xdot)))
\end{lstlisting}

\begin{figure}[ht]
    \begin{center}
    \begin{minipage}{0.33\textwidth}
        \centering
        \includegraphics[scale=0.41, trim = 100 60 60 70,clip]{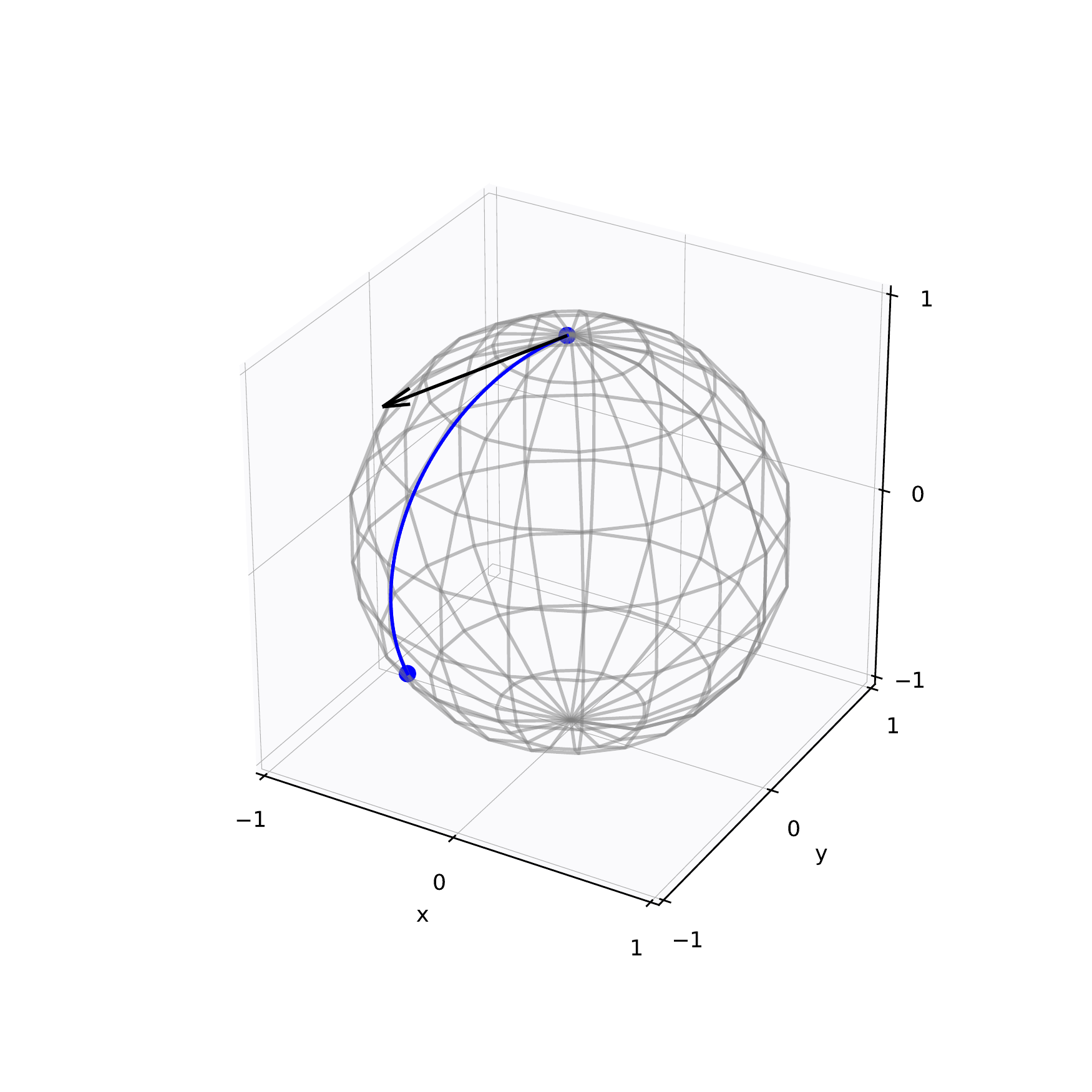}
    \end{minipage}%
    \begin{minipage}{0.33\textwidth}
        \centering
        \includegraphics[scale=0.41,trim = 100 70 70 80,clip]{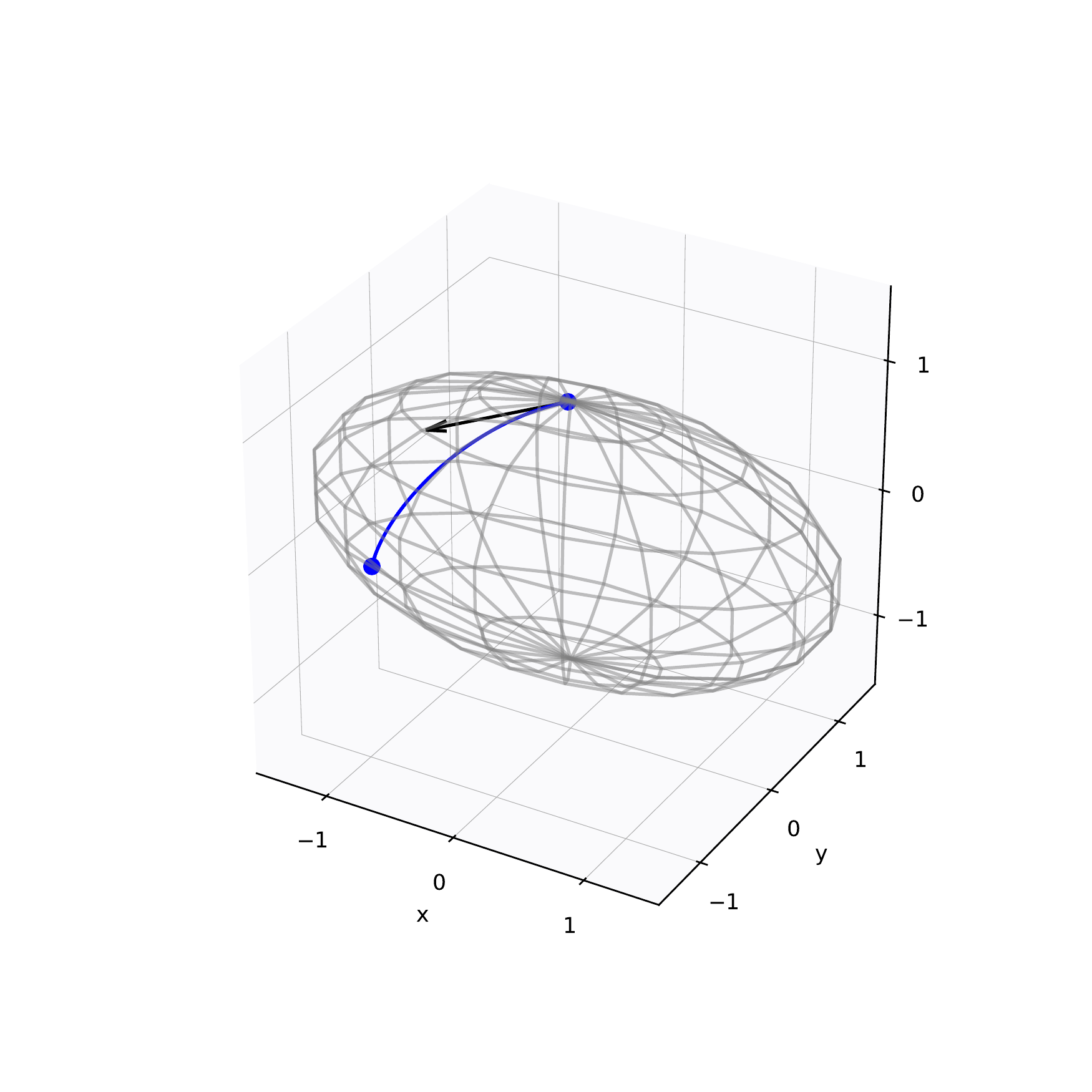}
    \end{minipage}%
    \begin{minipage}{0.33\textwidth}
        \centering
        \includegraphics[scale=0.31,trim = 20 20 30 20,clip]{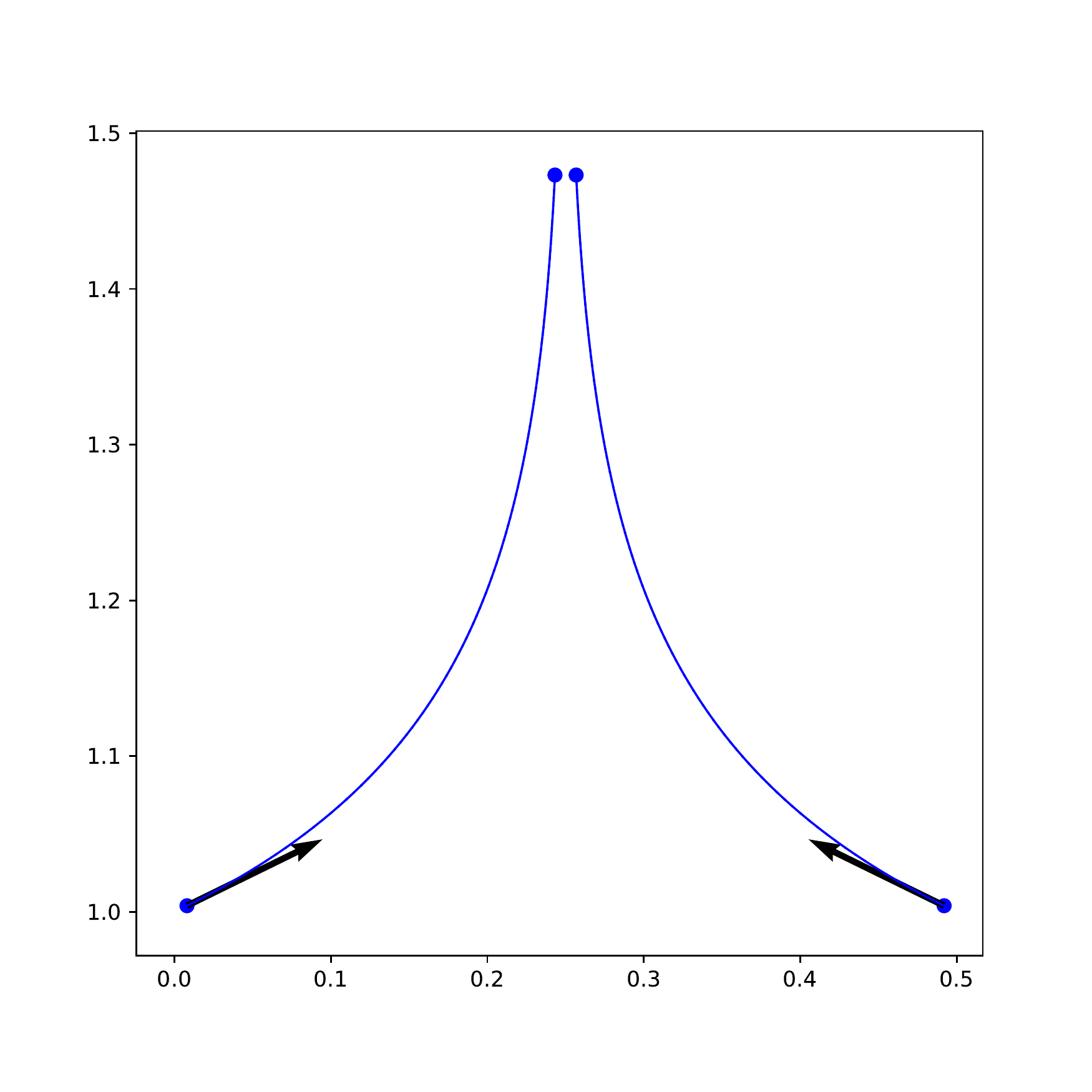}
    \end{minipage}
    \caption{The solution of the geodesic equations for three different manifolds; the sphere $S^2$, an ellipsoid, and landmark manifold defined in the LDDMM framework. The arrows symbolizes the initial tangent vector $v_0$.}
    \label{fig:geoeq}
    \end{center}
\end{figure}

Figure \ref{fig:geoeq} shows examples of geodesics on three different manifolds obtained as the solution to the geodesic equations in \eqref{eq:geoeq} using the above code.

\subsection{The Exponential and Logarithm Maps}

For a geodesic $\gamma_t^v$, $t\in [0,1]$ with initial velocity $\dot\gamma_0^v = v$, the exponential map, $\text{Exp}_x\colon T_x\M\to\M$, $x\in\M$ is defined by 
\begin{align}
    \text{Exp}_x(v) = \gamma_1^v\, , 
\label{eq:exp}
\end{align}
and can be numericaly computed from the earlier presented geodesic equation. 

\begin{lstlisting}
"""
Exponential map 

Args:
    x: Initial point of geodesic
    v: Velocity vector

Returns:
    y: Endpoint of geodesic
"""
Exp = lambda x,v: geodesic(x,v)[1][-1,0]
\end{lstlisting}

Where defined, the inverse of the exponential map is denoted the logarithm map. For computational purposes, we can define the logarithm map as finding a minimizing geodesic between $x_1,x_2\in\M$, that is 
\begin{align}
    \text{Log}(x_1,x_2) = \argmin_v \|\text{Exp}_{x_1}(v) - x_2\|_{\M}^2\,, 
  \label{loss-log}
\end{align}
for a norm coming for example from the embedding of $\M$ in $\RR^3$. From the logarithm, we also get the geodesic distance by
\begin{align}
    d(x,y) = \|\text{Log}(x,y)\|\, .
\label{eq:distgeo}
\end{align}
The logarithm map can be implemented in Theano by using the symbolic calculations of derivatives by computing the gradient of the loss function \eqref{loss-log} with Theano function \lstinline!T.grad!, and then use it in a standard minimisation algorithm such as BFGS.  
An example implementation is given below, where we used the function \lstinline!minimize! from the Scipy package. 

\begin{lstlisting}
"""
Logarithm map

Args:
    v0: Initial tangent vector
    x1: Initial point for geodesic
    x2: Target point on the manifold

Return:
    Log: Tangent vector 
"""
# Loss function:
loss = lambda v,x1,x2: 1./d*T.sum(T.sqr(Exp(x1,v)-x2))
dloss = lambda v,x1,x2: T.grad(loss(v,x1,x2),v)
# Logarithm map: (v0 initial guess)
Log = minimize(loss, v0, jac=dloss, args=(x1,x2))
\end{lstlisting}

\subsection{Geodesics in Hamiltonian Form}
\label{sec:geoHam}

In section \ref{sec:geoeq}, geodesics were computed as solutions to the standard second order geodesic equations. 
We now compute geodesics from a Hamiltonian viewpoint.
Let the manifold $\M$ be equipped with a cometric $g^*$ and consider a connection $\nabla$ on $\M$. 
Given a point $x\in\M$ and a covector $p\in T_x^*\M$, geodesics can be obtained as the solution to Hamilton's equations, given by the derivative of the Hamiltonian, which in our case is
\begin{align}
  H(x,p) = \frac{1}{2}\langle p , g_x^*(p)\rangle_{T_x^*\M\times T_x^*\M}\,.
\label{eq:ham}
\end{align}
Hamilton's equations are then
\begin{align}
  \frac{d}{dt}x &= \nabla_p H(x,p) \nonumber\\
  \frac{d}{dt}p &= -\nabla_x H(x,p)\, ,
\label{eq:Hamilton}
\end{align}
and describe the movement of a particle at position $x\in\M$ with momentum $p\in T_x^*\M$.

Depending on the form of the Hamiltonian and in particular of the metric, the implementation of Hamilton's equations \eqref{eq:Hamilton} can be difficult.
In the present case, the metric on $\M$ is inherited from an embedding $F$, hence $g^*$ is defined only via derivatives of $F$, which makes the computation possible with Theano. 

\begin{lstlisting}
"""
Calculate the Exponential map defined by Hamilton's equations.

Args:
    x: Point on manifold
    p: Momentum vector at x
    gsharp(x): Matrix representation of the cometric at x
    
Returns:
    Exp: Tensor (t,xt)
               t: Time evolution
               xt: Geodesic path
"""
# Hamiltonian:
H = lambda x,p: 0.5*T.dot(p,T.dot(gsharp(x),p))

# Hamilton's equation
dx = lambda x,p: T.grad(H(x,p),p)
dp = lambda x,p: -T.grad(H(x,p),x)
def ode_Hamiltonian(t,x):
    dxt = dx(x[0],x[1])
    dpt = dp(x[0],x[1])
    return T.stack((dxt,dpt))

# Geodesic:
Exp = lambda x,v: integrate(ode_Ham,T.stack((x,g(v))))
\end{lstlisting}

 Calculating geodesics on a Riemannian manifold $\M$ by solving Hamilton's equations can be generalized to manifolds for which only a sub-Riemannian structure is available. 
An example of such geodesics is given in section \ref{sec:FM} on a different construction, the frame bundle.

\begin{example}[Geodesic on the sphere]
\label{ex:geoSphere}
Consider the sphere $S^2\subset\RR^3$ in stereographic coordinates such that for $(x,y)\in\RR^2$, a point on the sphere is given by $F(x,y)$ with $F$ defined in \eqref{eq:steo}. 
Equip $S^2$ with the metric $g$ defined in \eqref{eq:metric} and let $x_0 = F(0,0)\in S^2$ and $v_0 = dF(1,-1)\in T_{x_0}S^2$. 
The initial momentum vector is chosen as the corresponding covector of $v_0$ defined by the flat map $\flat\colon T\M\to T^*\M$, i.e. $p_0=v_0^\flat$. 
The geodesic, or the solution to Hamilton's equations can be seen in the left plot of Figure \ref{fig:geodesic}.

\begin{figure}[ht]
    \begin{center}
    \begin{minipage}{0.5\textwidth}
        \centering
        \includegraphics[scale=0.5, trim = 100 60 60 70,clip]{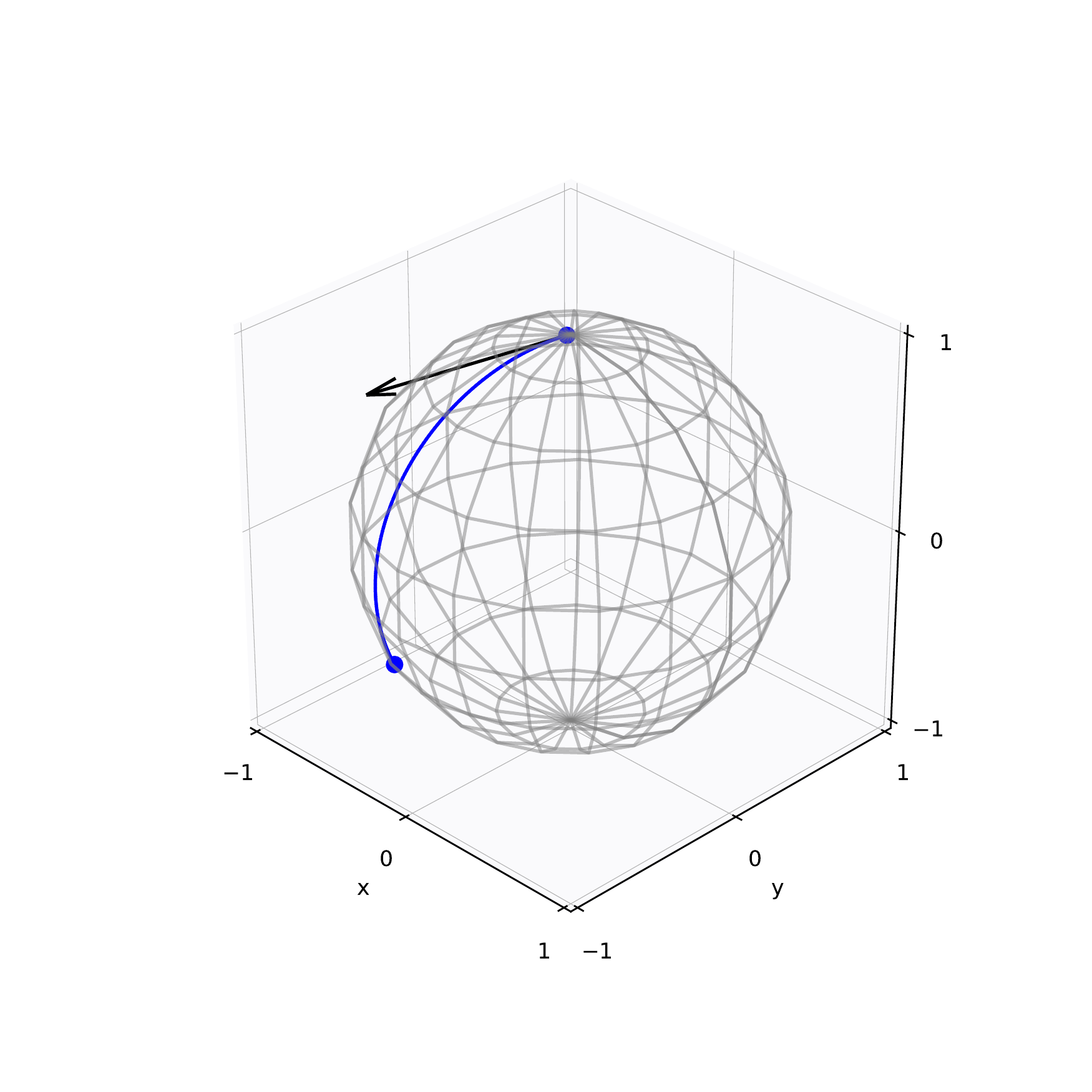}
    \end{minipage}%
    \begin{minipage}{0.5\textwidth}
        \centering
        \includegraphics[scale=0.5,trim = 180 60 130 70,clip]{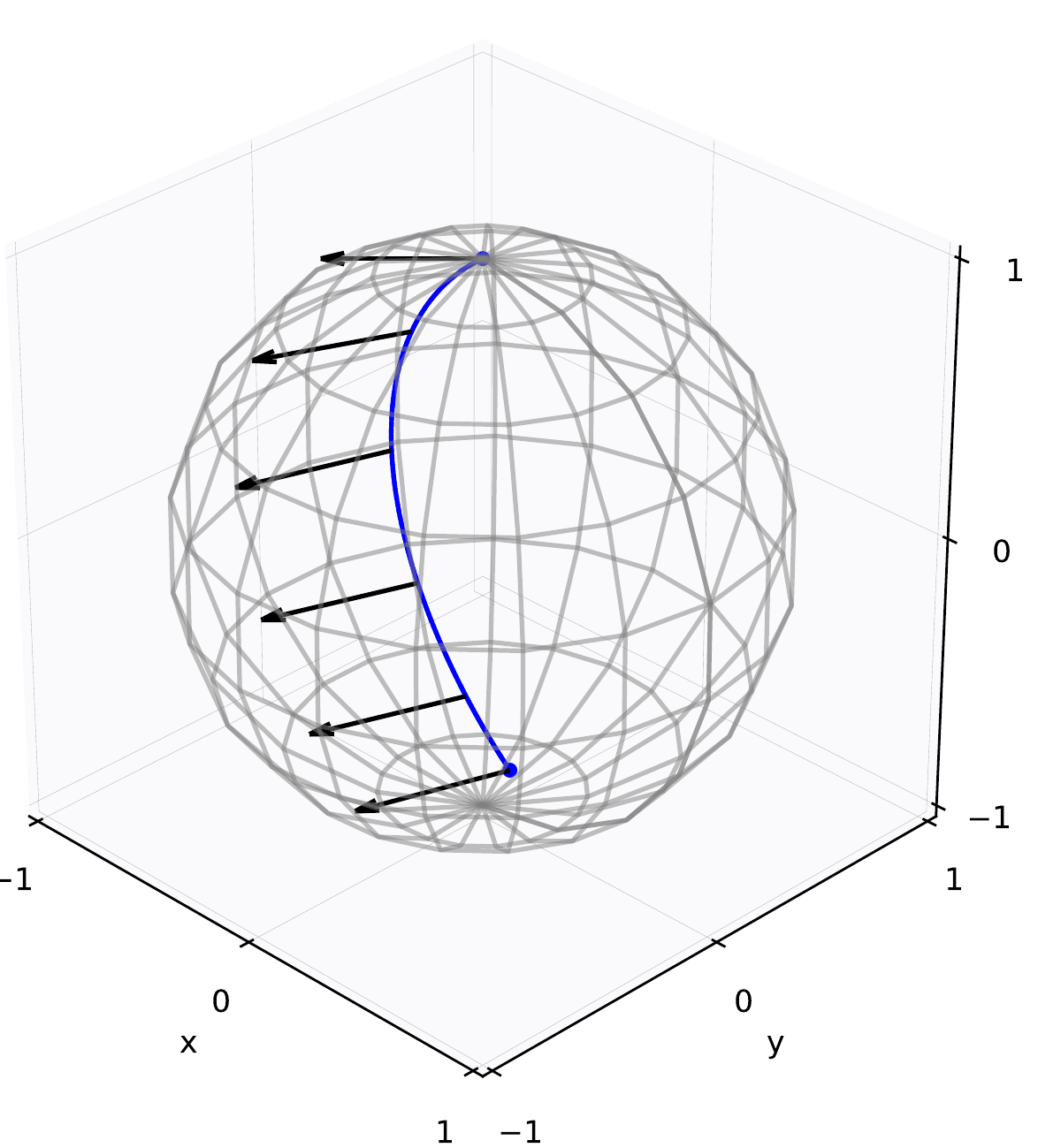}
    \end{minipage}
    \caption{(left) Geodesic defined by the solution to Hamilton's equations \eqref{eq:Hamilton} with initial point $x_0 = F(0,0)\in S^2$ and velocity $v_0 = dF(1,-1)\in T_{x_0}S^2$. See Example \ref{ex:geoSphere}. (right) Parallel transport of vector $v = dF \left (-\frac{1}{2},-\frac{1}{2}\right )$ along the curve $\gamma_t = F(t^2,-\sin(t))$. See Example \ref{ex:parsphere}.}
    \label{fig:geodesic}
    \end{center}
\end{figure}

\end{example}
 
\subsection{Parallel Transport}

Let again $\M$ be a $d$-dimensional manifold with an affine connection $\nabla$ and let $(U,\varphi)$ denote a local chart on $\M$ with coordinate basis $\partial_i = \frac{\partial}{\partial x^i}$ for $i=1,\ldots,d$. 
A vector field $V$ along a curve $\gamma_t$, is said to be parallel if the covariant derivative of $V$ along $\gamma_t$ is zero, i.e. $\nabla_{\dot{\gamma}_t} V = 0$. 
It can be shown that given a curve $\gamma\colon I\to\M$ and a tangent vector $v\in T_{\gamma_{t_0}}\M$ there exists a unique parallel vector field $V$ along $\gamma$ such that $V_{t_0} = v$. We further assume that $\gamma_t = (\gamma_t^i)_{i=1,\ldots,d}$ in local coordinates and we let $V_{t} = v^i(t)\partial_i$ be a vector field.
$V$ is then parallel to the curve $\gamma_t$ if the coefficients $v^i(t)$ solve the following differential equation,
\begin{align}
    \dot{v}^k(t) + \Gamma^k_{ij}(\gamma_t)\dot{\gamma}^i_t v^j(t) = 0\, .
\end{align}
The parallel transport can be implemented in an almost similar manner as the geodesic equations introduced in section \ref{sec:geoeq}.

\begin{lstlisting}
"""
Parallel Transport

Args:
    gamma: Discretized curve
    dgamma: Tangent vector of gamma to each time point
    v: Tangent vector that will be parallel transported.

Returns:
    pt: Tensor (t,vt)
             t: Time Evolution
             vt: Parallel transported tangent vector at each time point
"""
def ode_partrans(gamma,dgamma,t,v):
    dpt = - T.tensordot(T.tensordot(dgamma, Gamma_g(gamma), axes = [0,1]), 
                          v, axes = [1,0])
    return dpt

# Parallel transport
pt = lambda v,gamma,dgamma: integrate(ode_partrans,v,gamma,dgamma)
\end{lstlisting}

\begin{example}
\label{ex:parsphere}
    In this example, we consider a tangent vector $v = dF\left (-\frac{1}{2},-\frac{1}{2}\right)\in T_x S^2$ for $x = F(0,0)\in S^2$ that we want to parallel transport along the curve $\gamma\colon[0,1]\to S^2$ given by $\gamma_t = F\left (t^2,-\sin(t)\right )$. 
    The solution of the problem is illustrated in the right panel of Figure~\ref{fig:geodesic}.
\end{example}

\subsection{Curvature}
\label{sec:curv}

The curvature of a Riemannian manifold $\M$ is described by the Riemannian curvature tensor, a $(3,1)$-tensor $R\colon\mathcal{T}(\M)\times\mathcal{T}(\M)\times\mathcal{T}(\M)\to\mathcal{T}(\M)$ defined as
\begin{equation}
    R(X,Y)Z = \nabla_X\nabla_Y Z - \nabla_Y\nabla_X Z -\nabla_{[X,Y]}Z\, .
\end{equation}
Let $(U,\varphi)$ be a local chart on $\M$ and let $\partial_i$ for $i=1,\ldots,d$ denote the local coordinate basis with $dx^i$ being the dual basis. 
Given this local basis, the curvature tensor is, in coordinates, given as 
\begin{equation}
  R = R\indices{_{ijk}^m} dx^i\otimes dx^j\otimes dx^k\otimes\partial_m\, ,
\end{equation}
where the components $R\indices{_{ijk}^m}$ depend on the Christoffel symbols as follow
\begin{equation}
\label{eq:curv}
R(\partial_i,\partial_j)\partial_k = R\indices{_{ijk}^m}\partial_m = (\Gamma\indices{^l_{jk}}\Gamma\indices{^m_{il}}-\Gamma\indices{^l_{ik}}\Gamma\indices{^m_{jl}}+\partial_i\Gamma\indices{^m_{jk}}-\partial_j\Gamma\indices{^m_{ik}})\partial_m\, .
\end{equation}
In Theano, the Riemannian curvature tensor can be computed in coordinates as follow.

\begin{lstlisting}
"""
Riemannian curvature tensor in coordinates

Args:
    x: point on manifold

Returns:
    4-tensor R_ijk^m in with order i,j,k,m
"""
def R(x):
 return (T.tensordot(Gamma_g(x),Gamma_g(x),(0,2)).dimshuffle(3,0,1,2) 
       - T.tensordot(Gamma_g(x),Gamma_g(x),(0,2)).dimshuffle(0,3,1,2) 
       +  T.jacobian(Gamma_g(x).flatten(),x).reshape((d,d,d,d)).dimshuffle(3,1,2,0)
       -  T.jacobian(Gamma_g(x).flatten(),x).reshape((d,d,d,d)).dimshuffle(1,3,2,0))
\end{lstlisting}

In addition to the curvature tensor $R\indices{_{ijk}^m}$, the Ricci and scalar curvature can be computed by contracting the indices as
\begin{equation}
  R_{ij} = R\indices{_{kij}^k} \ ,\  S = g^{ij}R_{ij} \,.
  \label{eq:Ricci_scalar}
\end{equation}

The sectional curvature can also be computed and describes the curvature of a Riemannian manifold by the curvature of a two-dimensional sub-manifold. 
Let $\Pi$ be a two-dimensional sub-plane of the tangent space at a point $x\in\M$. 
Let $e_1,e_2$ be two linearly independent tangent vectors spanning $\Pi$. 
The sectional curvature is the Gaussian curvature of the sub-space formed by geodesics passing $x$ and tangent to $\Pi$, that is 
\begin{align}
    \kappa(e_1,e_2) = \frac{\langle R(e_1,e_2)e_2,e_1\rangle}{\|e_1\|^2\|e_2\|^2 - \langle e_1,e_2\rangle^2}\, .
\end{align}

\begin{example}[Curvature of $S^2$]
We consider $x = F(0,0)\in S^2$ and the orthonormal basis vectors $e_1 = dF(0.5,0)$, $e_2 = dF(0,0.5)$ in the tangent space $T_x\M$ with respect to the metric $g$. 
As expected, we found that the Gaussian curvature of $S^2$ is $1$ and its scalar curvature is 2~\cite{lee2006riemannian}. 
\end{example}

The Ricci, scalar and sectional curvature have also been implemented in Theano as follow.

\begin{lstlisting}
"""
Different curvature measures

Args:
    x: point on manifold
    e1, e2: linearly independent tangent vectors
"""
# Ricci curvature:
Ricci_curv = lambda x: T.tensordot(R(x),T.eye(d),((0,3),(0,1)))
# Scalar curvature:
S_curv = lambda x: T.tensordot(Ricci_curv(x),gsharp(x),((0,1),(0,1)))
# Sectional curvature:
def sec_curv(x,e1,e2):
        Rflat = T.tensordot(R(x),g(x),[3,0])
        sec = T.tensordot(
                T.tensordot(
                    T.tensordot(
                        T.tensordot(Rflat, e1, [0,0]), 
                        e2, [0,0]),
                    e2, [0,0]), 
                e1, [0,0])
        return sec
\end{lstlisting}

\section{Dynamics on Lie Groups}
\label{sec:Lie}

\begin{figure}
    \begin{center}
        \includegraphics[scale=0.5, trim = 100 60 60 70,clip]{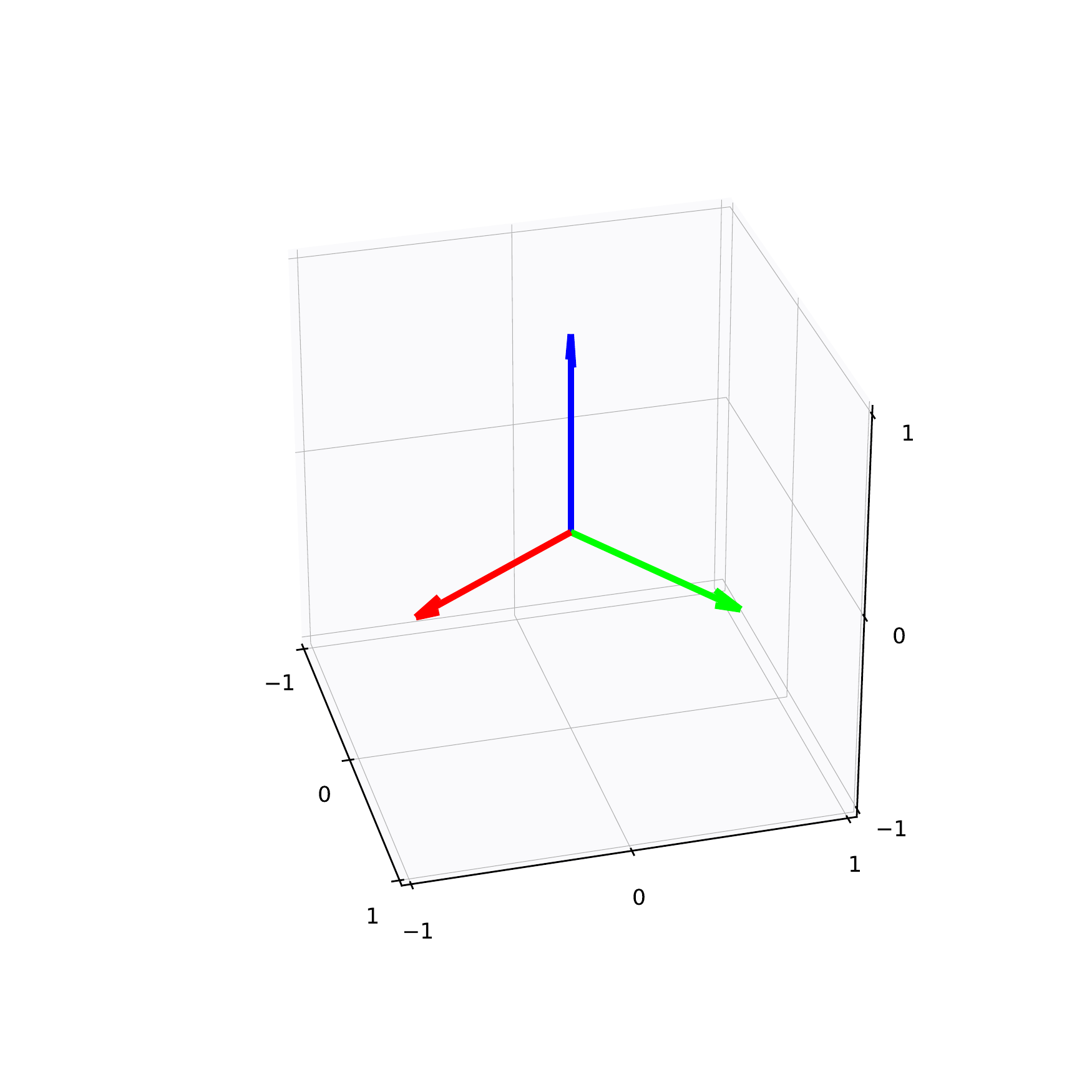}
        \caption{We show an example of an element of $\SO(3)$ represented as a matrix $g\in\RR^{3\times 3}$. The vectors represent each column of $g$.}
    \label{fig:exso3}
    \end{center}
\end{figure}

In this section, we consider a manifold equipped with a smooth group structure, that is $\M=G$ is a Lie group. 
As the most interesting finite dimensional Lie groups are isomorphic to matrix groups, we can without loss of generalities represent elements of Lie group $G$ by matrices. 
We will give examples of how various fundamental Lie group constructions can be written with Theano and how to compute geodesics in the Hamiltonian and Lagrangian setting. 
We will mostly follow \cite{marsden_introduction_1999} for the notation and definitions. 
We will use $G =  \SO(3)$, the three dimensional rotation group acting on $\mathbb R^3$ as an illustration, where an element of $G$ is represented by a coordinate basis as for example in Figure~\ref{fig:exso3}.

The group operation on $G$ defines the left and right translation maps $L_a(g)=ag$ and $R_a(g)=ga$ for $a,g\in G$. 
As elements of $G$ are represented by matrices, these maps are in Theano computed by matrix multiplications. 
Their corresponding tangent maps $dL$ and $dR$ can be directly obtained by taking symbolic derivatives. 
The left and right translation maps relate elements of the Lie algebra $\mathfrak g$ of the group with the left (and right) invariant vector fields $X_\eta(g):=dL_g(\eta)$ on $TG$, where $\eta\in\mathfrak g$. 
The algebra structure on $\mathfrak g$ is then defined from the Jacobi-Lie bracket of vector fields $[\xi,\eta]=[X_\xi,X_\eta]$, $\xi,\eta\in\mathfrak g$.

Using invariance under the group action, either left or right, an inner product on $\mathfrak g=T_eG$ can be extended to a Riemannian metric on $G$ by setting $\left<v,w\right>_g= \left<dL_av,dL_aw\right>_{L_a(g)}$ for $v,w\in T_g G$.
Invariant metrics can thus be identified with a symmetric positive definite inner product $\langle \cdot, \cdot \rangle_A$ on $\mathfrak g$, where after fixing a basis for $\mathfrak g$, we can consider that $A\in \mathrm{Sym}^+(\mathfrak g)$ and $\langle\cdot, \cdot \rangle_A= \langle \cdot, A\cdot \rangle$.
Hence, $A^{-1}$ is the corresponding co-metric.

In Theano, these constructions can be formulated as shown below. 
A basis $e_i$ for $\mathfrak g$ is fixed, and \lstinline!LAtoV! is the inverse of the mapping $v\rightarrow e_iv^i$ between $V=\mathbb R^d$ and the Lie algebra $\mathfrak g$.

\begin{lstlisting}
"""
General functions for Lie groups

Args:
    g,h: Elements of G
    v: Tangent vector
    xi,eta: Elements of the Lie Algebra
    d: Dimension of G
    vg,wg: Elements of tangent space at g
"""

L = lambda g,h: T.tensordot(g,h,(1,0)) # left translation L_g(h)=gh
R = lambda g,h: T.tensordot(h,g,(1,0)) # right translation R_g(h)=hg

# Derivative of L
def dL(g,h,v): 
    dL = T.jacobian(L(theano.gradient.disconnected_grad(g),h).flatten(),
                      h).reshape((N,N,N,N))
    return T.tensordot(dL,v,((2,3),(0,1)))

# Lie bracket
def bracket(xi,eta): 
    return T.tensordot(xi,eta,(1,0))-T.tensordot(eta,xi,(1,0))

# Left-invariant metric
def g(g,v,w): 
    xiv = dL(inv(g),g,v)
    xiw = dL(inv(g),g,w)
    v = LAtoV(xiv)
    w = LAtoV(xiw)
    return T.dot(v,T.dot(A,w))
\end{lstlisting}

\subsection{Euler-Poincar\'e Dynamics}

In the context of Lie groups, we can also derive the geodesic equations for a left-invariant metric. 
Geodesics on the Lie group can, similar to geodesics on manifolds defined in section \ref{sec:geoHam}, be described as solutions to Hamilton's equations for a Hamiltonian generated from the left-invariant metric. 
In this section, we will, however, present another method for calculating geodesics based on the Euler-Poincar\'e equations.

The conjugation map $h\mapsto aha^{-1}$ for fixed $a\in G$ has as a derivative the adjoint map $\Ad(a):\mathfrak g\rightarrow\mathfrak g$, $\Ad(a)X=(L_a)_*(R_{a^{-1}})_*(X)$. 
The derivative of $\Ad$ with respect to $a$ is the Lie bracket $\ad_\xi:\mathfrak g\rightarrow \mathfrak g$, $\ad_\xi(\eta)=[\xi,\eta]$. The coadjoint action is defined by
$\left<\ad_\xi^*(\alpha),\eta\right>=\left<\alpha,\ad_\xi(\eta)\right>$, $\alpha\in\mathfrak g^*$ with $\left<\cdot,\cdot\right>$ the standard pairing on the Lie algebra $\mathfrak g$.
For the kinetic Lagrangian $l(\xi)=\xi^TA\xi$, $\xi\in\mathfrak g$, a geodesic is a solution of the Euler-Poincar\'e equation
\begin{equation}
  \partial_t\frac{\delta l}{\delta \xi}
  =
  \ad^*_\xi\frac{\delta l}{\delta \xi}\, ,
  \label{eq:EP}
\end{equation}
together with the reconstruction equation $\partial_t g_t=g_t\xi_t$. 
This relatively abstract set of equations can be expressed in Theano with the following code.

\begin{lstlisting}
"""
Euler-Poincare Geodesic Equations

Args:
    a,g: Element of G
    xi,eta: Element of Lie Algebra
    p,pp,mu: Elements of the dual Lie Algebra

Returns:
    EPrec: Tensor (t,xt)
                 t: Time evolution
                 gt: Geodesic path in G
"""
# Adjoint functions:
Ad = lambda a,xi: dR(inv(a),a,dL(a,e,xi))
ad = lambda xi,eta: bracket(xi,eta)
coad = lambda p,pp: T.tensordot(T.tensordot(C,p,(0,0)),pp,(1,0))

# Euler-Poincare equations:
def ode_EP(t,mu):
    xi = T.tensordot(inv(A),mu,(1,0))
    dmut = -coad(xi,mu)
    return dmut
EP = lambda mu: integrate(ode_EP,mu)

# reconstruction
def ode_EPrec(mu,t,g):
    xi = T.tensordot(inv(A),mu,(1,0))
    dgt = dL(g,e,VtoLA(xi))
    return dgt
EPrec = lambda g,mus: integrate(ode_EPrec,g,mus)
\end{lstlisting}

\begin{figure}[ht]
    \begin{center}
    \begin{minipage}{0.5\textwidth}
        \centering
        \includegraphics[scale=0.5, trim = 100 60 60 70,clip]{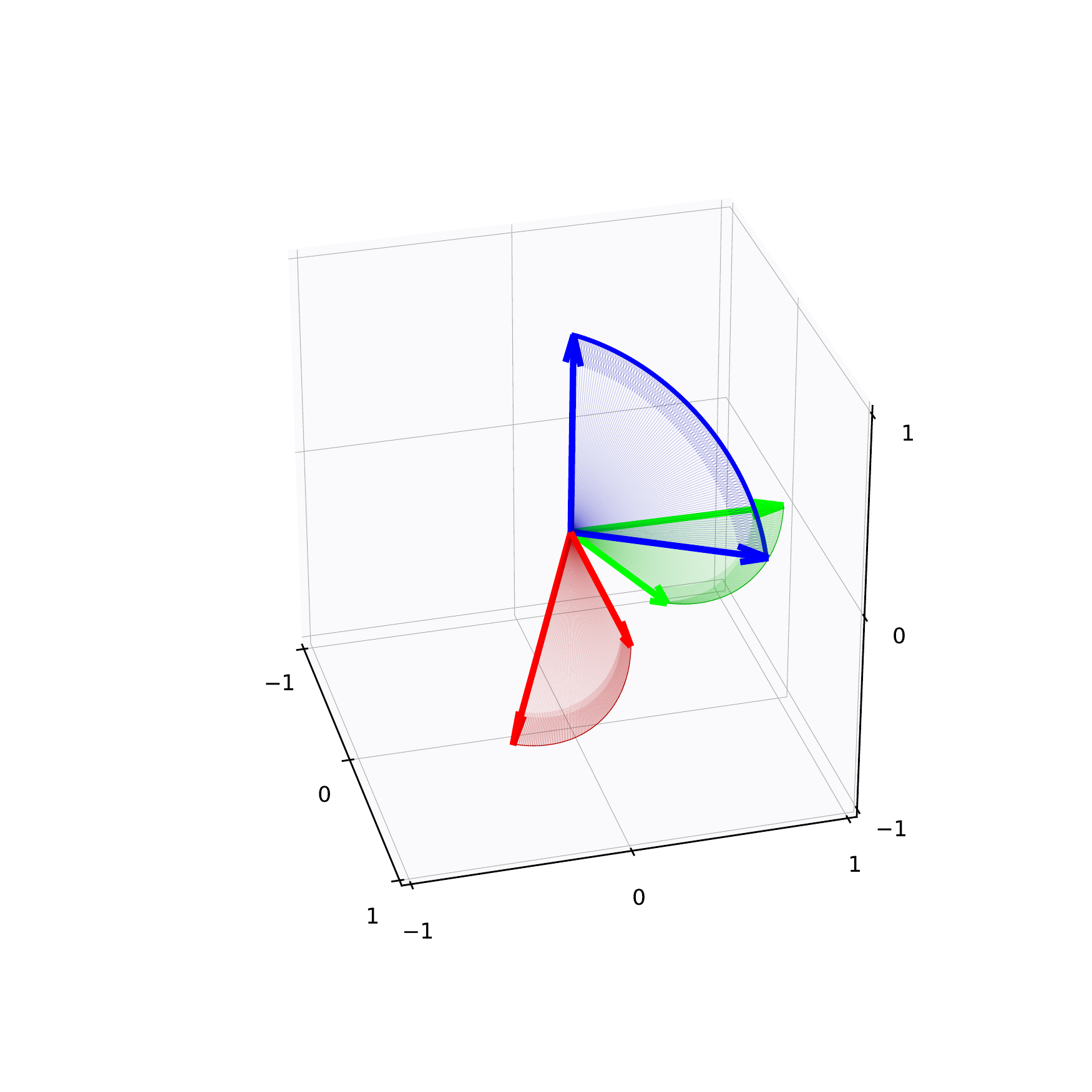}
    \end{minipage}%
    \begin{minipage}{0.5\textwidth}
        \centering
        \includegraphics[scale=0.5,trim = 100 60 60 70,clip]{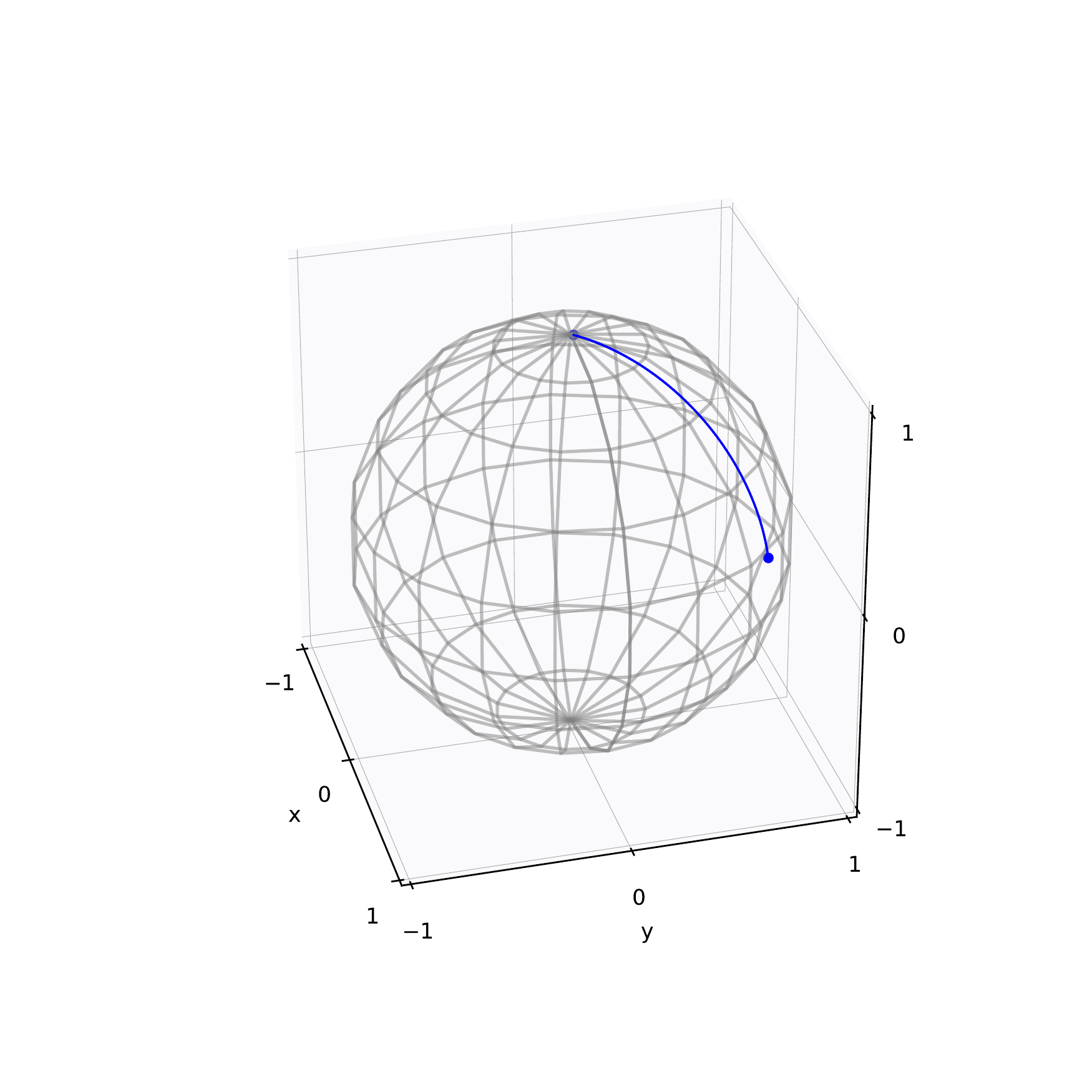}
    \end{minipage}
    \caption{(left) Geodesic on $\SO(3)$ found by the Euler-Poincar\'e equations. (right) The geodesic on $\SO(3)$ projected to the sphere using the left action $g.x=gx$ for $x\in S^2\subset\RR^3$.}
    \label{fig:geoso3}
    \end{center}
\end{figure}

\begin{example}[Geodesic on $SO(3)$]
Let $g_0\in G$ be the identity matrix. 
An example of a geodesic on $\SO(3)$ found as the solution to the Euler-Poincar\'e equation is shown in Figure \ref{fig:geoso3}.
\end{example}

\subsection{Brownian motion on $\boldsymbol{G}$}

In the following subsection, we will go through a construction of Brownian motions on a group $G$ where the evolution is given as a Stratonovich SDE. 
With a group structure, we can simulate a Brownian motion which remains in the group $G$. 
Using the inner product $A$, let $e_1,\ldots,e_d$ be an orthonormal basis for $\mathfrak g$, and construct an orthonormal set of vector fields on the group as $X_i(g)=dL_ge_i$, for $g\in G$. 
Recall that the structure constant of the Lie algebra $C\indices{^i_{jk}}$ are the same as in the commutator of these vector fields, that is
\begin{equation}
  [X_j,X_k] = C\indices{^i_{jk}}X_i\, . 
  \label{eq:structcoeffs}
\end{equation}
The corresponding Brownian motion on $G$ is the following Stratonovich SDE
\begin{equation}
  dg_t = - \frac{1}{2}\sum_{j,i}C\indices{^j_{ij}}X_i(g_t)dt + X_i(g_t)\circ dW_t^i\, , 
  \label{eq:lieabrowneq}
\end{equation}
where $W_t$ is an $\RR^d$-valued Wiener processes. 
We refer to \cite{liao_levy_2004} for more information on Brownian motions on Lie groups.

In Theano, the stochastic process \eqref{eq:lieabrowneq} can be integrated with the following code. 

\begin{lstlisting}
"""
SDE for Brownian Motions on a Lie group G

Args:
    g: Starting point for the process
    dW: Steps of a Euclidean Brownian motion

Returns:
    Tensor (t,gt)
        t: Time evolution
        gt: Evolution of g    
"""
def sde_Brownian(dW,t,g):
    X = T.tensordot(dL(g,e,eiLA),sigma,(2,0))
    det = -.5*T.tensordot(T.diagonal(C,0,2).sum(1),X,(0,2))
    sto = T.tensordot(X,dW,(2,0))
    return (det,sto)
Brownian = lambda g,dWt: integrate_sde(sde_Brownian,integrator_stratonovich,g,dWt)
\end{lstlisting}

Here, we used \lstinline!integrate_sde! which is a discrete time stochastic integrator as described in section \ref{sec:stocnum}. 

\begin{figure}[ht]
    \begin{center}
    \begin{minipage}{0.5\textwidth}
        \centering
        \includegraphics[scale=0.5, trim = 100 60 60 70,clip]{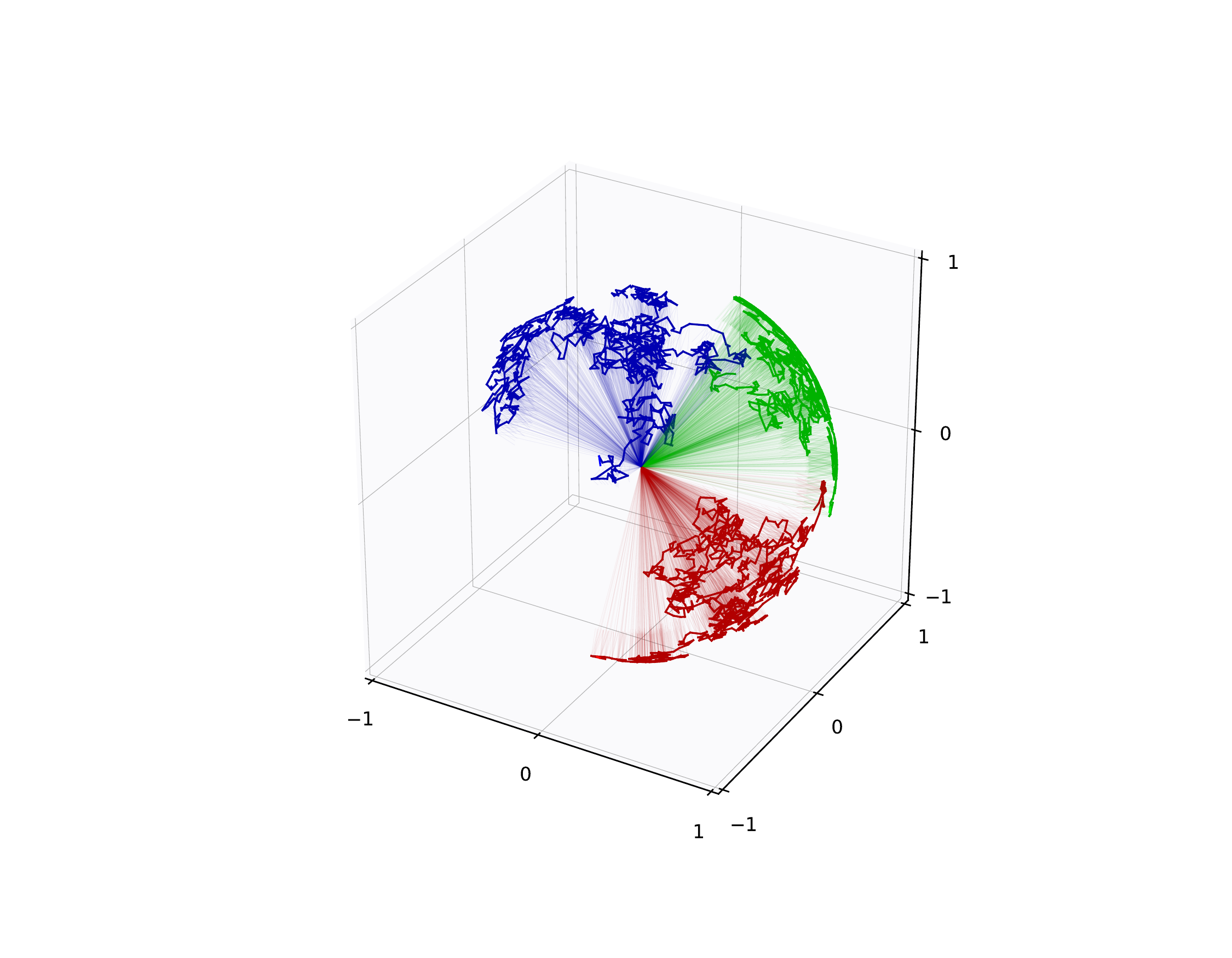}
    \end{minipage}%
    \begin{minipage}{0.5\textwidth}
        \centering
        \includegraphics[scale=0.5,trim = 100 60 60 70,clip]{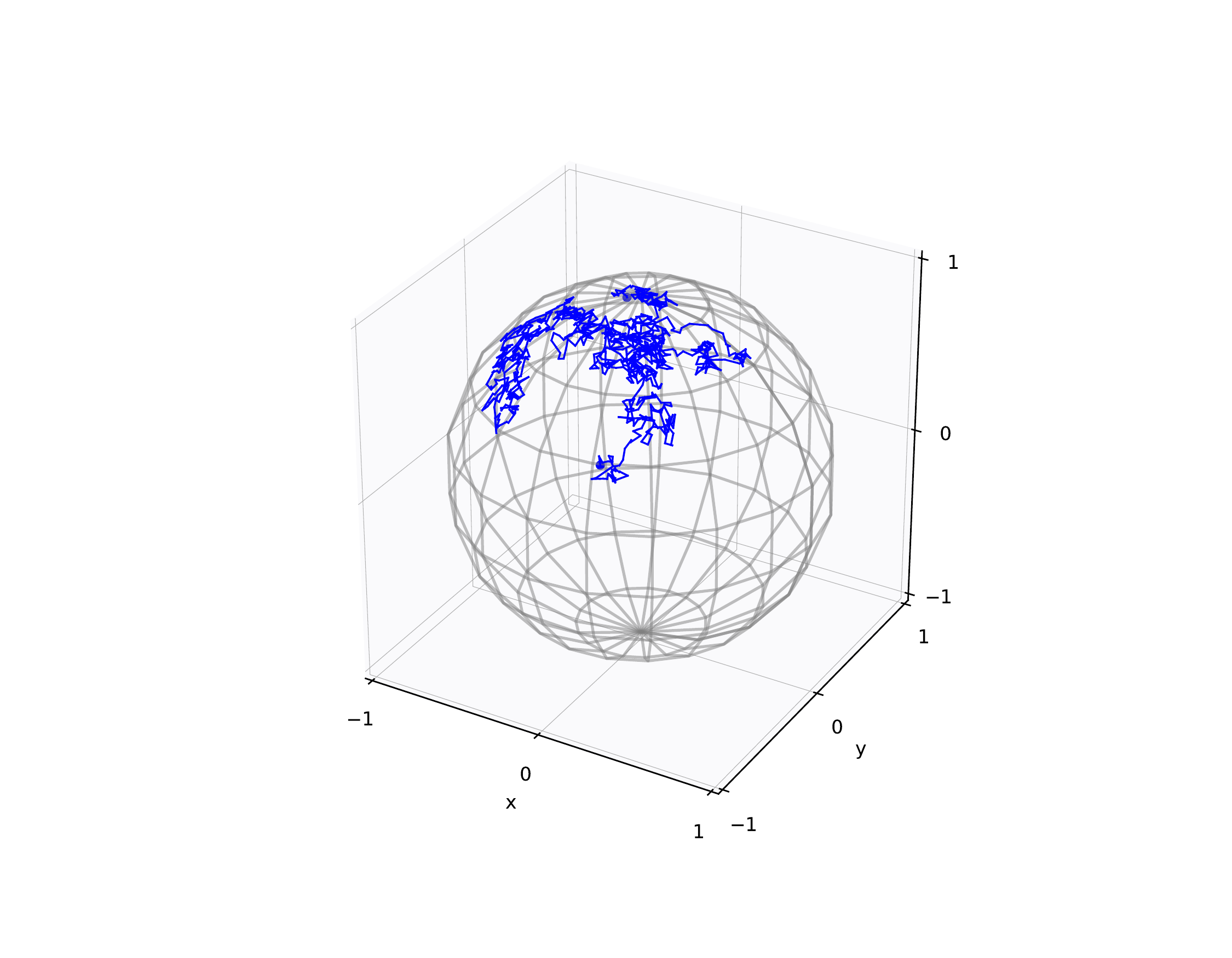}
    \end{minipage}
    \caption{(left) Brownian motion on the group $\SO(3)$ defined by \eqref{eq:lieabrowneq}. The initial point, $x_0\in\SO(3)$, was set to the identity matrix. (right) The projection by the left action of the Brownian motion on $\SO(3)$ to the sphere}
    \label{fig:brownso3}
    \end{center}
\end{figure}

\begin{example}[Brownian motion on $SO(3)$]
Figure \ref{fig:brownso3} shows an example of a Brownian motion on $\SO(3)$. The initial point $x_0\in\SO(3)$ for the Brownian motion was the 3-dimensional identity matrix.
\end{example}

There are other ways of defining stochastic processes on a Lie group $G$. 
An example can be found in \cite{arnaudon_noise_2016} for finite dimensional Lie groups and in \cite{holm_variational_2015} for infinite dimensions.
See also \cite{cruzeiro2017momentum} for the general derivation of these stochastic equations. 
In this theory, the noise is introduced in the reconstruction relation to form the motion on the dual of the Lie algebra to the Lie group and appears in the momentum formulation of the Euler-Poincar\'e equation given in \eqref{eq:EP}. 
This framework has also been implemented in Theano and can be found in the repository. 
Another interesting approach, not yet implemented in Theano, is the one of \cite{arnaudon2014stochastic}, where noise is introduced on the Lie group, and an expected reduction by symmetries results in a dissipative deterministic Euler-Poincar\'e equation. 

\section{Sub-Riemannian Frame Bundle Geometry}
\label{sec:FM}

We now consider dynamical equations on a more complicated geometric construction, a frame bundle or more generally fibre bundles.
A frame bundle $F\M = \{F_x\M\}_{x\in\M}$ is the union of the spaces $F_x\M$, the frames of the tangent space at $x\in\M$. 
A frame $\nu\colon\RR^d\to T_x\M$ is thus an ordered basis for the tangent space $T_x\M$. 
The frame bundle $F\M$ is a fibre bundle $\pi:FM\to M$ with projection $\pi$ and can be equipped with a natural sub-Riemannian structure induced by the metric $g$ on $\M$~\cite{mok_differential_1978}. 
Given a connection on $\M$ the tangent space $TF\M$ can be split into a horizontal and vertical subspace, $HF\M$ and $VF\M$, i.e. $TF\M = HF\M\oplus VF\M$. 
Consider a local trivialization $u = (x,\nu)$ of $F\M$ so that $\pi(u)=x$.
A path $u_t=(x_t,\nu_t)$ on $F\M$ is horizontal if $\dot{u}_t\in HFM$ for all $t$. 
A horizontal motion of $u_t$ corresponds to a parallel transport of the frame along the curve $\pi(u_t)$ on $\M$. 
Consequently, the parallel transport $\nu_t$ of a frame $\nu_0$ of $T_{x_0}\M$ along a curve $x_t$ on $\M$ is called a horizontal lift of $x_t$. 

Let $\partial_i = \frac{\partial}{\partial x^i}$, $i=1,\ldots,d$ be a coordinate frame and assume that the frame $\nu$ has basis vectors $\nu_\alpha$ for $\alpha=1,\ldots,d$ such that $(x,\nu)$ has coordinates $(x^i,\nu_\alpha^i)$ where $\nu_\alpha = \nu_\alpha^i\frac{\partial}{\partial x^i}$. 
In these coordinates, a matrix representation of a sub-Riemannian metric $g_{F\M}\colon TF\M^\star\to HF\M$ is given by
\begin{align}
    (g_{F\M})_{ij} = \begin{pmatrix}
W^{-1} & -W^{-1}\Gamma^T \\
-\Gamma W^{-1} & \Gamma W^{-1}\Gamma^T
\end{pmatrix}\, ,
\end{align}
where $(W^{-1})^{ij} = \delta^{\alpha\beta}\nu_\alpha^i\nu_\beta^j$ and the matrix $\Gamma = (\Gamma^{k_\alpha}_{i})$ has elements $\Gamma^{k_\alpha}_i = \Gamma\indices{^k_{ij}} \nu_\alpha^j$. 
We refer to \cite{strichartz_sub-riemannian_1986,mok_differential_1978,sommer_anisotropic_2015} for more details on sub-Riemannian structures and the derivation of the sub-Riemannian metric on $F\M$.
Using the sub-Riemannian metric $g_{F\M}$, normal geodesics on $F\M$ can be generated by solving Hamilton's equations described earlier in \eqref{eq:Hamilton}.

\begin{example}[Normal sub-Riemannian geodesics on $F\M$]
    With the same setup as in Example \ref{ex:geoSphere}, let $u_0 = (x_0,\nu_0)\in FS^2$ such that $x_0 = F(0,0)$ and $\nu_0$ has orthonormal frame vectors $\nu_1 = dF(0.5,0)$, $\nu_2 = dF(0,0.5)$. 
    Figure \ref{fig:FMgeo} shows two geodesics on $FS^2$ visualised on $S^2$ with different initial momenta.

\begin{figure}[ht]
    \begin{center}
    \begin{minipage}{0.5\textwidth}
        \centering
        \includegraphics[scale=0.5, trim = 190 60 170 70,clip]{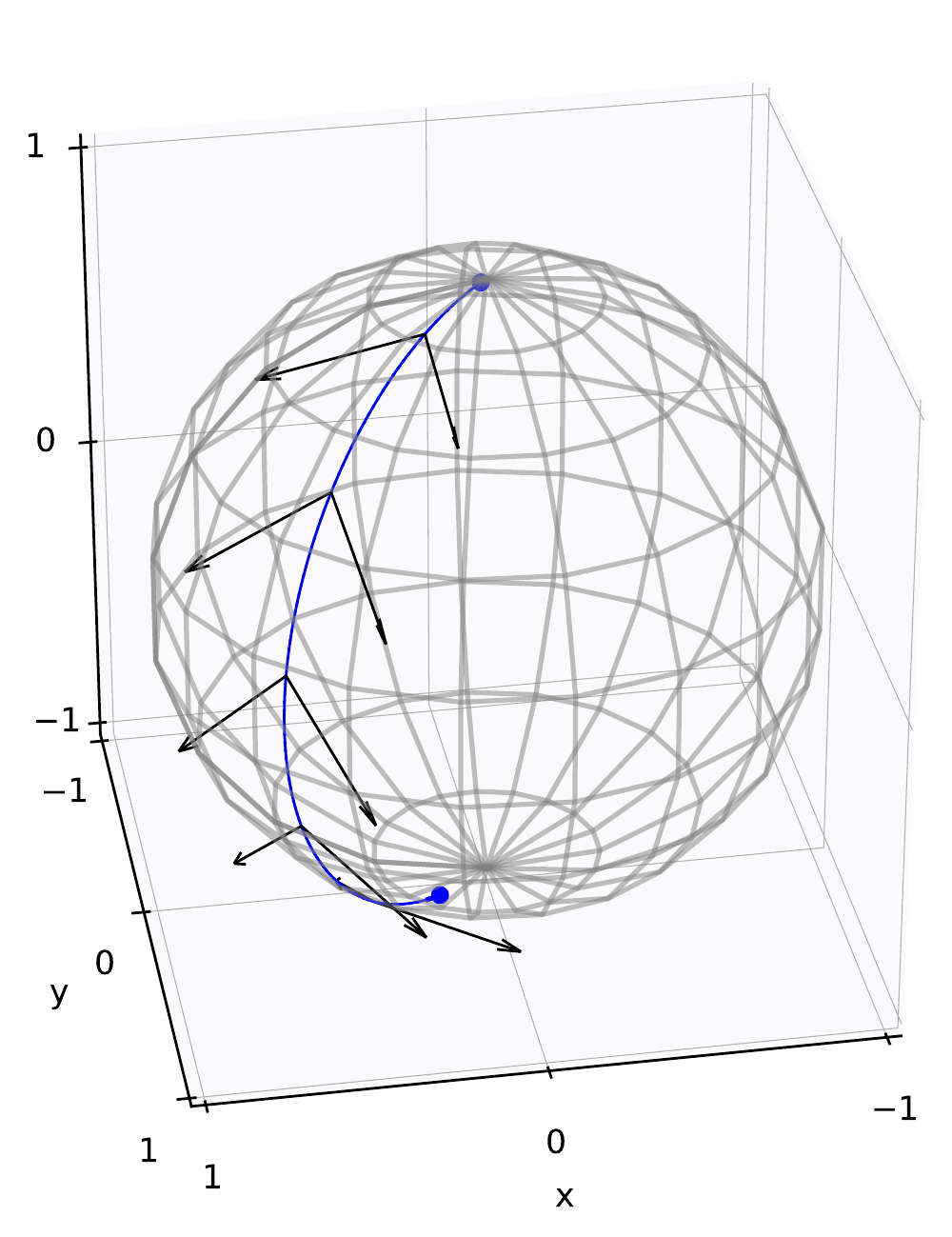}
    \end{minipage}%
    \begin{minipage}{0.5\textwidth}
        \centering
        \includegraphics[scale=0.5,trim = 190 60 170 70,clip]{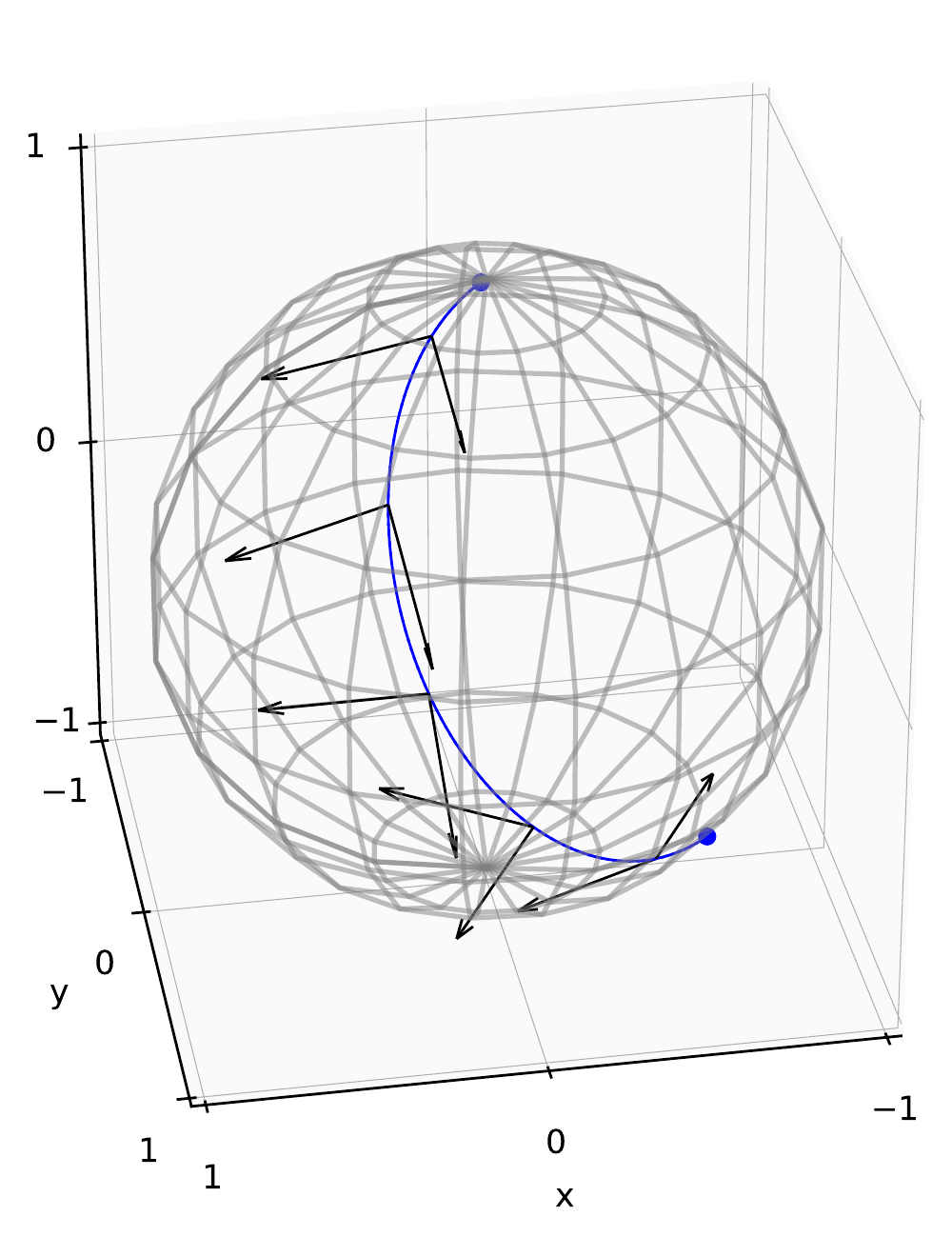}
    \end{minipage}
    \caption{Geodesics on $FS^2$ solving Hamiltion's equations for the sub-Riemannian metric $g_{F\M}$ with different initial momenta. 
    The curves on $S^2$ show the evolution of $x_t$ while the evolution of the frame $\nu_t$ is shown by the tangent vectors in $T_{x_t}S^2$.}
    \label{fig:FMgeo}
    \end{center}
\end{figure}

\end{example}

\subsection{Curvature}

The curvature form on the frame bundle is defined from the Riemannian curvature tensor $R\in\mathcal{T}_1^3(\M)$ described in section \ref{sec:curv}~\cite{kolar_natural_1993}. 
Let $u = (x,\nu)$ be a point in $F\M$, the curvature form $\Omega\colon TF\M\times TF\M\to\mathfrak{gl}(d)$ on the frame bundle is 
\begin{equation}
    \Omega(v_u,w_u) = u^{-1}R(\pi_*(v_u),\pi_*(w_u))u\, , \quad  v_u,w_u\in T_uF\M\, ,
\end{equation}
where $\pi_*\colon TF\M\to T\M$ is the projection of a tangent vector of $F\M$ to a tangent vector of $\M$. 
By applying the relation, $\Omega(v_u,w_u) = \Omega(h_u(\pi_*(v_u)),h_u(\pi_*(w_u)))$, where $h_u:T_{\pi(u)}\M\rightarrow H_uF\M$ denotes the horizontal lift, the curvature tensor $R$ can be considered as a $\mathfrak{gl}(d)$ valued map 
\begin{align}
  \begin{split}
    R_u&\colon \mathcal{T}^2(T_{\pi(u)}\M)\to\mathfrak{gl}(d)\\
    (v,w)&\mapsto\Omega(h_u(\pi_*(v_u)),h_u(\pi_*(w_u)))\, , 
  \end{split}
\end{align}
for $u\in F\M$. 
The implementation of the curvature form $R_u$ is shown in the code below.

\begin{lstlisting}
"""
Riemannian Curvature form
R_u (also denoted Omega) is the gl(n)-valued curvature form u^{-1}Ru for a frame u for T_xM

Args:
    x: point on the manifold

Returns:
    4-tensor (R_u)_ij^m_k with order i,j,m,k
"""
def R_u(x,u):
    return T.tensordot(T.nlinalg.matrix_inverse(u), 
                         T.tensordot(R(x),u,(2,0)),
                         (1,2)).dimshuffle(1,2,0,3)
\end{lstlisting}

\begin{figure}
    \centering
    \includegraphics[scale=0.6,trim = 100 60 60 70,clip]{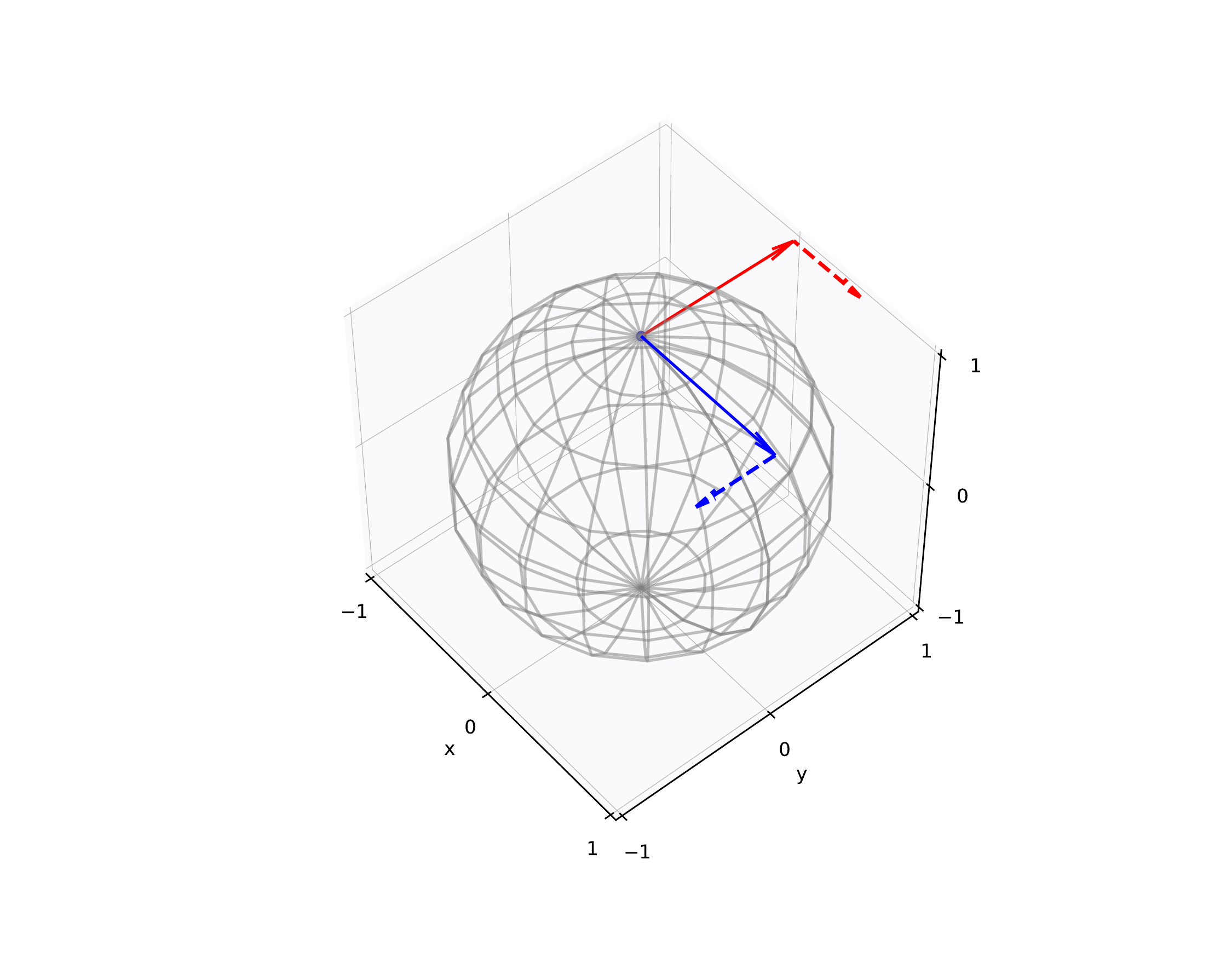}
    \caption{Curvature of each basis vector of $\nu$. The solid arrows represents the basis vectors, while the dashed arrows are the curvature form $\Omega(\nu_1,\nu_2)$. 
The figure shows in which direction the basis vectors would change if they were parallel transported around an infinitesimal parallelogram spanned by the basis vectors of $\nu$.}
    \label{fig:curv}
\end{figure}

\begin{example}[Curvature on $S^2$]
Let $u = (x,\nu)\in F\M$ with $x = F(0,0)$ and $\nu$ as shown in Figure \ref{fig:curv} (solid arrows). 
We visualize the curvature at $u$ by the curvature form $\Omega(\nu_1,\nu_2)$, calculated by applying $R$ to the basis vectors of $\nu$. 
The curvature is represented in Figure \ref{fig:curv} by the dashed vectors showing the direction for which each basis vector change by parallel transporting the vectors around an infinitesimal parallelogram spanned by $\nu$. 
\end{example}

\subsection{Development and Stochastic Development}
\label{sec:development}

The short description of the development process in this section is based on the book~\cite{hsu_stochastic_2002}. 
The presented approach has also been described in~\cite{elworthy_geometric_1988,sommer_anisotropic_2015,sommer_modelling_2017}, where the method is used for generalisation of Brownian motions to manifolds.

Using the frame bundle and its horizontal and vertical splitting, deterministic paths and stochastic processes on $F\M$ can be constructed from paths and stochastic processes on $\RR^d$. 
In the deterministic case, this process is called development and when mapping Euclidean semi-martingales to $\M$-valued semi-martingales, the corresponding mapping is stochastic development. 
The development unrolls paths on $F\M$ by taking infinitesimal steps corresponding to a curve in $\RR^d$ along a basis of $HF\M$.
Let $e\in\RR^d$ and $u=(x,\nu)\in F\M$, then a horizontal vector field $H_e\in H_uF\M$ can be defined by the horizontal lift of the vector $\nu e\in T_x\M$, that is 
\begin{align*}
H_e(x) = h_u(\nu e)\,.
\end{align*}
If $e_1,\ldots,e_d$ is the canonical basis of $\RR^d$, then for any $u\in F\M$, a basis for the horizontal subspace $H_uF\M$ is represented by the horizontal vector fields $H_i(x) = H_{e_i}(x)$, $i = 1,\ldots,d$. 
Consider a local chart $(U,\varphi)$ on $\M$, the coordinate basis $\partial_i = \frac{\partial}{\partial x^i}$ on $U$, and the projection map $\pi\colon F\M\to\M$, then the coordinate basis $\partial_i$ induces a local basis on the subset $\tilde{U} = \pi^{-1}(U)\subseteq F\M$. 
Notice that the basis vectors $\nu e_1,\ldots,\nu e_d$ of $T_x\M$ can be written as $\nu e_j = \nu_j^i\partial_i$ for each $j=1,\ldots,d$. 
Hence $(x^i,\nu_j^i)$ is a chart for $\tilde{U}$ and $\left (\frac{\partial}{\partial x^i}, \frac{\partial}{\partial \nu_j^i}\right )$ spans the tangent space $T_uF\M$. 
The horizontal vector fields can be written in this local coordinate basis as
\begin{align}
\label{eq:Hori}
    H_i(q) = \nu_i^j\frac{\partial}{\partial x^j} - \nu_i^j\nu_m^l\Gamma_{jl}^k\frac{\partial}{\partial \nu_m^k}\, .
\end{align}
The code below shows how these horizontal vector fields in the local basis can be implemented in Theano.

\begin{lstlisting}
"""
Horizontal Vector Field Basis

Args:
    x: Point on the manifold
    nu: Frame for the tangent space at x
    Gamma_g(x): Christoffel symbols at x

Returns:
    Matrix of coordinates for each basis vector
"""
def Hori(x,nu): 
    dnu = - T.tensordot(nu, T.tensordot(nu,Gamma_g(x),axes = [0,2]),
                          axes = [0,2])
    dnu = dnu.reshape((nu.shape[1],dnu.shape[1]*dnu.shape[2]))
    return T.concatenate([nu,dnu.T], axis = 0)
\end{lstlisting}

\begin{figure}[ht]
    \centering
    \includegraphics[scale=0.6,trim = 100 40 60 70,clip]{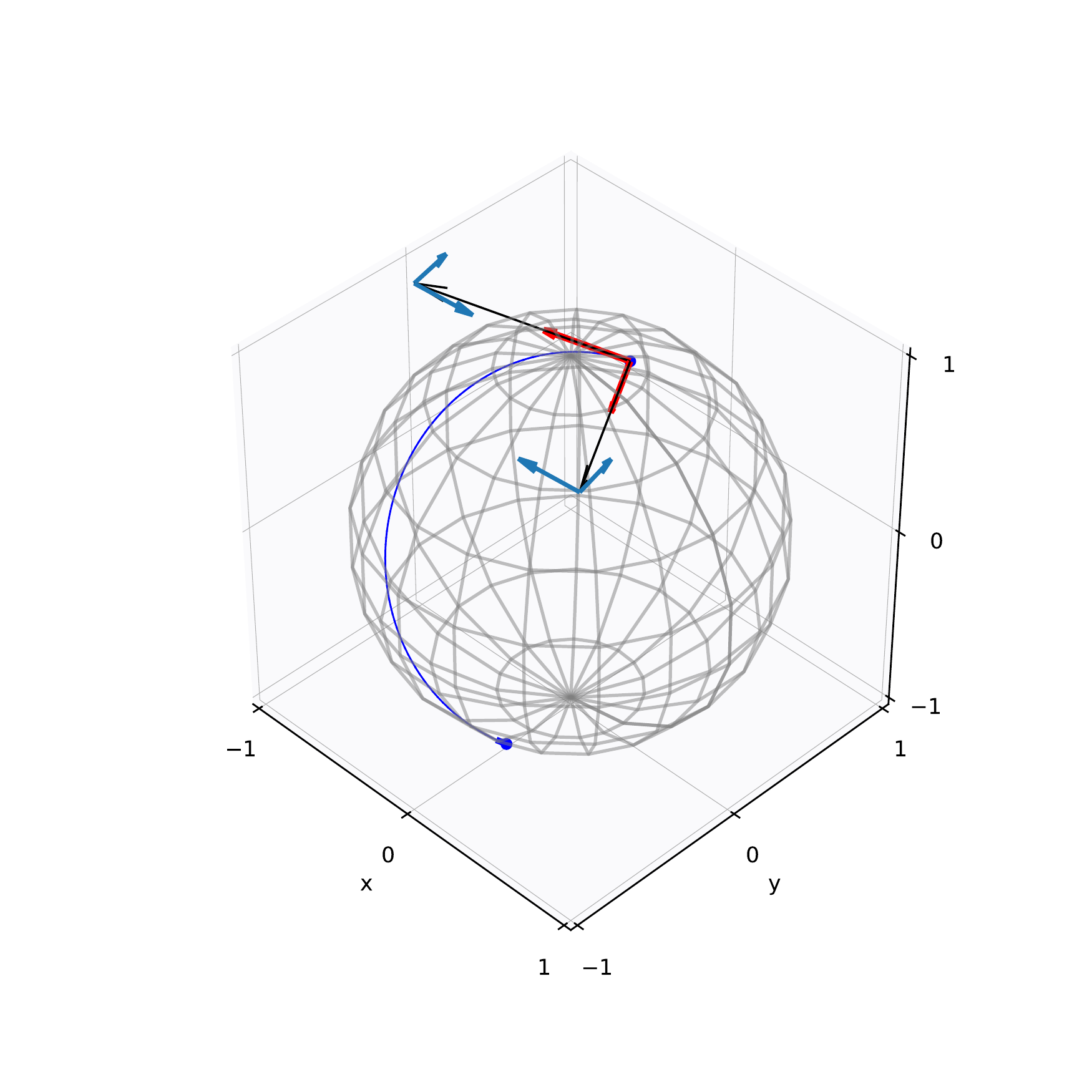}
    \caption{Horizontal vector fields for the point $u = (x,\nu)\in F\M$ with $x = F(0.1,0.1)$ and the frame $\nu$ visualized with  black arrows. 
    The horizontal tangent vectors at $x$ is shown in red and the horizontal tangent vectors for each tangent vector at $\nu$ is shown in blue.}
    \label{fig:Hori}
\end{figure}

\begin{example}[Horizontal vector fields]
\label{ex:hori}
Figure~\ref{fig:Hori} illustrates the horizontal vector fields $H_i$ at a point $u\in FS^2$. 
Let $u = (x,\nu)$ with $x = F(0.1,0.1)\in S^2$ and $\nu$ being the black frame shown in the figure. 
The horizontal basis for $u$ is then found by \eqref{eq:Hori} and is plotted in Figure \ref{fig:Hori} with the red frame being the horizontal basis vectors for $x$ and the blue frames are the horizontal basis vectors for each frame vector in $\nu$. 
The horizontal basis vectors describe how the point $x$ and the frame $\nu$ change horizontally.
\end{example}

Let now $W_t$ be a $\RR^d$-valued Euclidean semi-martingale, e.g. a Brownian motion. 
The stochastic version of the development maps $W_t$ to $F\M$ by the solution to the Stratonovich stochastic differential equation
\begin{align}
\label{eq:stocdev}
    dU_t = \sum_{i=1}^d H_i(U_t)\circ dW^i_t\, .
\end{align}
The solution $U_t$ to this stochastic differential equation is a path in $F\M$ for which a stochastic path on $\M$ can be obtained by the natural projection $\pi:U_t\to \M$.
The stochastic development of $W_t$ will be denoted $\varphi_{u_0}(W_t)$ where $u_0\in F\M$ is the initial point on $F\M$.
In Theano this Stratonovich stochastic differential equation can be implemented as follow.

\begin{lstlisting}
"""
Stochastic Development

Args:
    dW: Steps of stochastic process
    u: Point in FM
    drift: Vector of constant drift of W

Returns:
    det: Matrix of deterministic evolution of process on FM
    sto: Matrix of stochastic evolution of the process
"""
def stoc_dev(dW,u,drift):
    x = u[0:d]
    nu = u[d:(d+rank*d)].reshape((d,rank))
    det = T.tensordot(Hori(x,nu), drift, axes = [1,0]) 
    sto = T.tensordot(Hori(x,nu), dW, axes = [1,0])
    return det, sto
\end{lstlisting}

The variable \lstinline!drift! can be used to find the stochastic development of a process with defined drift.
The numerical solution to this SDE requires the use of stochastic numerical integration methods, described in the appendix \ref{sec:stoc}, such as the Euler-Heun scheme, used in the example below. 

\begin{example}[Deterministic and stochastic Development]
\label{ex:dev}
Let $\gamma_t$ be a curve in $\RR^2$ defined by 
\begin{align*}
    \gamma(t) = (20\sin(t),t^2 + 2t), \quad t\in [0,10]\, ,
\end{align*}
and $x = F(0,0)\in S^2$. 
Consider the orthonormal frame for $T_x\M$ given by the Gram-Schmidt decomposition based on the metric $g$ of the vectors $v_1 = dF(-1,1), \ v_2 = dF(1,1)$. 
The curve $\gamma_t$ is a deterministic process in $\RR^2$ and hence \eqref{eq:stocdev} can be applied to obtain the development of $\gamma_t$ to $S^2$. 
In Figure \ref{fig:dev} is shown the curve $\gamma_t$ and its development on the sphere.
\begin{figure}
    \begin{center}
    \begin{minipage}{0.5\textwidth}
        \centering
        \includegraphics[scale=0.5, trim = 20 20 30 20,clip]{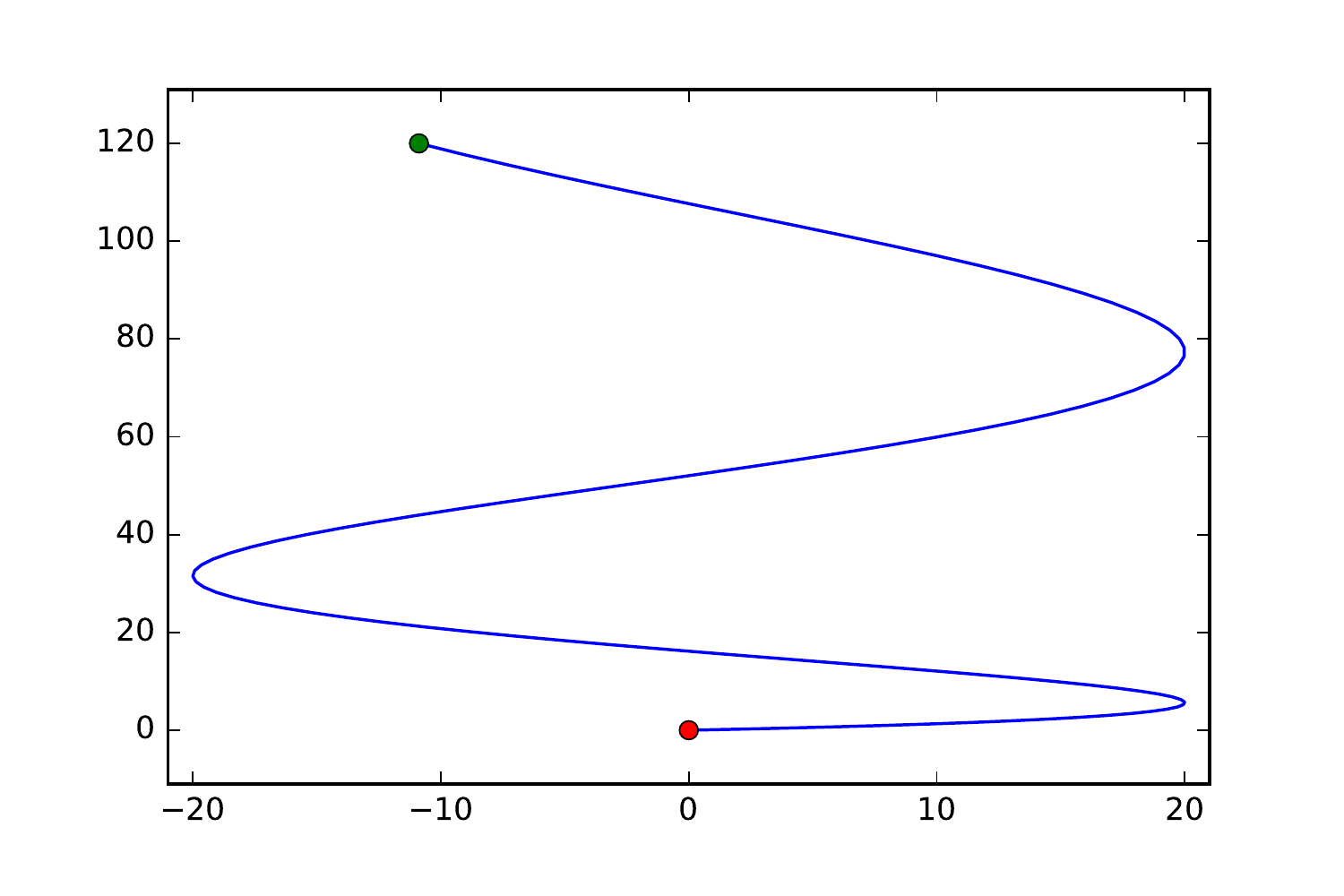}
    \end{minipage}%
    \begin{minipage}{0.5\textwidth}
        \centering
        \includegraphics[scale=0.5,trim = 100 60 60 70,clip]{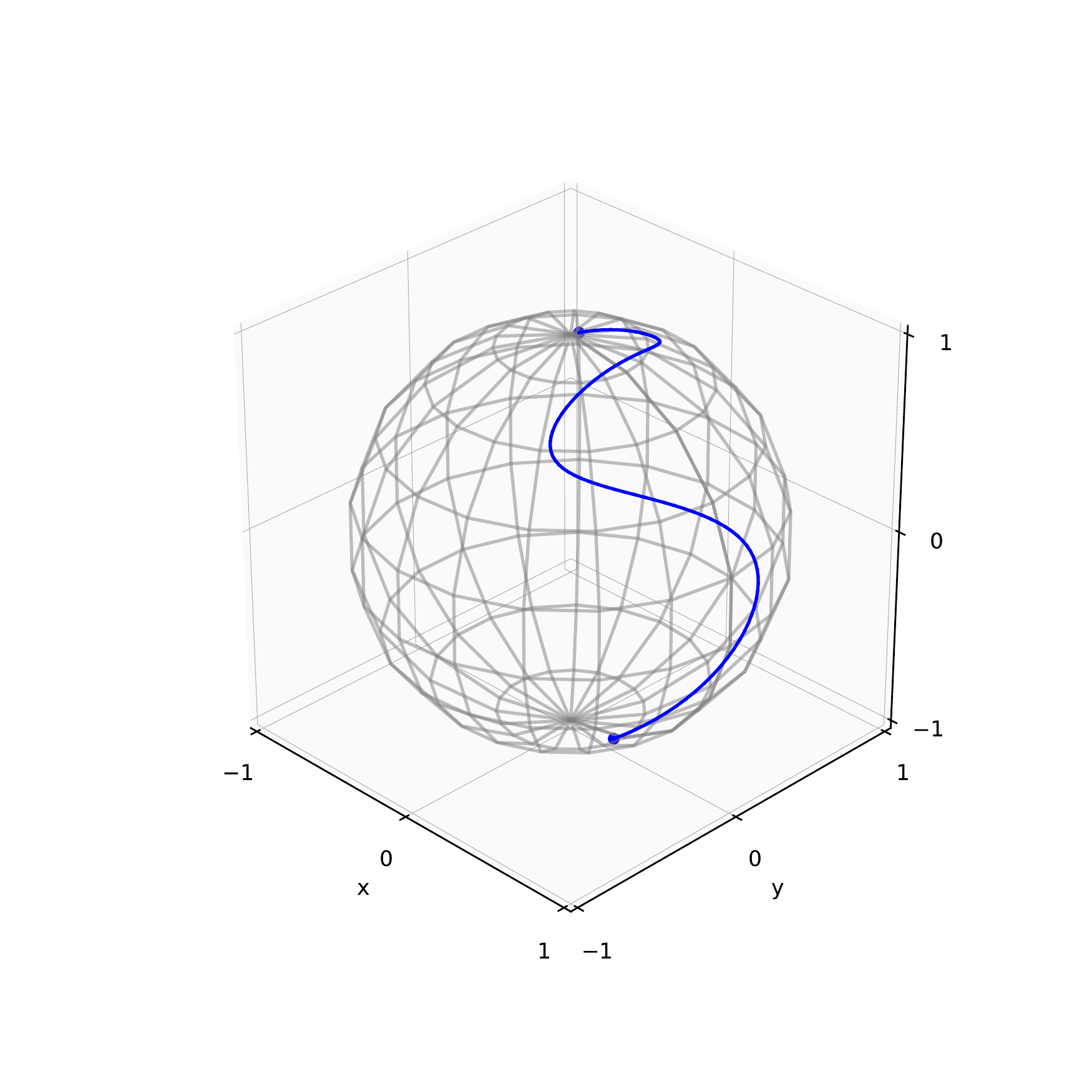}
    \end{minipage}
    \caption{(left) The curve $\gamma_t$ defined in Example \ref{ex:dev}. The red and green point denotes the start and endpoint respectively. (right) The development of $\gamma_t$ on the sphere.}
    \label{fig:dev}
    \end{center}
\end{figure}

Let then $X_t$ be a stochastic process in $\RR^2$ defined from a Brownian motion, $W_t$, with drift, $\beta$. 
Discretizing in time, the increments $dW_t$ follow the normal distribution $\mathcal{N}(0,dt I_2)$, here with $dt = 0.0001$. 
Let $\beta = (0.5,0.5)$ such that 
\begin{align*}
    dX_t = dW_t + \beta dt\, .
\end{align*}
A sample path of $X_t$ is shown in Figure \ref{fig:stocdev}. 
The stochastic development of $X_t$ is obtained as the solution to the Stratonovich stochastic differential equation defined in \eqref{eq:stocdev}. 
The resulting stochastic development on $S^2$ is shown in the right plot of Figure \ref{fig:stocdev}.
\begin{figure}[ht]
    \begin{center}
    \begin{minipage}{0.5\textwidth}
        \centering
        \includegraphics[scale=0.5, trim = 20 20 30 20,clip]{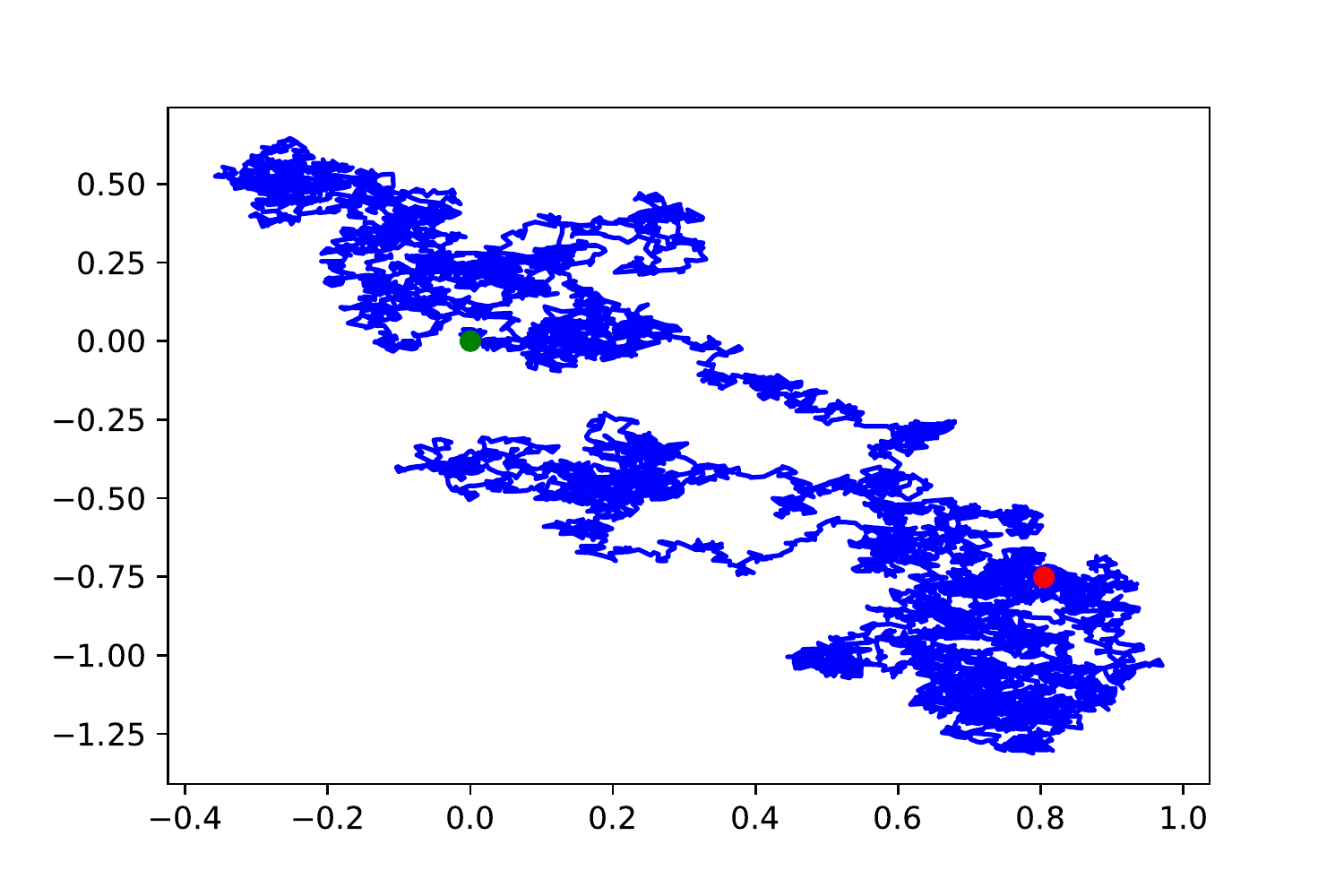}
    \end{minipage}%
    \begin{minipage}{0.5\textwidth}
        \centering
        \includegraphics[scale=0.5,trim = 100 60 60 70,clip]{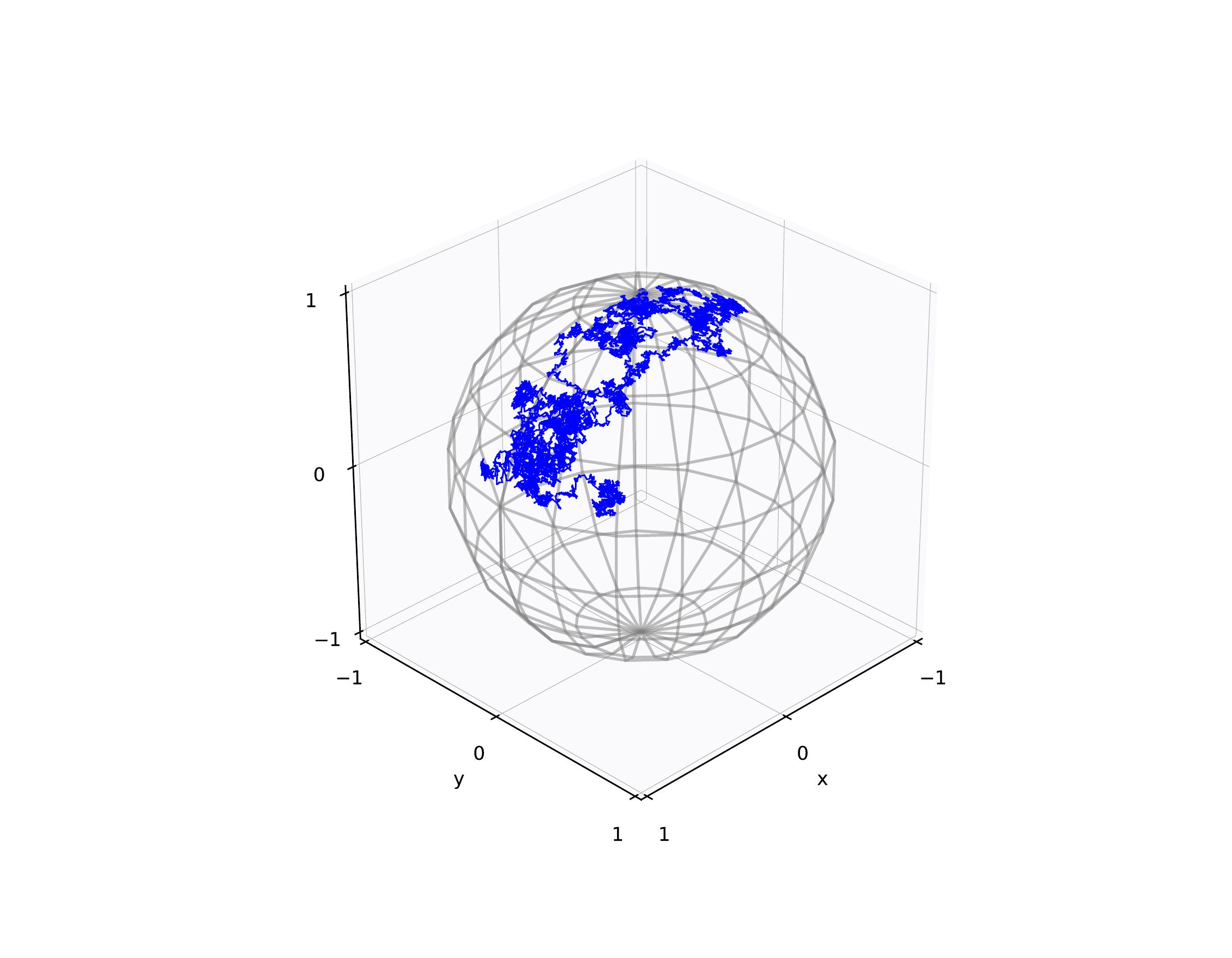}
    \end{minipage}
    \caption{(left) The stochastic process $X_t$ defined in Example \ref{ex:dev}. The red and green point denotes the start--and endpoint respectively of the process. (right) The stochastic development of $X_t$ on $S^2$.}
    \label{fig:stocdev}
    \end{center}
\end{figure}
\end{example}

\subsection{Most Probable Path equations}

The most common distance measure on Riemannian manifolds is the geodesic distance. 
However, in contexts where data exhibit non-trivial covariance, it is argued in \cite{sommer_anisotropic_2015,sommer_modelling_2017} that weighting the geodesic energy by the inverse of the covariance, the precision, gives a useful generalization of the geodesic distance. Extremal paths for the corresponding variational problem are precisely projections of $F\M$ geodesics with respect to the sub-Riemannian metric $g_{F\M}$ constructed earlier. 
These paths also have an interpretation as being most probable for a specific measure on the path space.
 
More formally, let $X_t$ be a stochastic process with $X_0=x_0$. Most probable paths in the sense of Onsager-Machlup \cite{fujita_onsager-machlup_1982} between $x_0,y\in\M$ are curves $\gamma_t\colon [0,1]\to\M$, $\gamma_0=x_0$ maximizing
\begin{equation}
    \mu_\varepsilon^M(\gamma_t) = P(d_g(X_t,\gamma_t)<\varepsilon,\ \forall t\in[0,1])\, ,
\end{equation}
for $\varepsilon\to 0$ and with the Riemannian distance $d_g$. 
Most probable paths are in general not geodesics but rather extremal paths for the Onsager-Machlup functional 
\begin{equation}
   \int_0^1 L_M(\gamma_t,\dot\gamma_t) \ dt = -E[\gamma_t] + \frac{1}{12}\int_0^1 S(\gamma_t)\ dt\,.
\label{eq:OMf}
\end{equation}
Here, $S$ denotes the scalar curvature of $\M$ and the geodesic energy is given by $E[\gamma_t] = \frac{1}{2}\int_0^1 \|\dot\gamma_t\|^2_g \ dt$. In comparison, geodesics only minimize the energy $E[\gamma_t]$. 

Instead of calculating the MPPs based on the Onsager-Machlup functional on the manifold, the MPPs for the driving process $W_t$ can be found. It has been shown in~\cite{sommer_modelling_2017} that under reasonable conditions, the MPPs of the driving process exist and coincide with projections of the sub-Riemannian geodesics on $F\M$ obtained from \eqref{eq:Hamilton} with the sub-Riemannian metric $g_{F\M}$. 
The implementation of the MPPs shown below is based on this result and hence returns the tangent vector in $T_uF\M$ which leads to the sub-Riemannian geodesic on $F\M$ starting at $u$ and hitting the fibre at $y$.

Let $W_t$ be a standard Brownian motion and $X_t = \varphi_{u_0}(W_t)$, the stochastic development of $W_t$ with initial point $u_0\in F\M$. 
Then, the most probable path of the driving process $W_t$ from $x_0=\pi(u_0)$ to $y\in\M$ is defined as a smooth curve $\gamma_t\colon [0,1]\to\M$ with $\gamma_0 = x_0$, $\gamma_1=y$ satisfying
\begin{equation}
    \argmin_{\gamma_t,\gamma_0=x_0,\gamma_1=y}\ \int_0^1 -L_{\RR^n}\left(\varphi_{u_0}^{-1}(\gamma_t),\frac{d}{dt}\varphi_{u_0}^{-1}(\gamma_t)\right)\ dt\, ,
\end{equation}
that is, the anti-development $\varphi_{u_0}^{-1}(\gamma_t)$ is the most probable path of $W_t$ in $\RR^n$.
The implementation of the MPPs is given below.

\begin{lstlisting}
"""
Most probable paths for the driving process

Args:
    u: Starting point in FM
    y: Point on M

Returns:
    MPP: vector in T_uFM for sub-Riemannian geodesic hitting fiber above y
"""
loss = lambda v,u,y: 1./d*T.sum((Expfm(u,g(u,v))[0:d]-y)**2)
dloss = lambda v,u,y: T.grad(loss(v,u,y),v)
# Returns the optimal horizontal tangent vector defining the MPP:
MPP = minimize(loss, np.zeros(d.eval()), jac=dloss, args=(u,y))
\end{lstlisting}

\begin{figure}[ht]
    \centering
    \includegraphics[scale=0.6,trim = 100 60 60 70,clip]{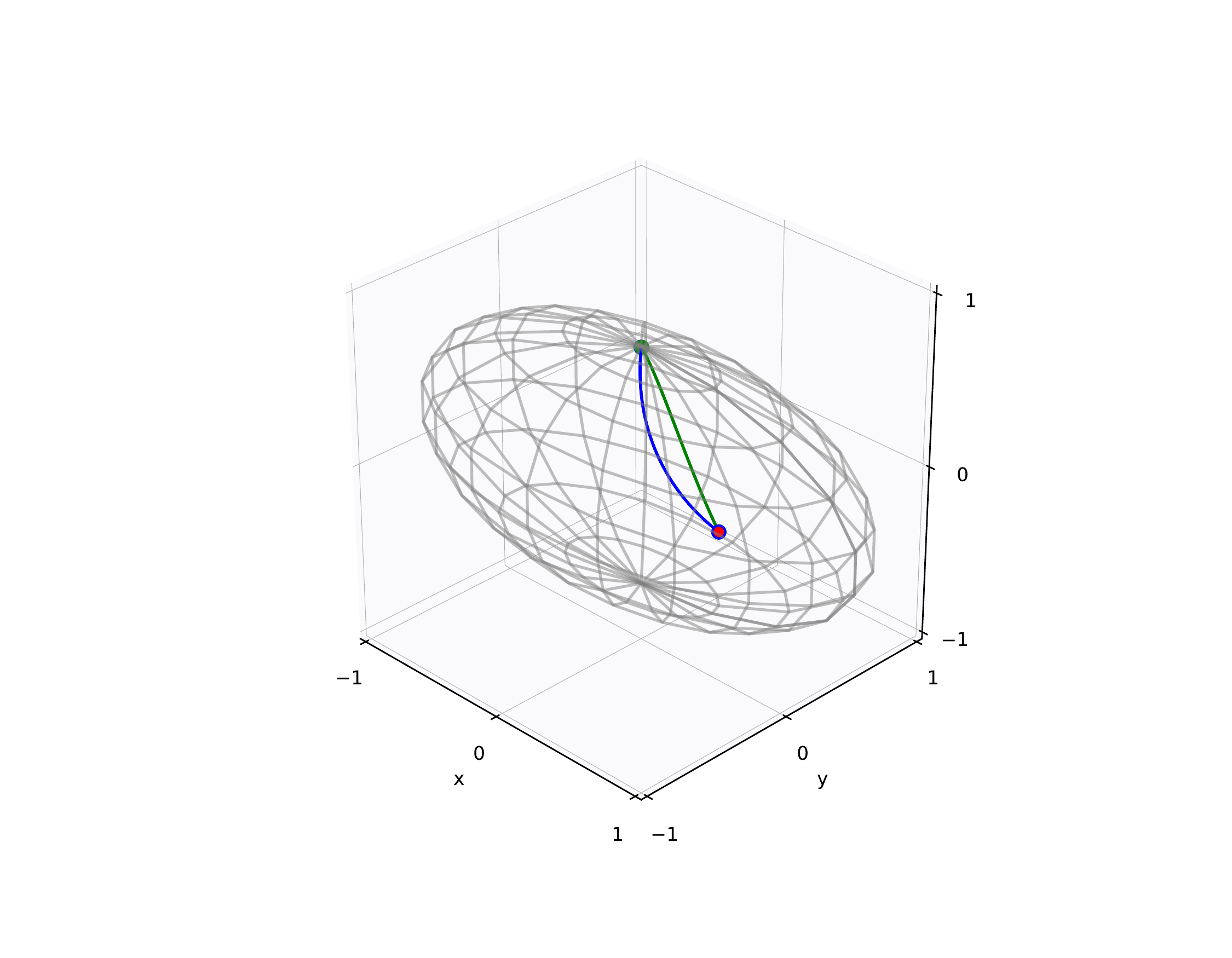}
    \caption{A most probable path between $x_0 = F(0,0)$ and $y = F(0.5,0.5)$ (red point) on an ellipsoid. The blue curve is the MPP and the green the Riemannian geodesic between $x_0$ and $y$.}
    \label{fig:MPP}
\end{figure}

\begin{example}[Most Probable Path on ellipsoid]
Let $u_0 = (x_0,\nu_0)\in F\M$ for which $x_0 = F(0,0)$ and $\nu_0$ consists of the tangent vectors $dF(0.1,0.3)$, $dF(0.3,0.1)$ and $y = F(0.5,0.5)\in S^2$. 
We then obtain a tangent vector $v = (1.03,-5.8,0,0,0,0)\in H_{u_0}F\M$ which leads to the MPP shown in Figure \ref{fig:MPP} as the blue curve. 
For comparison, the Riemannian geodesic between $x_0$ and $y$ is shown in green.
\end{example}

\section{Landmark Dynamics}
\label{sec:landmark}

In this section, we will apply the previous generic algorithms to the example of the manifold of landmarks, seen as a finite dimensional representation of shapes in the Large Deformation Diffeomorphic Metric Mapping (LDDMM).
We will not review this theory in details here but only show how to adapt the previous code to this example. 
We refer to the book \cite{beg2005computing} for more details and LDDMM and landmark dynamics.  

Let $\M\cong \mathbb R^{dn}$ be the manifold of $n$ landmarks with positions $\boldsymbol x_i\in \mathbb R^d$ on a $d$-dimensional space.  
From now on, we will only consider landmarks in a plane, that is $d=2$.
In the LDDMM framework, deformations of shapes are modelled as flows on the group of diffeomorphisms acting on any data structure, which in this case are landmarks. 
To apply this theory, we need to have a special space, a reproducing kernel Hilbert space (RKHS), denoted by $V$.
In general, an RKHS is a Hilbert space of functions for which evaluations of a function $v\in V$ at a point $x\in\M$ can be performed as an inner product of $v$ with a kernel evaluated at $x$. 
In particular, for $v\in V$, $v(x) = \langle K_x,v\rangle_V$ for all $x\in M$, for which $K_x = K(.,x)$. 
In all the examples of this paper, we will use a Gaussian kernel given by
\begin{align}
    K(\boldsymbol x_i,\boldsymbol x_j) = \alpha\cdot\text{exp}\left(\frac{\|\boldsymbol x_i - \boldsymbol x_j\|^2}{2\sigma^2}\right)\, , 
\end{align}
 with standard deviation $\sigma = 0.1$ and a scaling parameter $\alpha\in\RR^d$.  

The diffeomorphisms modelling the deformation of shapes in $\M$ is defined by the flow
\begin{equation}
    \partial_t\varphi(t) = v_t\circ\varphi(t), \ \text{for} \ v_t\in V\, ,
\label{eq:diff}
\end{equation}
where $\varphi:\M \to \M$ and $\circ$ means evaluation $v_t(\varphi)$ for a time-dependent vector field $v_t$. 
Given a shape $x_1\in\M$, a deformation of $x_1$ can be obtained by applying to $x_1$ a diffeomorphism $\varphi$ obtained as a solution of $\eqref{eq:diff}$ for times bteween $0$ and $1$. 
We write $x_2 = \varphi(1)\cdot x_1$, the resulting deformed shape.

Let a shape $x$ in the landmark manifold $\M$ be given as the vector of positions $x=(x^1_1,x^2_1,\ldots,x^1_n,x^2_n)$, where the upper indices are the positions of each landmark on the image. 
Consider $\xi,\eta\in T^*_x\M$. The cometric on $\M$ is thus
\begin{equation}
    g^*_x(\xi,\eta ) = \sum_{i,j=1}^n \xi_i K(\boldsymbol{x}_i,\boldsymbol{x}_j) \eta_j\, ,
\label{eq:landmetric}
\end{equation}
where the components of the cometric are $g^{ij} = K(\boldsymbol{x}_i,\boldsymbol{x}_j)$ for $\boldsymbol{x}_i=(x^1_i,x^2_i)$. 
The coordinates of the metric are the inverse kernel matrix $g_{ij}=K^{-1}(\boldsymbol{x}_i,\boldsymbol{x}_j)$ and the cometric \eqref{eq:landmetric} corresponds to the standard landmark Hamiltonian when $\xi=\eta= p$, the momentum vector of the landmarks. 

\begin{figure}[ht]
    \begin{center}
    \begin{minipage}{0.5\textwidth}
        \centering
        \includegraphics[scale=0.55, trim = 20 20 30 20,clip]{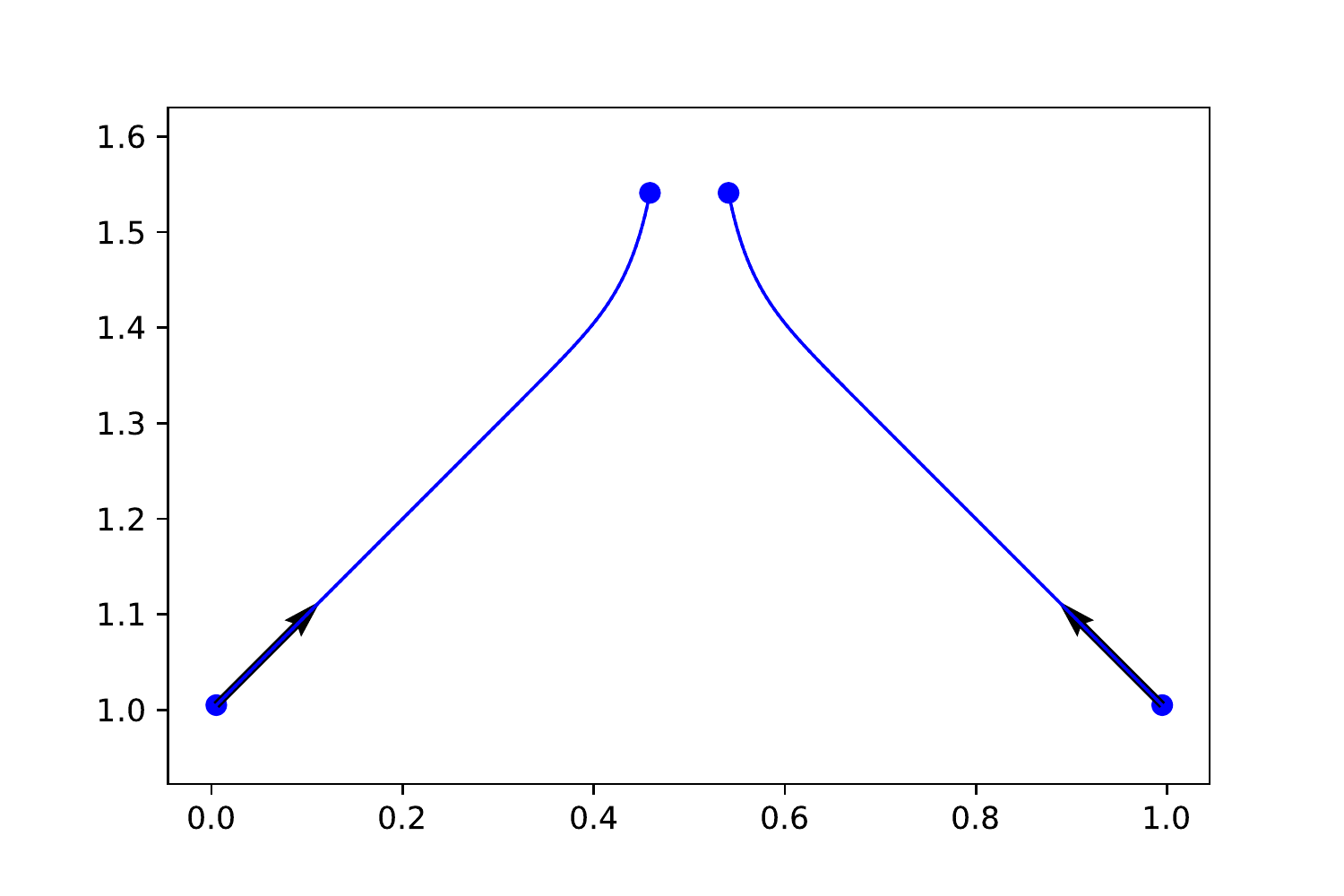}
    \end{minipage}%
    \begin{minipage}{0.5\textwidth}
        \centering
        \includegraphics[scale=0.55,trim = 20 20 30 20,clip]{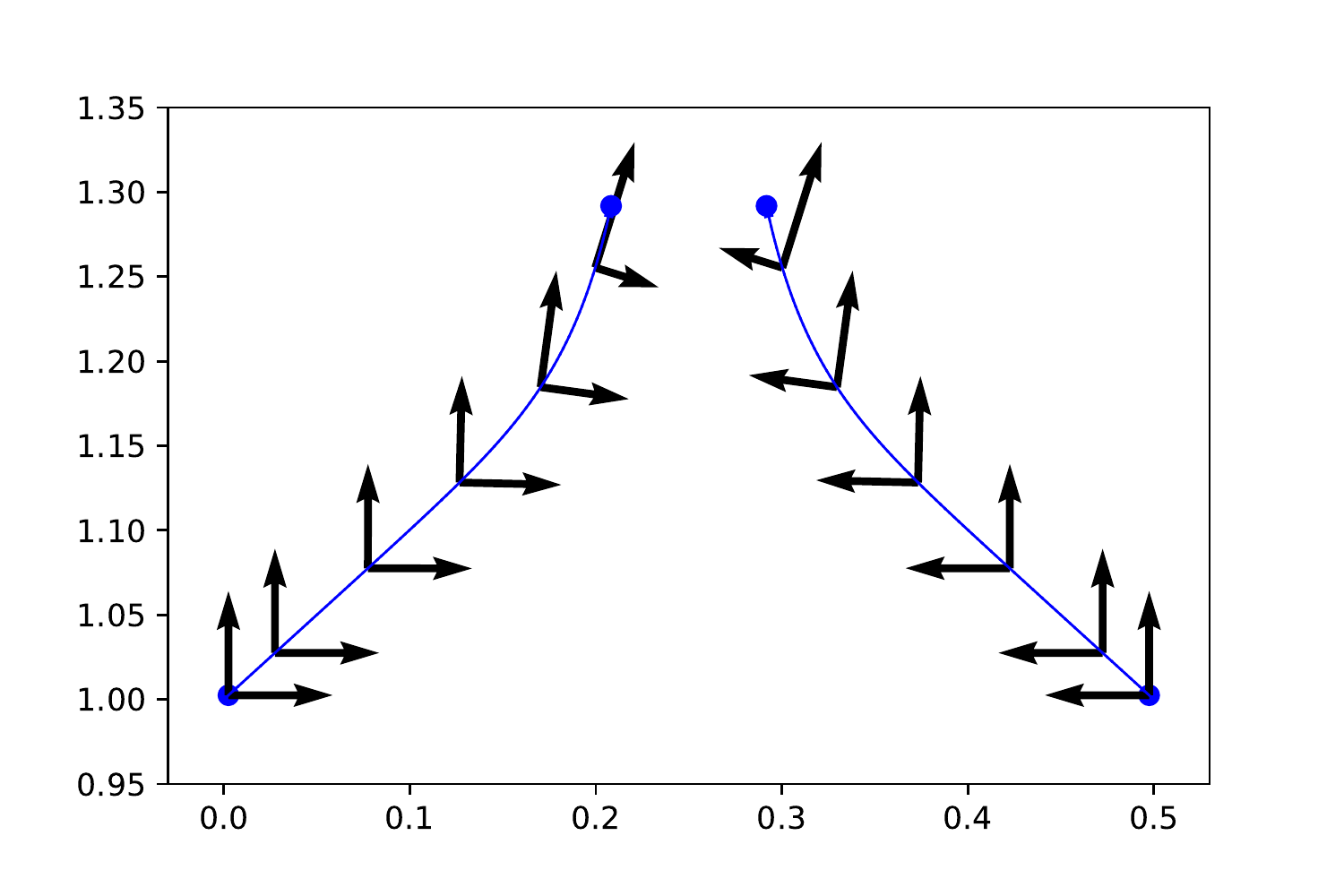}
    \end{minipage}
    \caption{Geodesics on the landmark manifold. (left) Geodesic on $\M$ found with Hamilton's equations. (right) Geodesic on $F\M$ as the solution to Hamilton's equations generated from the sub-Riemannian structure on $F\M$.}
    \label{fig:landgeo}
    \end{center}
\end{figure}

Recall that the Christoffel symbols depend only on the metric, hence they can be obtained by the general equation \eqref{eq:Chris}.
Geodesics on $\M$ can then be obtained as solutions of Hamilton's equations described in section \ref{sec:geoHam} with this landmark Hamiltonian.
An example of geodesics for two landmarks is shown in Figure \ref{fig:landgeo} along with an example of a geodesic on the frame bundle $F\M$, obtained as the solution to Hamilton's equations generated from the sub-Riemannian structure on $F\M$ described in section~\ref{sec:FM}.

\begin{figure}[ht]
    \begin{center}
    \begin{minipage}{0.5\textwidth}
        \centering
        \includegraphics[scale=0.55, trim = 20 20 30 20,clip]{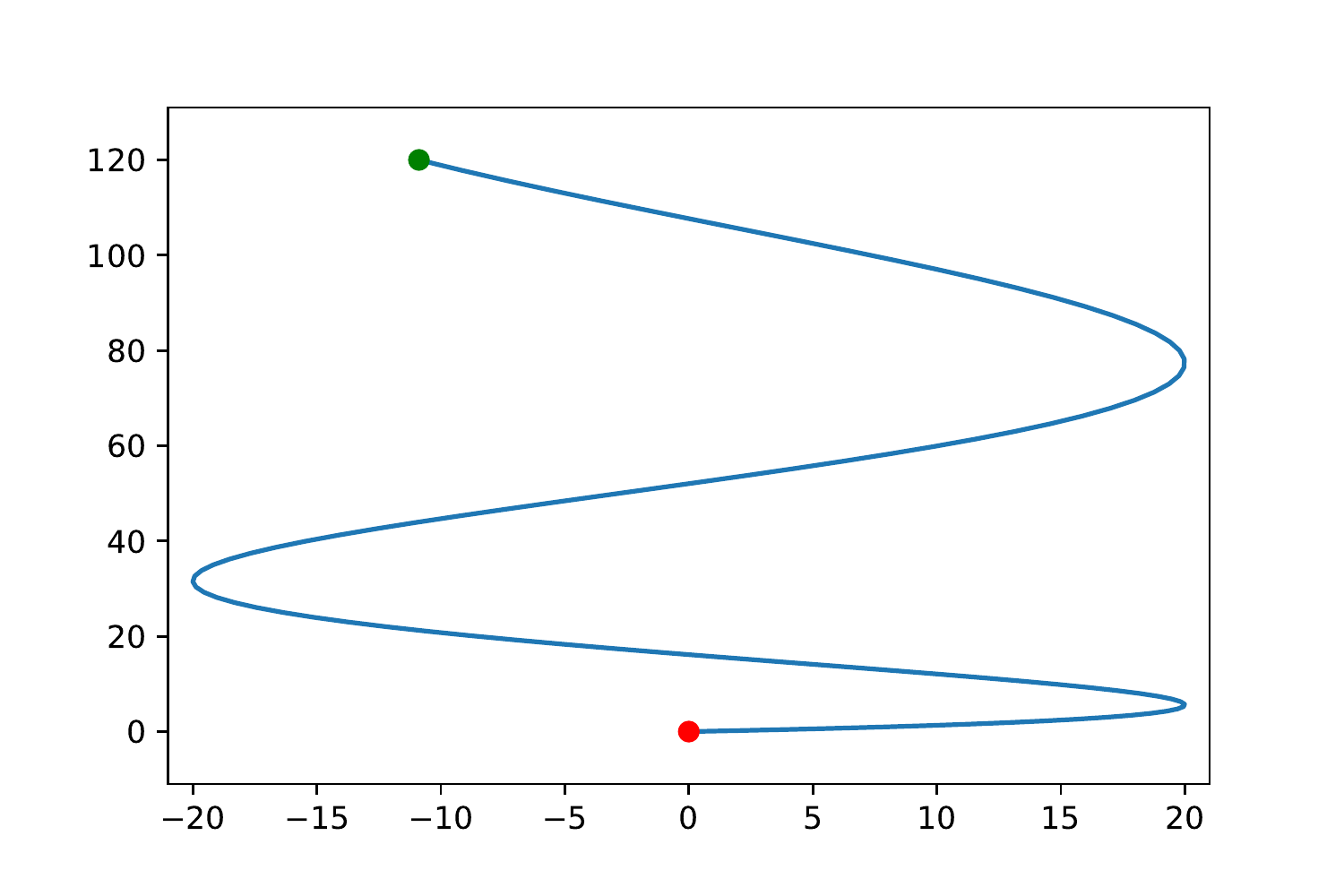}
    \end{minipage}%
    \begin{minipage}{0.5\textwidth}
        \centering
        \includegraphics[scale=0.55,trim = 20 20 30 20,clip]{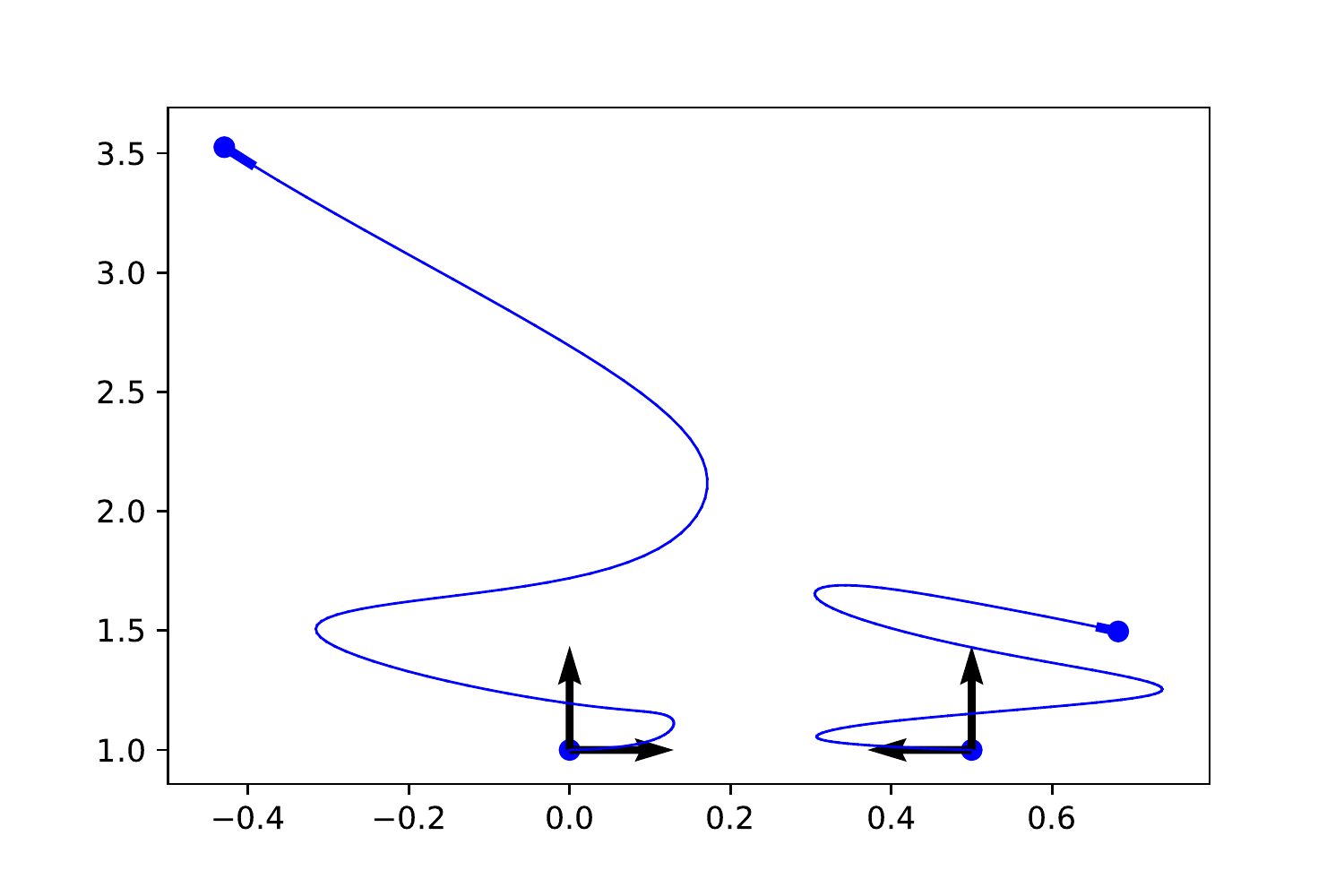}
    \end{minipage}
    \begin{minipage}{0.5\textwidth}
        \centering
        \includegraphics[scale=0.55, trim = 20 20 30 20,clip]{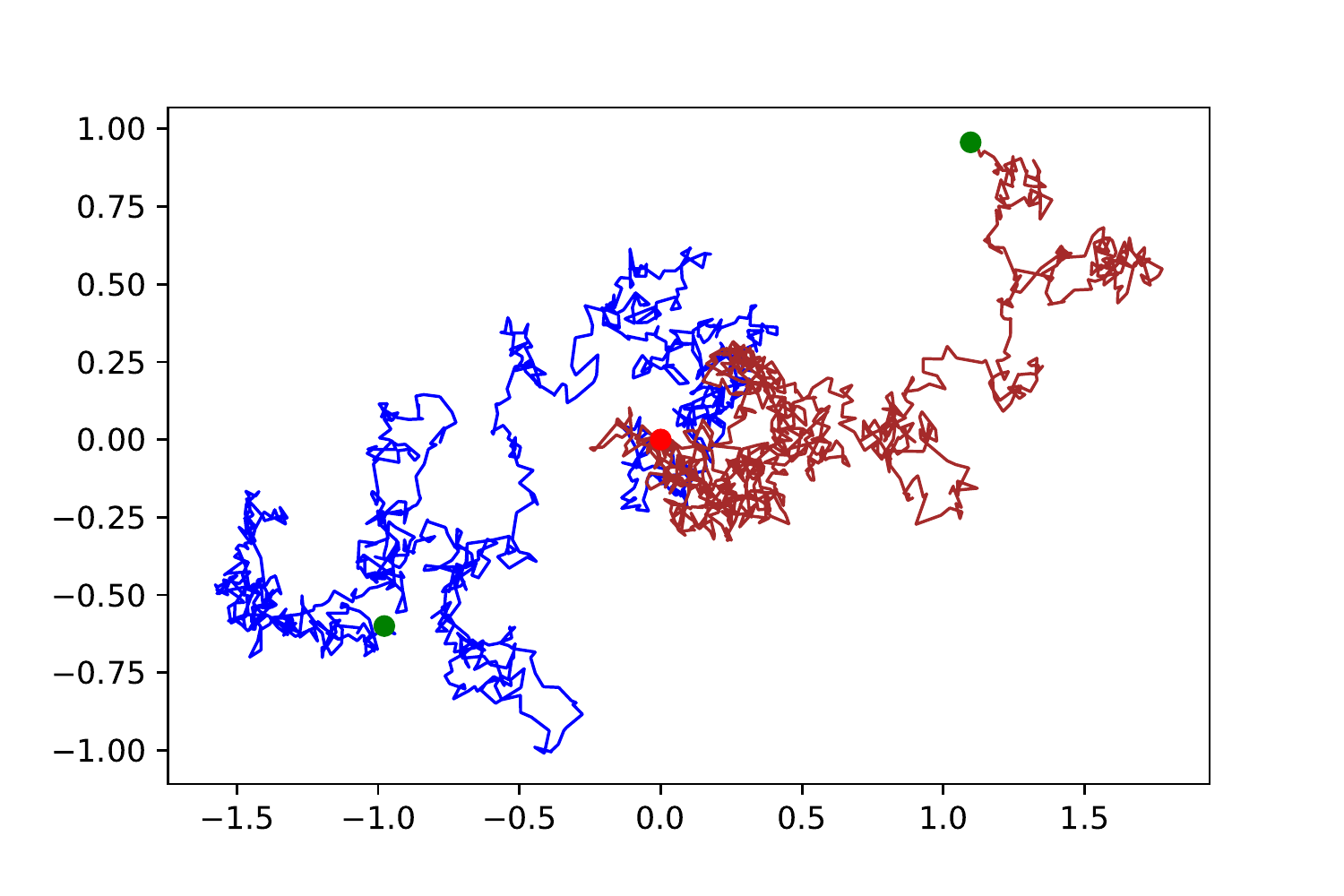}
    \end{minipage}%
    \begin{minipage}{0.5\textwidth}
        \centering
        \includegraphics[scale=0.55,trim = 20 20 30 20,clip]{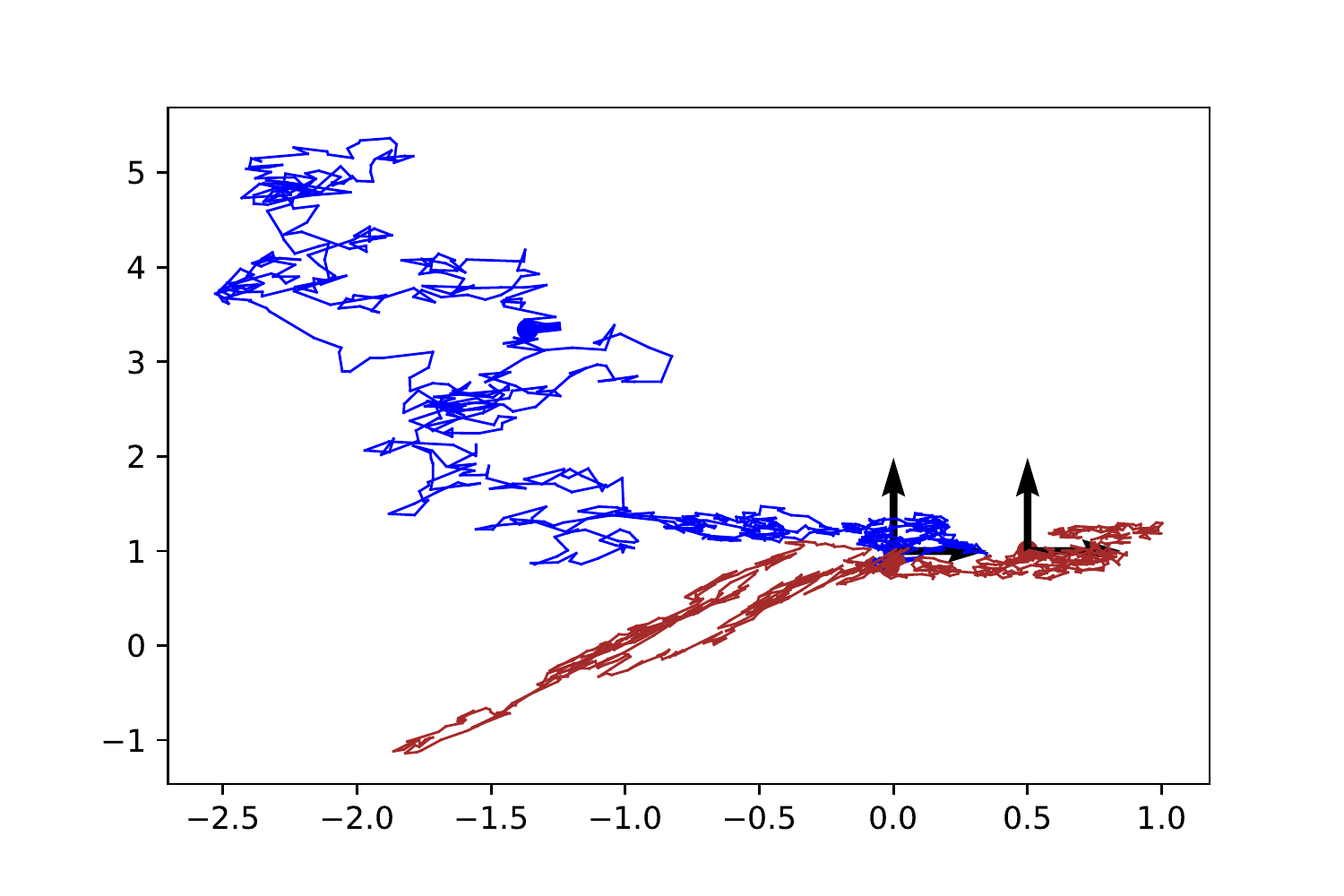}
    \end{minipage}
    \caption{Deterministic development (top left) The curve $\gamma_t$ defined in Example \ref{ex:landdev}. The red and green point denotes the start--and endpoint of the process respectively using the displayed frame. (right) The development of $\gamma_t$ on each landmark. Bottom row: Stochastic development (left) Brownian motion, $W_t$, in $\RR^4$ plotted as two processes. (right) The stochastic development of $W_t$ to the manifold.}
    \label{fig:landdev}
    \end{center}
\end{figure}

\begin{example}[Stochastic Development]
\label{ex:landdev}
We use a two landmarks manifold $\M$, that is $\mathrm{dim}(\M)= 4$. 
Then, as in Example \ref{ex:dev}, we consider the curve $\gamma_t = (20\sin(t),t^2+2t)$, $t\in [0,10]$ (Figure~\ref{fig:landdev} top left panel) and a point $x = (0,1,0.5,1)\in\M$.
The initial frame for each landmark is given as the canonical basis vectors $e_1=(1,0)$, $e_2=(0,1)$ shown in Figure~\ref{fig:landdev} (top right panel) as well as the deterministic development of $\gamma_t$ to $\M$.
Figure~\ref{fig:landdev} (bottom right panel) shows an example of a stochastic development for a $4$-dimensional stochastic process $W_t$ displayed on the bottom left panel.
Notice that in the deterministic case, a single curve was used for both landmarks, thus their trajectories are similar and only affected by the correlation between landmarks. 
In the stochastic case, the landmarks follow different stochastic paths, also affected by the landmarks interaction.
\end{example}

The examples shown in this section can, in addition, be applied to a higher dimensional landmark manifold as seen in figure \ref{fig:landmatch}. 
For more examples on Theano code used with more landmarks on for example the Corpus Callosum shapes, we refer to \cite{kuehnel2017,arnaudon_stochastic_2017}.
For another stochastic deformation of shapes in the context of computational anatomy, with examples on landmarks, we refer to \cite{arnaudon_geometric_2017,arnaudon_stochastic_2017}, where the focus was on noise inference in these models. 
These works were inspired by \cite{holm_variational_2015}, where stochastic models for fluid dynamics were introduced such that geometrical quantities remain preserved, and applied for finite dimensions in \cite{arnaudon_noise_2016}.
In the same theme of stochastic landmark dynamics, \cite{marsland_langevin_2017} introduced noise and dissipation to also tackle noise inference problems. 

\section{Non-Linear Statistics}
\label{sec:non-lin}
This section focuses on a selection of basic statistical concepts generalized to manifolds and how these can be implemented in Theano. 
We refer to \cite{pennec_intrinsic_2006} for an overview of manifold valued statistics.

\subsection{Fr\'echet Mean}

The Fr\'echet Mean is an intrinsic generalization of the mean-value in Euclidean space~\cite{frechet_les_1948}. 
Consider a manifold $\M$ with a distance $d$ and let $P$ be a probability measure on $\M$. 
The Fr\'echet mean set is defined as the set of points minimizing the function
\begin{equation}
    F(y) = \argmin_{x\in\M} \mathbb{E}_P\left[ d(x,y)^2\right], \quad y\in\M\, .
\label{eq:FrechM}
\end{equation}
Unlike the Euclidean mean, the solution to \eqref{eq:FrechM} is not necessarily unique. 
If the minimum exists and is unique, the minimum is called the Fr\'echet mean. 
The Fr\'echet mean for a sample of data points $y_1,\ldots,y_n$ is estimated as
\begin{equation}
    F_{\bar{y}} = \argmin_{x\in\M}\frac{1}{n}\sum_{i=1}^n d(x,y_i)^2\, .
\end{equation}
When considering a Riemannian manifold, a natural choice of distance measure is the geodesic distance described in section \ref{sec:geoeq}. 
With this choice of distance, the empirical Fr\'echet mean reduces to
\begin{equation}
    F_{\bar{y}} = \argmin_{x\in\M}\frac{1}{n}\sum_{i=1}^n\|\text{Log}(x,y_i)\|^2\, ,
\end{equation}
which can be implemented in Theano as follow. 

\begin{lstlisting}
"""
Frechet Mean

Args:
    x: Point on the manifold
    y: Data points
    x0: Initial point for optimization

Returns:
    The average loss from x to data y

"""
def Frechet_mean(x,y):
    (cout,updates) = theano.scan(fn=loss, non_sequences=[v0,x], 
                                    sequences=[y], n_steps=n_samples)
    return 1./n_samples*T.sum(cout)
dFrechet_mean = lambda x,y: T.grad(Frechet_mean(x,y),x)
FMean = minimize(Frechet_mean, x0, jac=dFrechet_mean, args=y)
\end{lstlisting}

\begin{figure}[ht]
    \begin{center}
    \begin{minipage}{0.5\textwidth}
        \centering
        \includegraphics[scale=0.5, trim = 100 40 60 70,clip]{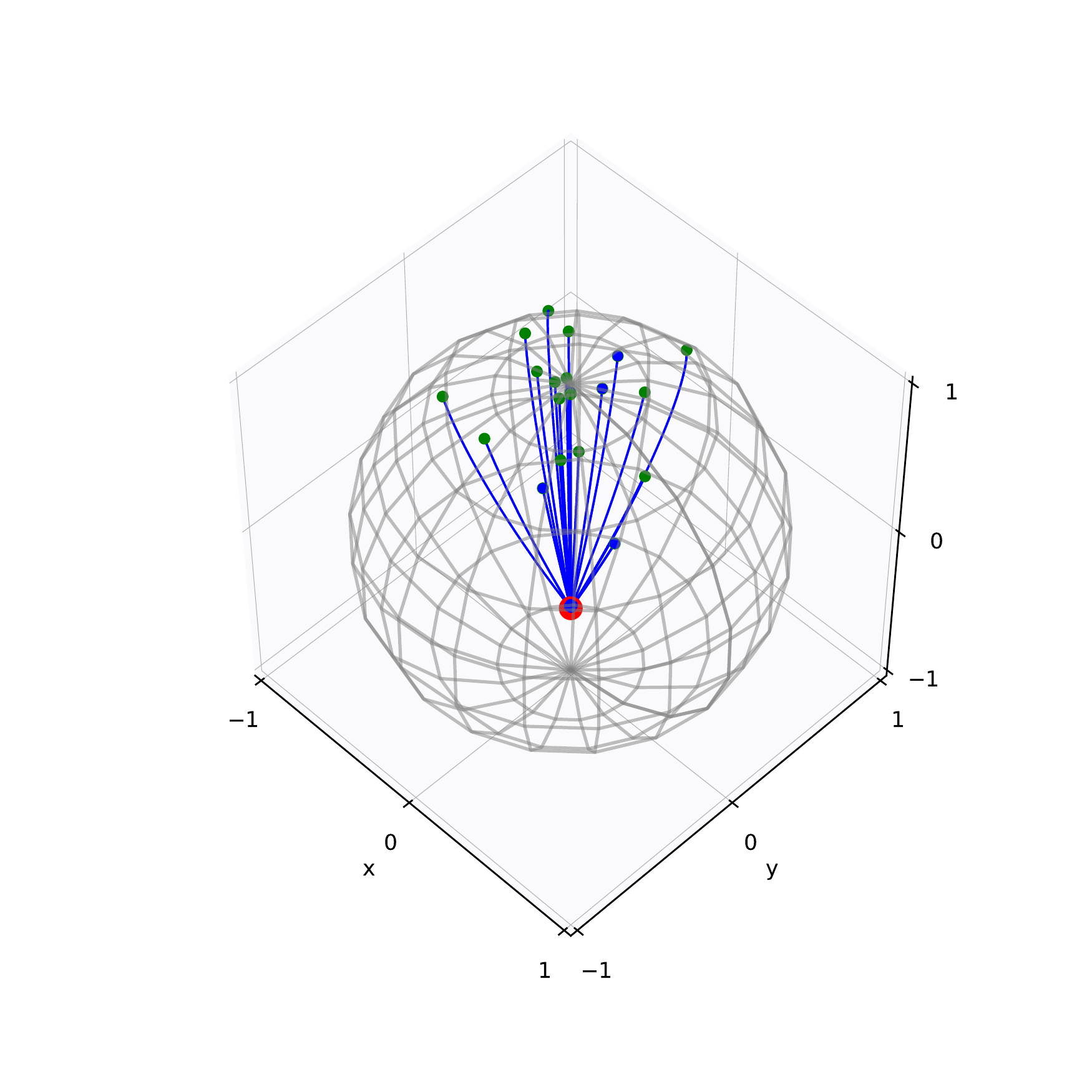}
    \end{minipage}%
    \begin{minipage}{0.5\textwidth}
        \centering
        \includegraphics[scale=0.5,trim = 100 40 60 70,clip]{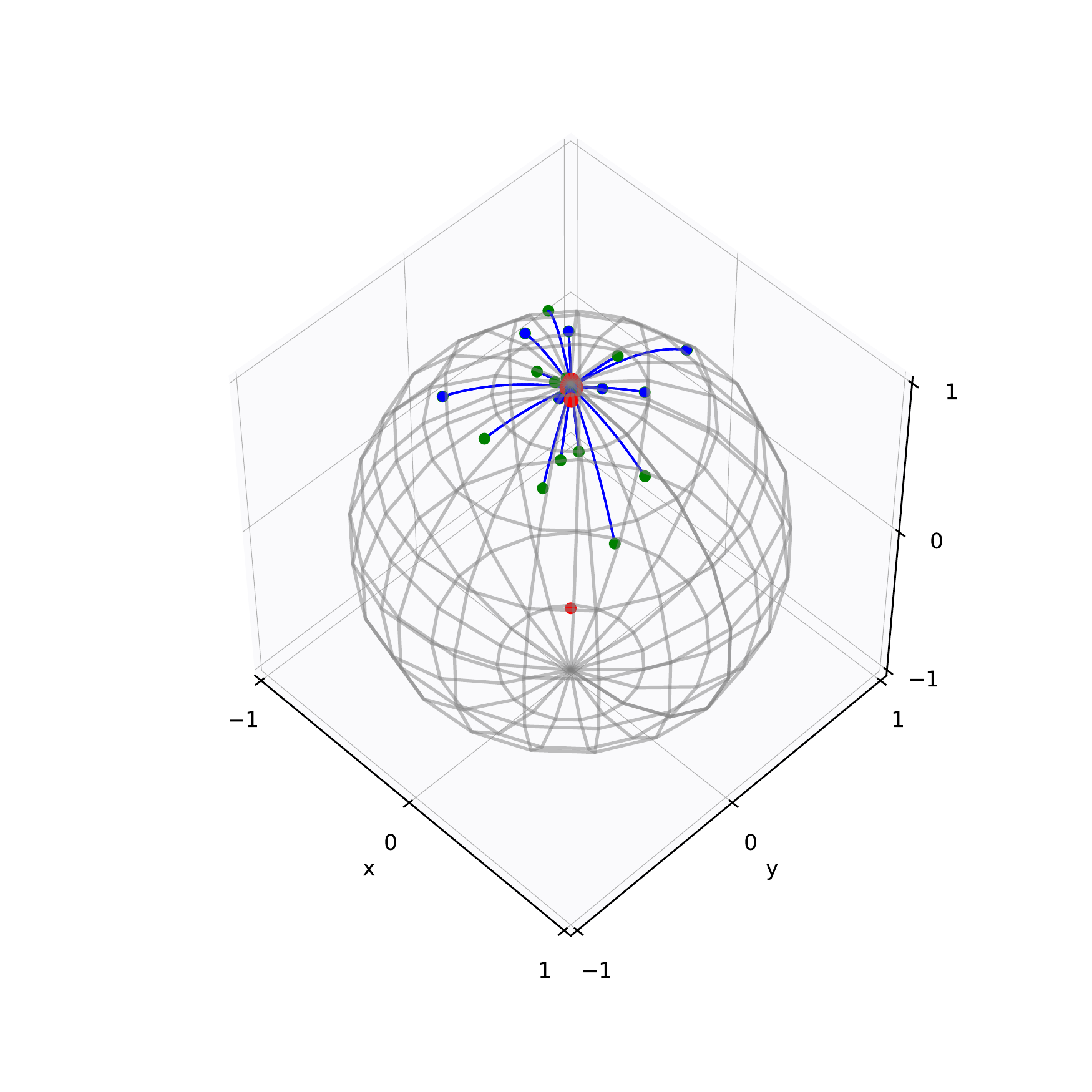}
    \end{minipage}
    \caption{(left) Sampled data points, with the red point being the initial guess of the mean. (right) The resulting empirical Frechet mean. The iterated results are visualized as red dots. The final result is the largest red dot with the distance minimizing geodesics to each datapoint.}
    \label{fig:Frechet}
    \end{center}
\end{figure}

\begin{example}[Fr\'echet mean on $S^2$]
\label{ex:frechet}
    Consider the Levi-Civita connection on $S^2$ and equip $S^2$ with the geodesic distance given in \eqref{eq:distgeo}. 
A sample set of size $20$ is generated on the northern hemisphere. Each coordinate of a sample point has been drawn from a normal distribution with mean $0$ and standard deviation $0.2$. 
The initial guess of the Fr\'echet mean is $F(0.4,-0.4)$. 
The sample set and initial mean are shown in the left plot of Figure \ref{fig:Frechet}. 
The resulting empirical Frechet mean found with the implementation above is visualized in Figure \ref{fig:Frechet}.
\end{example}

The Fr\'echet mean can not just be used to calculate the mean on manifolds. 
In~\cite{sommer_modelling_2017}, the authors presented a method for estimating the mean and covariance of normal distributions on manifolds by calculating the Fr\'echet mean on the frame bundle. The next section will describe a way to generalize normal distributions to manifolds.

\subsection{Normal Distributions}
\label{sec:normdist}

\begin{figure}[ht]
    \begin{center}
    \begin{minipage}{0.5\textwidth}
        \centering
        \includegraphics[scale=0.5, trim = 20 20 30 20,clip]{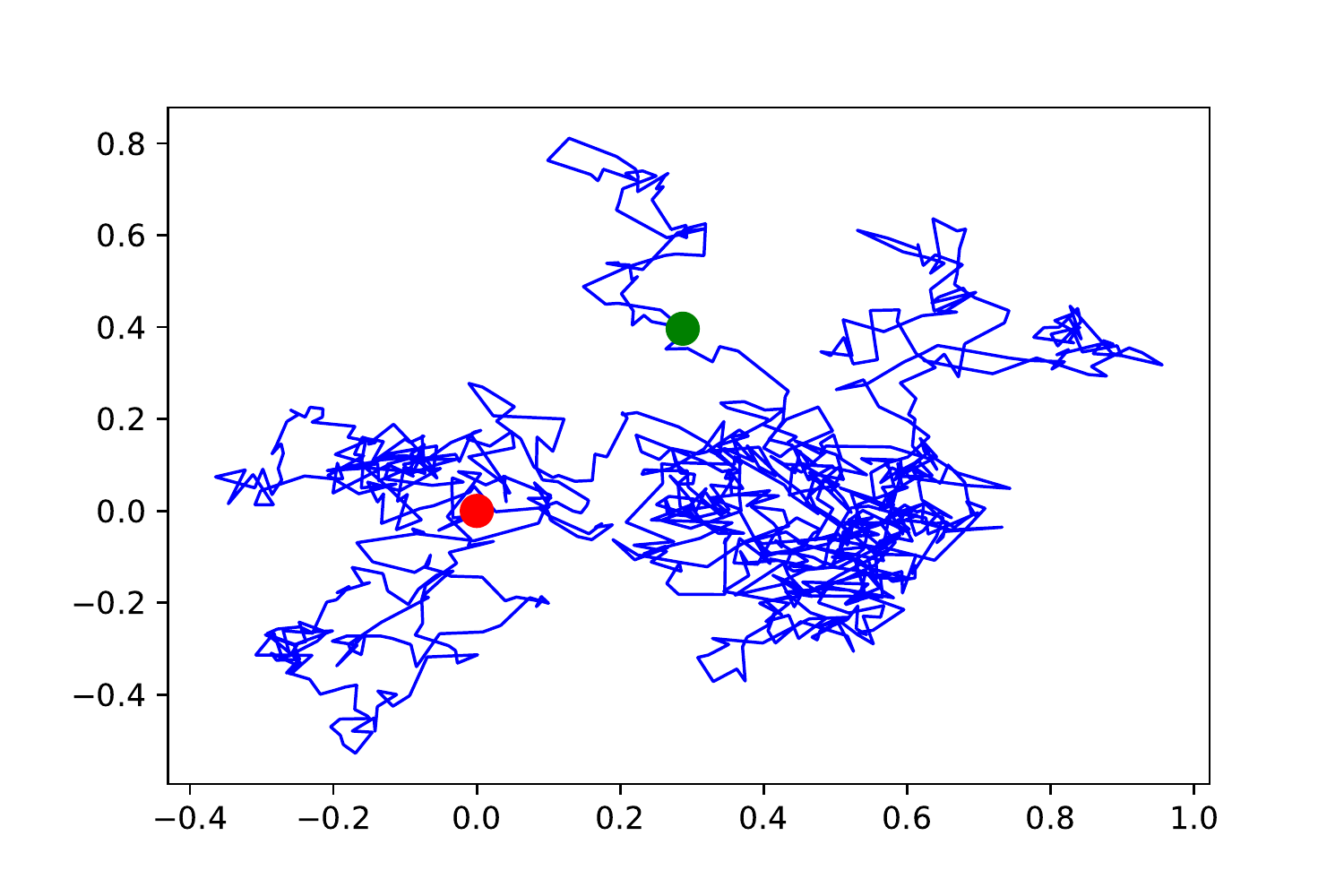}
    \end{minipage}%
    \begin{minipage}{0.5\textwidth}
        \centering
        \includegraphics[scale=0.35,trim = 100 40 60 70,clip]{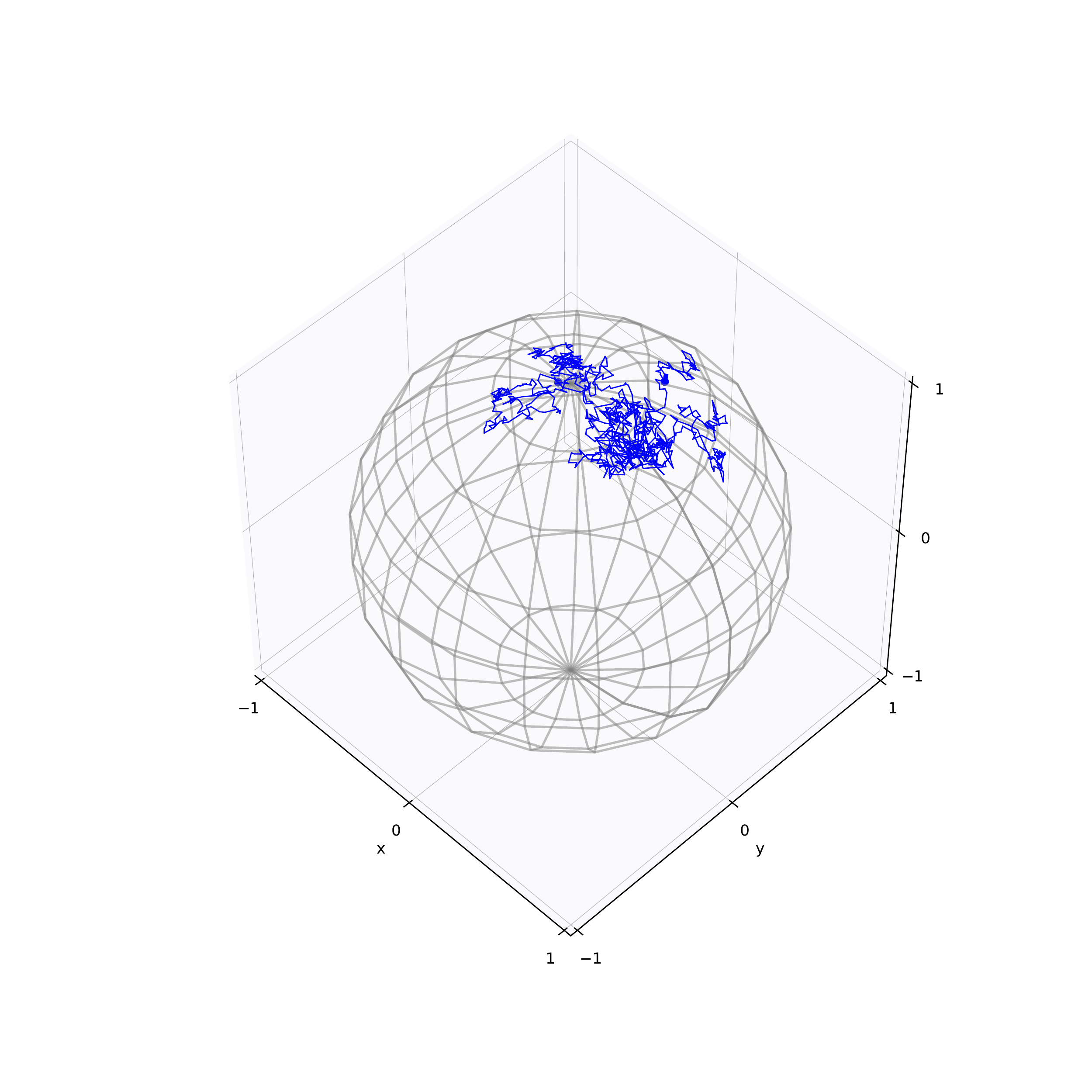}
    \end{minipage}
    \caption{(left) Brownian motion, $W_t$, in $\RR^2$. (right) The stochastic development of $W_t$ to the sphere with initial point $u=(x,\nu)$, for $x=F(0,0)$ and $\nu$ the frame consisting of the canonical basis vectors $e_1,e_2$.}
    \label{fig:Brown}
    \end{center}
\end{figure}

 Normal distributions in Euclidean spaces can be considered as the transition distribution of Brownian motions. 
The generalization of normal distributions to manifolds can be defined in a similar manner. 
In~\cite{elworthy_geometric_1988}, isotropic Brownian motions on $\M$ are constructed as the stochastic development of isotropic Brownian motions on $\RR^n$ based on an orthonormal frame. However, \cite{sommer_anisotropically_2016,sommer_modelling_2017} suggested performing stochastic development with non-orthonormal frames, which leads to anisotropic Brownian motions on $\M$. 
Let $W_t$ be a Brownian motion on $\RR^2$ and consider the initial point $u = (x,\nu)\in FS^2$, for $x=F(0,0)$ and $\nu$ the frame consisting of the canonical basis vectors $e_1,e_2$. 
An example of a Brownian motion path on the sphere, derived as the stochastic development of $W_t$ in $\RR^2$, is shown in Figure \ref{fig:Brown}.

 Based on the definition of Brownian motions on a manifold, normal distributions can be generalized as the transition distribution of Brownian motions on $\M$. 
Consider the generalization of the normal distribution $\mathcal{N}(\mu,\Sigma)$. 
When defining the normal distribution on $\M$ as the stochastic development of Brownian motions, the initial point on $\M$ is the mean and the initial frame represents the covariance of the resulting normal distribution.

\begin{example}[Normal distributions on $S^2$]
\label{ex:norm_dist}
 Let $W_t$ be a Brownian motion on $\RR^2$ and consider $x=F(0,0)\in S^2$ being the mean of the normal distributions in this example. 
Two normal distributions with different covariance matrices have been generated, one isotropic and one anisotropic distribution. 
The normal distributions are $\mathcal{N}(0,\Sigma_i)$ for $i=1,2$ with covariance matrices
\begin{align}
\label{eq:cov}
    \Sigma_1 = \begin{pmatrix}
0.15 & 0 \\
0 & 0.15
\end{pmatrix}, \quad \Sigma_2 = \begin{pmatrix}
0.2 & 0.1 \\
0.1 & 0.1
\end{pmatrix}\, . 
\end{align}

As explained above, the initial frame $\nu$ represents the covariance of the normal distribution on a manifold $\M$. Therefore, we chose $\nu_1$ with basis vectors being the columns of $\Sigma_1$ and $\nu_2$ with basis vectors represented by the columns of $\Sigma_2$. 
Density plots of the resulting normal distributions are shown in Figure \ref{fig:norm_dist}.

\begin{figure}[ht]
    \begin{center}
    \begin{minipage}{0.5\textwidth}
        \centering
        \includegraphics[scale=0.5, trim = 130 30 90 30,clip]{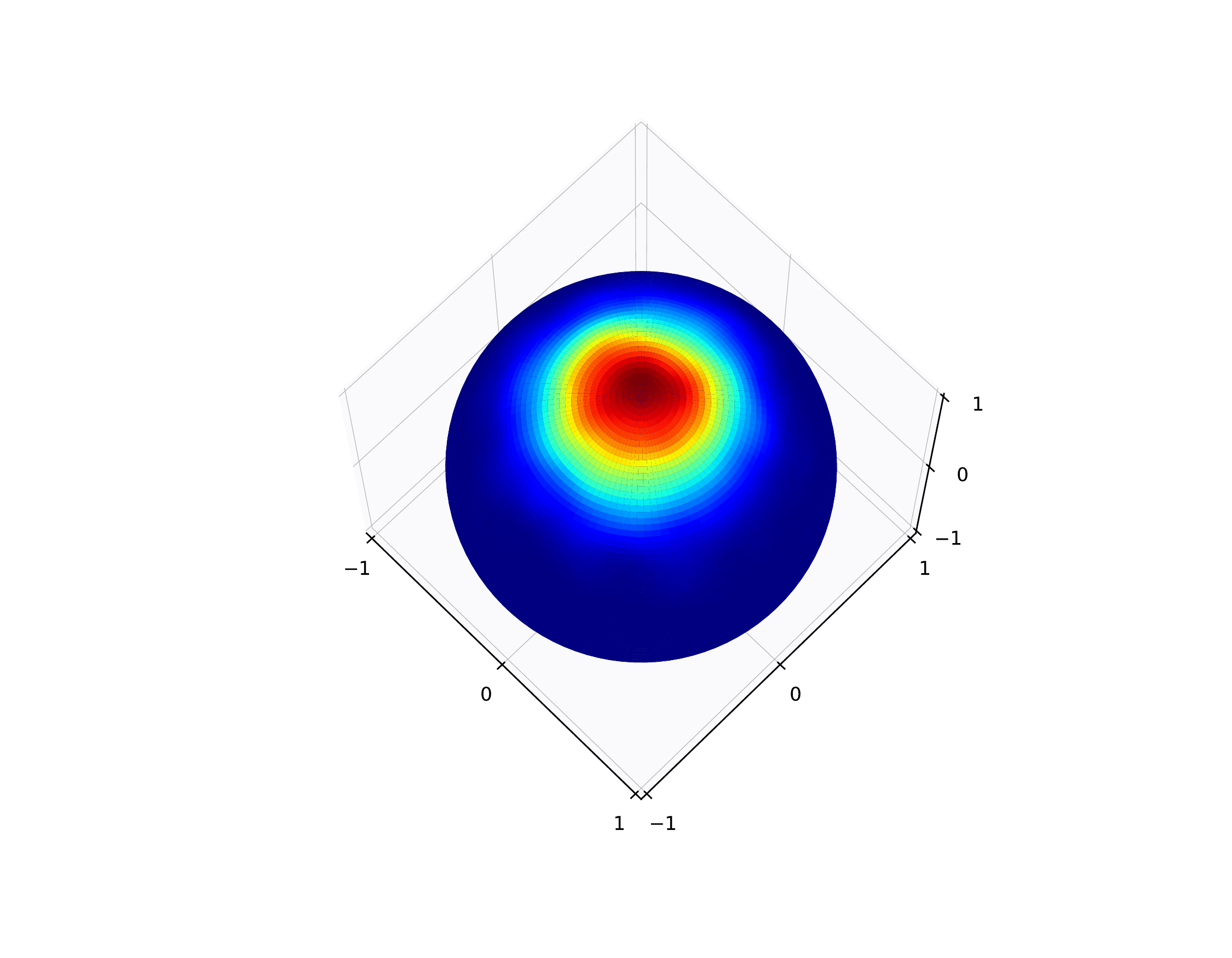}
    \end{minipage}%
    \begin{minipage}{0.5\textwidth}
        \centering
        \includegraphics[scale=0.5,trim = 130 30 90 30,clip]{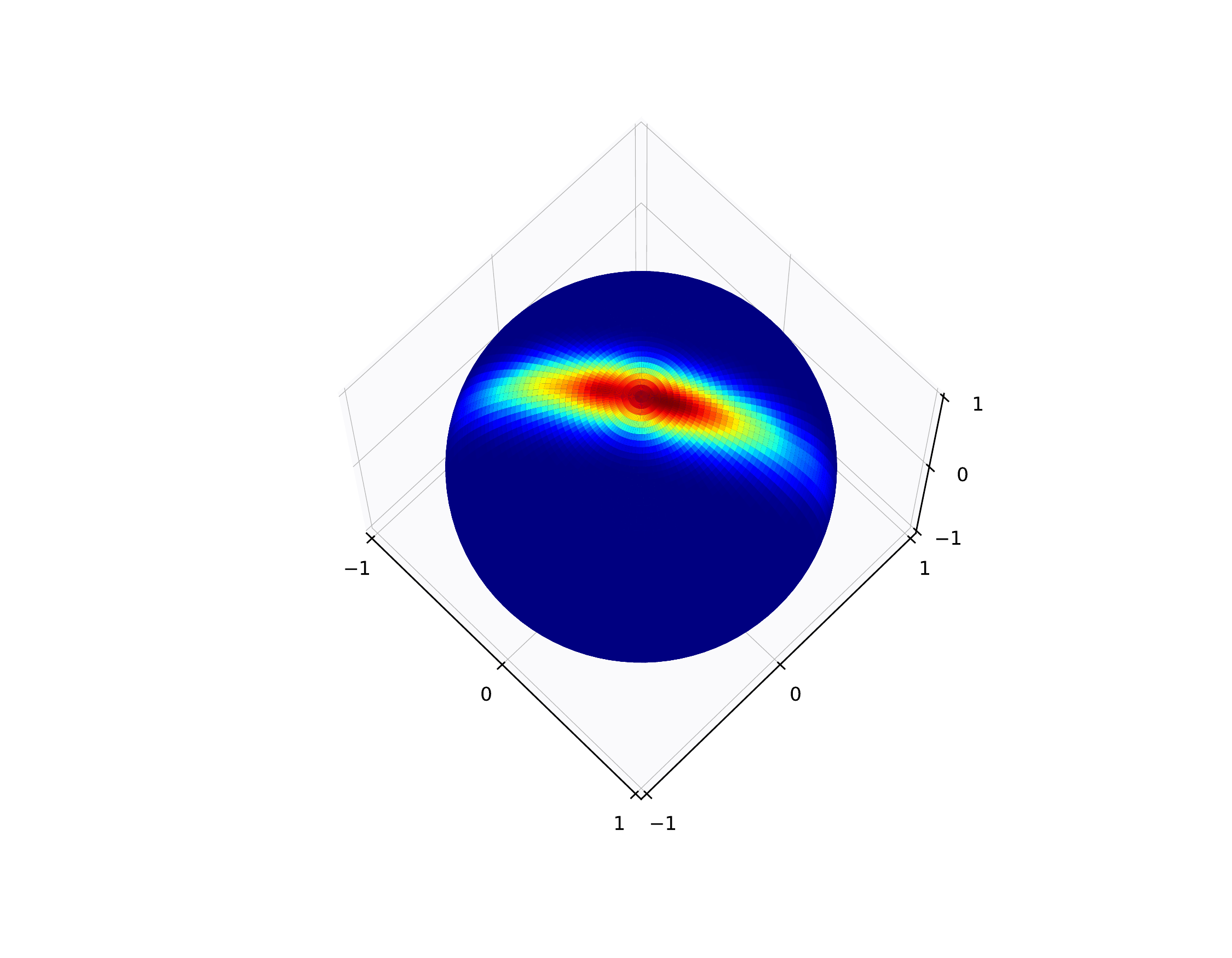}
    \end{minipage}
    \caption{(left) Density estimate of the isotropic normal distribution on $S^2$ with covariance $\Sigma_1$ given in \eqref{eq:cov}. (right) Density estimate of the anisotropic normal distribution on $S^2$ with covariance $\Sigma_2$.}
    \label{fig:norm_dist}
    \end{center}
\end{figure}

\end{example}  

\section{Conclusion}
\label{sec:con}

In this paper, we have shown how the Theano framework and Python can be used for implementation of concepts from differential geometry and non-linear statistics.  
The opportunity to perform symbolic calculations makes implementations of even complex concepts such as stochastic integration and fibre bundle geometry easy and concise. 
The symbolic representation is often of great practical value for the implementation process, leading to shorter code, fewer bugs, and faster implementations, and formulas can almost directly be translated to Theano code. 
As seen in the examples, the symbolic representation of functions allows taking derivatives of any variables and of any order. 
The task of calculating gradients for optimization procedures can be difficult and prone to errors while with symbolic calculations, only a few lines of code is needed to optimize over, for instance, the parameters of a stochastic integrator or the evolution of a sub-Riemannian geodesic.
This makes numerical testing of new ideas fast and efficient and easily scalable to useful applications if optimized for parallel computers of GPUs. 

We have just shown here a small fragment of mathematical problems which can be implemented with Theano and other similar software.
Other problems that could be solved using these methods can be found in statistical analysis on manifold-valued data, such as geodesic regression, longitudinal analysis, and PCA, or in computational anatomy, by solving registration problem on continuous shapes and images and analysing or modelling shape deformations. 
For example, we refer to \cite{kuehnel2017,arnaudon_stochastic_2017,arnaudon_geometric_2017} for further examples of Theano in the field of computational anatomy which were not treated here. 

Packages such as Theano have their limitations, and one must sometimes be careful in the implementation and aware of the limitations of the algorithms. 
For example, if equations are simple enough that derivatives can be written explicitly, the code can in some situations be faster when computing from the explicit formula rather than relying on the automatic differentiation.
For complicated constructions, the compilation step can be computationally intensive as well as memory demanding. 
Such limitations can be overcome by carefully writing the code in order to limit the compilation time and have the parameters of Theano properly adjusted to the machine at hand. 

With this paper and its accompanying code\footnote{\url{http://bitbucket.org/stefansommer/theanogeometry}}, we hope to stimulate the use of modern symbolic and numerical computation frameworks for experimental applications in mathematics, for computations in applied mathematics, and for data analysis by showing how the resulting code allows for flexibility and simplicity in implementing many experimental mathematics endeavours. 

\subsubsection*{Acknowledgements}
{\small 
LK and SS are supported by the CSGB Centre for Stochastic Geometry and Advanced Bioimaging funded by a grant from the Villum Foundation.
AA is supported by the EPSRC through award EP/N014529/1 funding the EPSRC Centre for Mathematics of Precision Healthcare.
}

\bibliographystyle{alpha}

\begin{thebibliography}{BMB{\etalchar{+}}01}

\bibitem[A{\etalchar{+}}16]{45381}
M.~Abadi et~al.
\newblock Tensorflow: A system for large-scale machine learning.
\newblock In {\em 12th USENIX Symposium on Operating Systems Design and
  Implementation (OSDI 16)}, pages 265--283, 2016.

\bibitem[ACC14]{arnaudon2014stochastic}
M.~Arnaudon, X.~Chen, and A.~B. Cruzeiro.
\newblock Stochastic euler-poincar{\'e} reduction.
\newblock {\em Journal of Mathematical Physics}, 55(8):081507, 2014.

\bibitem[ACH17]{arnaudon_noise_2016}
A.~Arnaudon, A.~L. Castro, and D.~D. Holm.
\newblock Noise and dissipation on coadjoint orbits.
\newblock {\em arXiv:1601.02249, To appear in JNLS}, 2017.

\bibitem[AHPS17]{arnaudon_stochastic_2017}
A.~Arnaudon, D.~D. Holm, A.~Pai, and S.~Sommer.
\newblock A {Stochastic} {Large} {Deformation} {Model} for {Computational}
  {Anatomy}.
\newblock In {\em Information {Processing} in {Medical} {Imaging}}, Lecture
  {Notes} in {Computer} {Science}, pages 571--582. Springer, 2017.

\bibitem[AHS17]{arnaudon_geometric_2017}
A.~Arnaudon, D.~D. Holm, and S.~Sommer.
\newblock A {Geometric} {Framework} for {Stochastic} {Shape} {Analysis}.
\newblock {\em submitted, arXiv:1703.09971 [cs, math]}, March 2017.

\bibitem[BEKS17]{bezanson_julia:_2017}
J.~Bezanson, A.~Edelman, S.~Karpinski, and V.~Shah.
\newblock Julia: {A} {Fresh} {Approach} to {Numerical} {Computing}.
\newblock {\em SIAM Review}, 59(1):65--98, January 2017.

\bibitem[Blo]{bloch2003nonholonomic}
A.~Bloch.
\newblock {\em Nonholonomic mechanics and control}, volume~24.
\newblock Springer.

\bibitem[BMB{\etalchar{+}}01]{blowey2013theory}
C.~Bernardi, Y.~Maday, J.~F. Blowey, J.~P. Coleman, and A.~W. Craig.
\newblock {\em Theory and numerics of differential equations: Durham 2000}.
\newblock Universitext. Springer-Verlag Berlin Heidelberg, 1 edition, 2001.

\bibitem[BMTY05]{beg2005computing}
M.~F. Beg, M.~I. Miller, A.~Trouv{\'e}, and L.~Younes.
\newblock Computing large deformation metric mappings via geodesic flows of
  diffeomorphisms.
\newblock {\em International journal of computer vision}, 61(2):139--157, 2005.

\bibitem[Chi09]{chirikjian2009stochastic}
G~Chirikjian.
\newblock {\em Stochastic Models, Information Theory, and Lie Groups, Volume 1:
  Classical Results and Geometric Methods. Applied and Numerical Harmonic
  Analysis}.
\newblock Birkh{\"a}user, 2009.

\bibitem[Chi11]{chirikjian2011stochastic}
G.~S Chirikjian.
\newblock {\em Stochastic Models, Information Theory, and Lie Groups, Volume 2:
  Analytic Methods and Modern Applications}, volume~2.
\newblock Springer Science \& Business Media, 2011.

\bibitem[CHR]{cruzeiro2017momentum}
A.~B. Cruzeiro, D.~D Holm, and T.~S Ratiu.
\newblock Momentum maps and stochastic clebsch action principles.
\newblock {\em Communications in Mathematical Physics}, pages 1--40.

\bibitem[Elw88]{elworthy_geometric_1988}
D.~Elworthy.
\newblock Geometric aspects of diffusions on manifolds.
\newblock In Paul-Louis Hennequin, editor, {\em {\'E}cole d'{\'E}t{\'e} de
  {Probabilit{\'e}s} de {Saint}-{Flour} {XV}--{XVII}, 1985--87}, number 1362 in
  Lecture {Notes} in {Mathematics}, pages 277--425. Springer Berlin Heidelberg,
  1988.

\bibitem[FK82]{fujita_onsager-machlup_1982}
T.~Fujita and S.-I. Kotani.
\newblock The {Onsager}-{Machlup} function for diffusion processes.
\newblock {\em Journal of Mathematics of Kyoto University}, 22(1):115--130,
  1982.

\bibitem[Fr{\'e}48]{frechet_les_1948}
M.~Fr{\'e}chet.
\newblock Les {\'e}l{\'e}ments al{\'e}atoires de nature quelconque dans un
  espace distancie.
\newblock {\em Ann. Inst. H. Poincar{\'e}}, 10:215--310, 1948.

\bibitem[Hol15]{holm_variational_2015}
D.~D. Holm.
\newblock Variational principles for stochastic fluid dynamics.
\newblock {\em Proc. Mathematical, Physical, and Engineering Sciences / The
  Royal Society}, 471(2176), April 2015.

\bibitem[Hsu02]{hsu_stochastic_2002}
E.~P. Hsu.
\newblock {\em Stochastic {Analysis} on {Manifolds}}.
\newblock American Mathematical Soc., 2002.

\bibitem[KS17]{kuehnel2017}
L.~K{\"u}hnel and S.~Sommer.
\newblock {\em Computational Anatomy in Theano}, pages 164--176.
\newblock Springer International Publishing, 2017.

\bibitem[KSM93]{kolar_natural_1993}
I.~Kol{\'a}{\v{r}}, J.~Slov{\'a}k, and P.~W. Michor.
\newblock {\em Natural {Operations} in {Differential} {Geometry}}.
\newblock Springer Berlin Heidelberg, Berlin, Heidelberg, 1993.

\bibitem[Lee03]{lee_introduction_2003}
J.~M Lee.
\newblock {\em Introduction to smooth manifolds}, volume 218 of {\em Graduate
  {Texts} in {Mathematics}}.
\newblock Springer-Verlag, New York, 2003.

\bibitem[Lee06]{lee2006riemannian}
J.~M. Lee.
\newblock {\em Riemannian manifolds: an introduction to curvature}, volume 176.
\newblock Springer Science \& Business Media, 2006.

\bibitem[Lia04]{liao_levy_2004}
M.~Liao.
\newblock {\em L{\'e}vy processes in {Lie} groups}.
\newblock Cambridge University Press, Cambridge; New York, 2004.

\bibitem[Mok78]{mok_differential_1978}
K.-P. Mok.
\newblock On the differential geometry of frame bundles of {Riemannian}
  manifolds.
\newblock {\em Journal F{\"u}r Die Reine Und Angewandte Mathematik},
  1978(302):16--31, 1978.

\bibitem[MR99]{marsden_introduction_1999}
J.~E. Marsden and T.~S. Ratiu.
\newblock {\em Introduction to {Mechanics} and {Symmetry}}, volume~17 of {\em
  Texts in {Applied} {Mathematics}}.
\newblock Springer New York, New York, NY, 1999.

\bibitem[MS17]{marsland_langevin_2017}
S.~Marsland and T.~Shardlow.
\newblock Langevin {Equations} for {Landmark} {Image} {Registration} with
  {Uncertainty}.
\newblock {\em SIAM Journal on Imaging Sciences}, 10(2):782--807, January 2017.

\bibitem[Pen06]{pennec_intrinsic_2006}
X.~Pennec.
\newblock Intrinsic {Statistics} on {Riemannian} {Manifolds}: {Basic} {Tools}
  for {Geometric} {Measurements}.
\newblock {\em J. Math. Imaging Vis.}, 25(1):127--154, 2006.

\bibitem[Sch10]{schaffter2010numerical}
T.~Schaffter.
\newblock Numerical integration of sdes: a short tutorial.
\newblock Technical report, 2010.

\bibitem[Som15]{sommer_anisotropic_2015}
S.~Sommer.
\newblock Anisotropic {Distributions} on {Manifolds}: {Template} {Estimation}
  and {Most} {Probable} {Paths}.
\newblock In {\em Information {Processing} in {Medical} {Imaging}}, volume 9123
  of {\em Lecture {Notes} in {Computer} {Science}}, pages 193--204. Springer,
  2015.

\bibitem[Som16]{sommer_anisotropically_2016}
S.~Sommer.
\newblock Anisotropically {Weighted} and {Nonholonomically} {Constrained}
  {Evolutions} on {Manifolds}.
\newblock {\em Entropy}, 18(12):425, November 2016.

\bibitem[SS17]{sommer_modelling_2017}
S.~Sommer and A.~M. Svane.
\newblock Modelling anisotropic covariance using stochastic development and
  sub-{Riemannian} frame bundle geometry.
\newblock {\em Journal of Geometric Mechanics}, 9(3):391--410, June 2017.

\bibitem[Str86]{strichartz_sub-riemannian_1986}
R.~S. Strichartz.
\newblock Sub-{Riemannian} geometry.
\newblock {\em Journal of Differential Geometry}, 24(2):221--263, 1986.

\bibitem[{The}16]{theano}
{Theano Development Team}.
\newblock {Theano: A {Python} framework for fast computation of mathematical
  expressions}.
\newblock {\em arXiv e-prints}, abs/1605.02688, 2016.

\bibitem[You10]{younes_shapes_2010}
L.~Younes.
\newblock {\em Shapes and {Diffeomorphisms}}.
\newblock Springer, 2010.

\end{thebibliography}
\newcommand{\etalchar}[1]{$^{#1}$}

\appendix

\section{Stochastic integration}
\label{sec:stoc}

In the following, we will give a brief description of some basic theory on stochastic differential equations and stochastic integration methods. The symbolic specification in Theano allows us to take derivatives of parameters specifying the stochastic evolutions, and the presented methods can, therefore, be used for e.g. maximum likelihood estimation over stochastic processes. The theory in this appendix is based on~\cite{schaffter2010numerical}. 

\subsection{Stochastic Differential Equations}

We consider here stochastic processes, $U_t$ in $\RR^n$, solutions to SDEs of the form
\begin{equation}
    dU_t = f(U_t,t)dt + g(U_t,t)dW_t, \quad t\in [0,T]\, ,
\label{eq:SDE}
\end{equation}
with drift $f(U_t,t)$ and diffusion field $g(U_t,t)$, functions from $\RR^n\times\RR$ to $\RR^n$.

There are two types of stochastic differential equations; It\^o and Stratonovich differential equations. The Stratonovich SDEs are usually denoted with $\circ$, such that \eqref{eq:SDE} reduces to
\begin{equation}
    dU_t = f(U_t,t)dt + g(U_t,t)\circ dW_t\, .
\end{equation}

For integration of deterministic ODEs, solutions to the integral equation can be defined as the limit of a sum of finite differences over the time interval. In this case, it does not matter in which point of the intervals the function is evaluated. For stochastic integrals, this is not the case. It\^o integrals are defined by evaluating at the left point of the interval, while Stratonovich integrals use the average between the value at the two endpoints of the interval. The two integrals do not result in equal solutions, but they are related by
\begin{equation}
    g(U_t,t) dW_t = \frac{1}{2}dg(U_t,t) g(U_t,dt)dt + g(U_t,t)\circ dW_t\, ,
\label{eq:cor}
\end{equation}
where $dg$ denotes the Jacobian of $g$~\cite{blowey2013theory}. Whether to choose It\^o or the Stratonovich framework depends on the problem to solve. One benefit from choosing the Stratonovich integral is that it obeys the chain rule making it easy to use in a geometric context.

\subsection{Discrete Stochastic Integrators}
\label{sec:stocnum}
We generally need numerical integration to find solutions to SDEs.
There are several versions of numerical integrators of different order of convergence. Two simple integrators are the Euler method for It\^o SDEs and the Euler-Heun for the Stratonovich SDEs.

\noindent\textbf{Euler Method.} Consider an It\^o SDE as defined in \eqref{eq:SDE}. Let $0 = t_0 < t_1 < \ldots < t_n = T$ be a discretization of the interval $[0,T]$ for which the stochastic process is defined and assume $\Delta t = T/n$. Initialize the stochastic process, $U_0 = u_0$ for some initial value $u_0$. The process $U_t$ is then recursively defined for each time point $t_i$ by,
\begin{equation}
    U_{t_{i+1}} = U_{t_i} + f(U_{t_i},t_i)\Delta t + g(U_{t_i},t_i)\Delta W_i\, ,
\end{equation}
in which $\Delta W_i = W_{t_{i+1}} - W_{t_i}$. Given an It\^o stochastic differential equation, \lstinline!sde_f!, the Euler method can be implemented in Theano by the following code example.

\begin{lstlisting}
"""
Euler Numerical Integration Method

Args:
    sde: Stochastic differential equation to solve
    integrator: Choice of integrator_ito or integrator_stratonovich
    x: Initial values for process
    dWt: Steps of stochastic process
    *ys: Additional arguments to define the sde

Returns:
    integrate_sde: Tensor (t,xt)
                        t: Time evolution
                        xt: Evolution of x
"""
def integrator_ito(sde_f):
    def euler(dW,t,x,*ys):
        (detx, stox, X, *dys) = sde_f(dW,t,x,*ys)
        ys_new = ()
        for (y,dy) in zip(ys,dys):
            ys_new = ys_new + (y+dt*dy,)
        return (t+dt,x + dt*detx + stox, *ys_new)
    return euler


# Integration:
def integrate_sde(sde,integrator,x,dWt,*ys):
    (cout, updates) = theano.scan(fn=integrator(sde),
            outputs_info=[T.constant(0.),x, *ys],
            sequences=[dWt],
            n_steps=n_steps)
    return cout
\end{lstlisting}

\noindent\textbf{Euler-Heun Method.} An equivalent integration method as the Euler method for It\^o\, SDEs, is the Euler-Heun method used to approximate the solution to Stratonovich SDEs. Consider a similar discretization as in the Euler method. The Euler-Heun numerical integration method is then defined as,
\begin{equation}
    U_{t_{i+1}} = U_{t_i} + f(U_{t_i},t_i)\Delta t + \frac{1}{2}\left(g(U_{t_i},t_i) + g(\hat{U}_{t_i},t_i)\right)\Delta W_i\, ,
\end{equation}
where $\hat{U}_{t_i} = U_{t_i} + g(U_{t_i},t_i)\Delta W_i$. The implementation of the Euler-Heun method is similar to the Euler method, such that based on a Stratonovich SDE, \lstinline!sde_f!, the implementation can be executed as follows,

\begin{lstlisting}
"""
Euler-Heun Numerical Integration Method

Args:
    sde: Stochastic differential equation to solve
    integrator: Choice of integrator_ito or integrator_stratonovich
    x: Initial values for process
    dWt: Steps of stochastic process
    *ys: Additional arguments to define the sde

Returns:
    integrate_sde: Tensor (t,xt)
                   t: Time evolution
                   xt: Evolution of x    
"""
def integrator_stratonovich(sde_f):
    def euler_heun(dW,t,x,*ys):
        (detx, stox, X, *dys) = sde_f(dW,t,x,*ys)
        tx = x + stox
        ys_new = ()
        for (y,dy) in zip(ys,dys):
            ys_new = ys_new + (y+dt*dy,)
        return (t+dt,
                 x + dt*detx + 0.5*(stox + sde_f(dW,t+dt,tx,*ys)[1]),
                 *ys_new)
    return euler_heun



# Integration:
def integrate_sde(sde,integrator,x,dWt,*ys):
    (cout, updates) = theano.scan(fn=integrator(sde),
            outputs_info=[T.constant(0.),x, *ys],
            sequences=[dWt],
            n_steps=n_steps)
    return cout
\end{lstlisting}

\end{document}